\documentclass[%
 reprint,
superscriptaddress,
nofootinbib,
 amsmath,amssymb,
 aps,
 prd,
]{revtex4-2}

\usepackage{graphicx}
\usepackage{dcolumn}
\usepackage{bm}
\usepackage{xcolor}
\usepackage{enumitem}

\usepackage[colorlinks=true,citecolor=blue,linkcolor=blue,urlcolor=blue, backref=false,pdfborder={0 0 0}]{hyperref}
\usepackage[nameinlink]{cleveref}
\usepackage{import}

\def\apj{ApJ}
\def\apjl{ApJL}
\def\apjs{ApJS}
\def\aj{AJ}

\def\mnras{MNRAS}
\def\physrep{PhysRep}
\def\prl{PhysRevLett}

\def\aap{{A\&A}}
\def\prd{PRD}
\def\jcap{JCAP}

\newcommand{\beq}{\begin{equation}}
\newcommand{\eeq}{\end{equation}}

\newcommand{\del}{\partial}

\newcommand{\HEALPIX}{{\texttt{Healpix}}}
\newcommand{\metacal}{{\texttt{Metacalibration}~}}
\newcommand{\decals}{{\texttt{DECaLS}}~}

\usepackage{xspace}

\newcommand{\namaster}{\texttt{NaMaster}~}

\newcommand{\Mpc}{\mathrm{Mpc}}

\newcommand{\photoz}{photo-$z$}
\newcommand{\bq}{\boldsymbol q}
\newcommand{\bx}{\boldsymbol x}

\newcommand{\bk}{\boldsymbol k}
\newcommand{\mo}{\mathcal{O}}

\newcommand{\cdk}{\ensuremath{C^{\delta_{g, \rm obs}^{i}, \gamma_{E}^{j}}}}

\newcommand{\edit}[1]{\textcolor{black}{#1}}

\usepackage{lineno}

\begin{document}


\title{Not all lensing is low: An analysis of DESI$\times$DES \\ using the Lagrangian Effective Theory of LSS}

\author{S.~Chen}
\thanks{Both authors contributed equally to this work.\\
        \url{sfschen@ias.edu} \\
        \url{jderose@lbl.gov}
        }
\affiliation{Institute for Advanced Study, 1 Einstein Drive, Princeton, NJ 08540, USA}

\author{J.~DeRose}
\thanks{Both authors contributed equally to this work.\\
        \url{sfschen@ias.edu} \\
        \url{jderose@lbl.gov}
        }
\affiliation{Lawrence Berkeley National Laboratory, 1 Cyclotron Road, Berkeley, CA 94720, USA}

\author{R.~Zhou}
\affiliation{Lawrence Berkeley National Laboratory, 1 Cyclotron Road, Berkeley, CA 94720, USA}

\author{M.~White}
\affiliation{Department of Physics, University of California, Berkeley, 366 LeConte Hall MC 7300, Berkeley, CA 94720-7300, USA}
\affiliation{University of California, Berkeley, 110 Sproul Hall \#5800 Berkeley, CA 94720, USA}

\author{S.~Ferraro}
\affiliation{Lawrence Berkeley National Laboratory, 1 Cyclotron Road, Berkeley, CA 94720, USA}
\affiliation{University of California, Berkeley, 110 Sproul Hall \#5800 Berkeley, CA 94720, USA}

\author{C.~Blake}
\affiliation{Centre for Astrophysics \& Supercomputing, Swinburne University of Technology, P.O. Box 218, Hawthorn, VIC 3122, Australia}

\author{J.~U.~Lange}
\affiliation{Department of Physics, University of Michigan, Ann Arbor, MI 48109, USA}
\affiliation{University of Michigan, Ann Arbor, MI 48109, USA}

\author{R.~H.~Wechsler}
\affiliation{Kavli Institute for Particle Astrophysics and Cosmology, Stanford University, Menlo Park, CA 94305, USA}
\affiliation{Physics Department, Stanford University, Stanford, CA 93405, USA}
\affiliation{SLAC National Accelerator Laboratory, Menlo Park, CA 94305, USA}

\author{J.~Aguilar}
\affiliation{Lawrence Berkeley National Laboratory, 1 Cyclotron Road, Berkeley, CA 94720, USA}

\author{S.~Ahlen}
\affiliation{Physics Dept., Boston University, 590 Commonwealth Avenue, Boston, MA 02215, USA}

\author{D.~Brooks}
\affiliation{Department of Physics \& Astronomy, University College London, Gower Street, London, WC1E 6BT, UK}

\author{T.~Claybaugh}
\affiliation{Lawrence Berkeley National Laboratory, 1 Cyclotron Road, Berkeley, CA 94720, USA}

\author{K.~Dawson}
\affiliation{Department of Physics and Astronomy, The University of Utah, 115 South 1400 East, Salt Lake City, UT 84112, USA}

\author{A.~de la Macorra}
\affiliation{Instituto de F\'{\i}sica, Universidad Nacional Aut\'{o}noma de M\'{e}xico,  Cd. de M\'{e}xico  C.P. 04510,  M\'{e}xico}

\author{P.~Doel}
\affiliation{Department of Physics \& Astronomy, University College London, Gower Street, London, WC1E 6BT, UK}

\author{A.~Font-Ribera}
\affiliation{Department of Physics \& Astronomy, University College London, Gower Street, London, WC1E 6BT, UK}
\affiliation{Institut de F\'{i}sica d’Altes Energies (IFAE), The Barcelona Institute of Science and Technology, Campus UAB, 08193 Bellaterra Barcelona, Spain}

\author{E.~Gaztañaga}
\affiliation{Institut d'Estudis Espacials de Catalunya (IEEC), 08034 Barcelona, Spain}
\affiliation{Institute of Cosmology and Gravitation, University of Portsmouth, Dennis Sciama Building, Portsmouth, PO1 3FX, UK}
\affiliation{Institute of Space Sciences, ICE-CSIC, Campus UAB, Carrer de Can Magrans s/n, 08913 Bellaterra, Barcelona, Spain}

\author{S.~Gontcho A Gontcho}
\affiliation{Lawrence Berkeley National Laboratory, 1 Cyclotron Road, Berkeley, CA 94720, USA}

\author{G.~Gutierrez}
\affiliation{Fermi National Accelerator Laboratory, PO Box 500, Batavia, IL 60510, USA}

\author{K.~Honscheid}
\affiliation{Center for Cosmology and AstroParticle Physics, The Ohio State University, 191 West Woodruff Avenue, Columbus, OH 43210, USA}
\affiliation{Department of Physics, The Ohio State University, 191 West Woodruff Avenue, Columbus, OH 43210, USA}
\affiliation{The Ohio State University, Columbus, 43210 OH, USA}

\author{C.~Howlett}
\affiliation{School of Mathematics and Physics, University of Queensland, 4072, Australia}

\author{R.~Kehoe}
\affiliation{Department of Physics, Southern Methodist University, 3215 Daniel Avenue, Dallas, TX 75275, USA}

\author{D.~Kirkby}
\affiliation{Department of Physics and Astronomy, University of California, Irvine, 92697, USA}

\author{T.~Kisner}
\affiliation{Lawrence Berkeley National Laboratory, 1 Cyclotron Road, Berkeley, CA 94720, USA}

\author{A.~Kremin}
\affiliation{Lawrence Berkeley National Laboratory, 1 Cyclotron Road, Berkeley, CA 94720, USA}

\author{M.~Landriau}
\affiliation{Lawrence Berkeley National Laboratory, 1 Cyclotron Road, Berkeley, CA 94720, USA}

\author{L.~Le~Guillou}
\affiliation{Sorbonne Universit\'{e}, CNRS/IN2P3, Laboratoire de Physique Nucl\'{e}aire et de Hautes Energies (LPNHE), FR-75005 Paris, France}

\author{M.~Manera}
\affiliation{Departament de F\'{i}sica, Serra H\'{u}nter, Universitat Aut\`{o}noma de Barcelona, 08193 Bellaterra (Barcelona), Spain}
\affiliation{Institut de F\'{i}sica d’Altes Energies (IFAE), The Barcelona Institute of Science and Technology, Campus UAB, 08193 Bellaterra Barcelona, Spain}

\author{A.~Meisner}
\affiliation{NSF NOIRLab, 950 N. Cherry Ave., Tucson, AZ 85719, USA}

\author{R.~Miquel}
\affiliation{Instituci\'{o} Catalana de Recerca i Estudis Avan\c{c}ats, Passeig de Llu\'{\i}s Companys, 23, 08010 Barcelona, Spain}
\affiliation{Institut de F\'{i}sica d’Altes Energies (IFAE), The Barcelona Institute of Science and Technology, Campus UAB, 08193 Bellaterra Barcelona, Spain}

\author{J.~ A.~Newman}
\affiliation{Department of Physics \& Astronomy and Pittsburgh Particle Physics, Astrophysics, and Cosmology Center (PITT PACC), University of Pittsburgh, 3941 O'Hara Street, Pittsburgh, PA 15260, USA}

\author{G.~Niz}
\affiliation{Departamento de F\'{i}sica, Universidad de Guanajuato - DCI, C.P. 37150, Leon, Guanajuato, M\'{e}xico}
\affiliation{Instituto Avanzado de Cosmolog\'{\i}a A.~C., San Marcos 11 - Atenas 202. Magdalena Contreras, 10720. Ciudad de M\'{e}xico, M\'{e}xico}

\author{N.~Palanque-Delabrouille}
\affiliation{IRFU, CEA, Universit\'{e} Paris-Saclay, F-91191 Gif-sur-Yvette, France}
\affiliation{Lawrence Berkeley National Laboratory, 1 Cyclotron Road, Berkeley, CA 94720, USA}

\author{W.~J.~Percival}
\affiliation{Department of Physics and Astronomy, University of Waterloo, 200 University Ave W, Waterloo, ON N2L 3G1, Canada}
\affiliation{Perimeter Institute for Theoretical Physics, 31 Caroline St. North, Waterloo, ON N2L 2Y5, Canada}
\affiliation{Waterloo Centre for Astrophysics, University of Waterloo, 200 University Ave W, Waterloo, ON N2L 3G1, Canada}

\author{F.~Prada}
\affiliation{Instituto de Astrof\'{i}sica de Andaluc\'{i}a (CSIC), Glorieta de la Astronom\'{i}a, s/n, E-18008 Granada, Spain}

\author{G.~Rossi}
\affiliation{Department of Physics and Astronomy, Sejong University, Seoul, 143-747, Korea}

\author{E.~Sanchez}
\affiliation{CIEMAT, Avenida Complutense 40, E-28040 Madrid, Spain}

\author{D.~Schlegel}
\affiliation{Lawrence Berkeley National Laboratory, 1 Cyclotron Road, Berkeley, CA 94720, USA}

\author{M.~Schubnell}
\affiliation{Department of Physics, University of Michigan, Ann Arbor, MI 48109, USA}
\affiliation{University of Michigan, Ann Arbor, MI 48109, USA}

\author{D.~Sprayberry}
\affiliation{NSF NOIRLab, 950 N. Cherry Ave., Tucson, AZ 85719, USA}

\author{G.~Tarl\'{e}}
\affiliation{University of Michigan, Ann Arbor, MI 48109, USA}

\author{B.~A.~Weaver}
\affiliation{NSF NOIRLab, 950 N. Cherry Ave., Tucson, AZ 85719, USA}


\date{\today}             

\begin{abstract}
In this work we use Lagrangian perturbation theory to analyze the harmonic space galaxy clustering signal of Bright Galaxy Survey (BGS) and Luminous Red Galaxies (LRGs) targeted by the Dark Energy Spectroscopic Instrument (DESI), combined with the galaxy--galaxy lensing signal measured around these galaxies using Dark Energy Survey Year 3 source galaxies. The BGS and LRG galaxies are extremely well characterized by DESI spectroscopy and, as a result, lens galaxy redshift uncertainty and photometric systematics contribute negligibly to the error budget of our ``$2\times2$-point'' analysis. On the modeling side, this work represents the first application of the \texttt{spinosaurus} code, implementing an effective field theory model for galaxy intrinsic alignments, and we additionally introduce a new scheme (\texttt{MAIAR}) for marginalizing over the large uncertainties in the redshift evolution of the intrinsic alignment signal. Furthermore, this is the first application of a hybrid effective field theory (HEFT) model for galaxy bias based on the $\texttt{Aemulus}\, \nu$ simulations. Our main result is a measurement of the amplitude of the lensing signal, $S_8=\sigma_8 \left(\Omega_m/0.3\right)^{0.5} = 0.850^{+0.042}_{-0.050}$, consistent with values of this parameter derived from the primary CMB. This constraint is artificially improved by a factor of $51\%$ if we assume a more standard, but restrictive parameterization for the redshift evolution and sample dependence of the intrinsic alignment signal, and $63\%$ if we additionally assume the nonlinear alignment model. We show that when fixing the cosmological model to the best-fit values from Planck PR4 there is $> 5 \sigma$ evidence for a deviation of the evolution of the intrinsic alignment signal from the functional form that is usually assumed in cosmic shear and galaxy--galaxy lensing studies.
\end{abstract}

\maketitle

\tableofcontents

\section{Introduction}

The weak lensing of photons by the gravity of intervening matter is one of the premier probes of the large scale structure of the universe. Since the lensing deflection is a consequence of general relativity given the cosmological distribution of matter, weak lensing in principle provides one of the few direct measurements of matter clustering on these scales. The amplitude of the lensing signal, frequently expressed in terms of the compressed parameter $S_8 = \sigma_8 (\Omega_m / 0.3)^{0.5}$, allows us to test the the standard $\Lambda$CDM model of cosmology---and its extensions---which tie the large-scale structure of the universe to the primordial fluctuations measured in the CMB as well as the expansion history of the universe.

Perhaps the most well-established method of measuring the weak lensing signal is through the distortion of galaxy shapes due to the deflection of photons by foreground matter. These deflections lead to changes in the ellipticities of the source galaxies correlated on large scales known as galaxy weak lensing. The galaxy lensing signal is in addition correlated with the clustering of foreground, or lens, galaxies which serve as biased tracers of the lensing matter. Combining the auto- and cross-correlations of lensing and galaxy clustering substantially increases the total signal to noise and, as a result, so-called ``$3\times2$-point'' analyses utilizing this full set of correlations have become a standard in the literature \cite{Kuijken:2015, Abbott:2017, desy3, kids1000, Sugiyama:2023}.

The current generation of galaxy lensing surveys like the Dark Energy Survey (DES) \cite{desy3}, the Kilo Degree Survey \cite{kids1000} and Hyper Suprime Cam (HSC) \cite{Miyatake2022} are able to constrain the lensing amplitude down to the few-percent level. Intriguingly, these constraints have tended to be not only comparable in precision to the value of $S_8$ inferred from Planck satellite measurements of the cosmic microwave background \cite{Aghanim:2018eyx} but also lower at the roughly $2\sigma$ level. This ``$S_8$ tension'' has also been observed in the cross-correlation of galaxy clustering and the weak lensing of the CMB \cite{white2022cosmological,Sailer24}, though higher values more consistent with the CMB, especially through using the auto-spectrum of CMB lensing, have also been measured \cite{Carron22,Madhavacheril2023,Farren24}. This tension also manifests itself on smaller scales, where it is often referred to as the ``lensing is low'' problem, and where interpretations are more degenerate with complex galaxy formation physics \cite{Leauthaud2016,Lange2021,Amon2022,Chavez-Montero2023,Contreras2023}. As a robust detection of this tension would signal a deviation of the growth of structure away from the predictions of $\Lambda$CDM and the need for physics beyond the standard model, it is critical to examine all steps of the modeling from first principles.

In this paper we focus on refining one particular aspect---the dynamical modeling---of standard galaxy--galaxy lensing (GGL) analyses. While recent years have seen significant advances in the perturbative modeling of galaxy clustering \cite{Ivanov:2019pdj,DAmico:2019fhj,Chen22a}, particularly in re-formulating perturbation theory and galaxy biasing as effective theories, these techniques have not yet become the norm in galaxy lensing analyses. In this work we will in particular explore the application of Lagrangian perturbation theory (LPT) and Hybrid Effective Field Theory (HEFT), its extension using dark matter dynamics from simulations, to model galaxy galaxy-lensing measurements \cite{Vlah15,Vlah_2016,modichenwhite19,Chen_2020,Chen_2021,Kokron_2021,Chen22b,DeRose2023,Chen24c}. In parallel, significant advances have also been made in emulating the predictions of N-body simulations of dark matter, removing the need for approximate schemes based for example on the halo model when constraining matter clustering through lensing. This work is the first application, along with \cite{Sailer24}, of state-of-the-art emulators based on the \texttt{Aemulus} $\nu$ simulations, which accurately interpolate between a broad set of $w$CDM and massive neutrino cosmologies, both to predict matter clustering directly and galaxy clustering through HEFT \cite{DeRose2023}. In future work we may extend this emulator to $w_0w_a$CDM models given the potential preference for this model by recent DESI BAO data \cite{DESI2024.VI.KP7A}, although see also \cite{Chen2024} which shows a significantly decreased preference for non-cosmological constant dark energy when analyzing these data alongside BOSS two- and three-point functions, CMB lensing, and Type Ia supernovae.

In addition to matter and galaxy densities, a particularly relevant aspect of our dynamical model will be the perturbative treatment of the \textit{shapes} of galaxies from which galaxy lensing is measured. Like their densities, the shapes of galaxies are biased tracers of the underlying matter distribution and exhibit large-scale correlations that can be confused with weak lensing \cite{Catelan01,Mackey02,Schmitz18,Blazek19}. The effective-theory formalism for describing this phenomenon, with galaxy shapes acting as a spin-2 biased tracer, was developed in ref.~\cite{Vlah20}, and the equivalent effective theory within the Lagrangian formalism, which we use in this work, was developed in ref.~\cite{Chen24} following earlier work in refs.~\cite{Taruya21,Matsubara22a,Matsubara22b}. At leading order, this \textit{intrinsic alignment} (IA) signal is proportional to the local tidal field; when projected along the line-of-sight, this is exactly proportional to the leading local contribution to weak lensing, making the careful treatment of IAs particularly important for correctly extracting the lensing amplitude \cite{Hirata04}. While IAs are thus a significant contaminant in galaxy lensing surveys, their effect is not catastrophic for two reasons: firstly, simulations and direct measurements have found their amplitude to be small, \edit{with linear and higher-order dimensionless bias parameters at the level of a few percent \cite{Singh2015,Chisari2015,Hilbert2017,Fortuna2021,Samuroff2023,Bakx23,Akitsu23,Delgado2023}, compared to the order-one bias parameters typically observed in galaxy densities in cosmological surveys. Secondly,} they are sensitive to the local matter distribution at the position of the lenses, as opposed to projected along the line of sight as is the lensing signal. Thus, for example, the cross correlation with a lens galaxy sample totally separated from the source sample is sensitive to the lensing signal but not the IA one. This makes a sufficiently flexible prescription for the redshift evolution of IAs particularly important, lest the lensing signal be confused with that of IAs. Other works have pointed out the importance of correctly modeling the complex redshift dependence of the IA signal for galaxy lensing studies \cite{Krause2016}, and some of the strengths of galaxy--galaxy lensing in mitigating the sensitivity to this dependence \cite{Samuroff2024}. In this work we propose a maximally flexible parameterization for these degrees of freedom, which we call $\texttt{MAIAR}$, putting the perturbative modeling of IAs on the same footing as that of galaxy densities and fully immunizing our analysis to biases due to their redshift dependence in a model agnostic manner.

The aim of this work is to consistently apply the theoretical models described above to analyze galaxy galaxy-lensing measurements using the photometrically selected Dark Energy Spectroscopic Instrument (DESI) target samples for the Bright Galaxy Survey (BGS) \cite{BGS.TS.Hahn.2023} and Luminous Red Galaxies (LRG) \cite{LRG.TS.Zhou.2023,Zhou2023b} as lenses and the year-three release of the \metacal catalog from the Dark Energy Survey (DES Y3) as sources to measure lensing \cite{Gatti2021}. \edit{Although the BGS and LRG samples are photmetrically selected, we can calibrate the redshift distributions of these samples nearly perfectly, as they are DESI target samples with greater than $99\%$ spectroscopic completeness. W}e use the harmonic-space 2-point auto-power spectrum of the DESI galaxies, and cross-power spectrum of the galaxies and lensing (``$2\times2$-point''), which, as we explain below, are particularly amenable to these techniques. The DESI imaging data has the largest overlap with the DES Y3 catalog of all Stage III lensing catalogs, and so we use the DES data rather than KiDS or HSC for this analysis. 
DESI is a Stage IV ground-based spectroscopic survey operated through the 4m Mayall Telescope at Kitt Peak National Observatory \cite{Snowmass2013.Levi,DESI2016a.Science,DESI2016b.Instr,DESI2022.KP1.Instr,FocalPlane.Silber.2023,Corrector.Miller.2023,Spectro.Pipeline.Guy.2023,SurveyOps.Schlafly.2023}. As of writing DESI has completed its survey validation and an early data release \cite{DESI2023a.KP1.SV,DESI2023b.KP1.EDR}, and the analysis of the Y1 data is well underway, including already-published results on the highest signal-to-noise measurements of the baryon acoustic oscillations feature to date and their cosmological implications \cite{DESI2024.III.KP4,DESI2024.IV.KP6,DESI2024.VI.KP7A}.

The combination of DESI galaxy and DES lensing data provides us competitive signal-to-noise measurements of the GGL signal compared to other state-of-the-art surveys \cite{Porredon2021,Pandey2021,Sugiyama:2023,Miyatake2023,Faga24}, and, more importantly, the spectroscopic calibration of the DESI target samples allows us to avoid lens photometric-redshift uncertainties and cleanly localize the distance scales associated with clustering measurements, making a direct application of perturbative techniques to a ``$2\times2$-point'' analysis particularly straightforward. While photometrically selected lens samples may provide greater raw signal-to-noise, a careful treatment of theoretical uncertainties renders this less important, motivating the use of less dense, but better-calibrated spectroscopically characterized galaxy samples. We envision this will continue into the next generation of surveys with DESI2 \cite{Schlegel2022a}, providing ideal lens samples for analogous analyses joint with Stage IV galaxy lensing data (e.g. Rubin and Euclid \cite{Ivezic:2008fe,laureijs2011euclid}).  We leave a full ``$3\times2$-point'' analysis to future work, as the modeling of the shape--shape auto-spectrum, i.e., cosmic shear, requires additional model complexity beyond that presented in this work. Furthermore, these analyses can straightforwardly be combined with the redshift-space distortion and CMB lensing signals measured with the same lens samples, providing a powerful combined probe of the growth of cosmic structure. 

The rest of the paper is structured as follows. The data and modeling, including an extensive discussion of the degrees of freedom in GGL analyses, are described in Sections~\ref{sec:data} and \ref{sec:model}. We describe the likelihood and analysis pipeline briefly in Section~\ref{sec:sampling} before validating them against mocks based on the Buzzard simulations \cite{DeRose2019,DeRose2022} in Section~\ref{sec:simval}. Finally, we apply our pipeline to the actual data in Section~\ref{sec:results} before concluding in Section~\ref{sec:conclusions}.

\section{Data}
\label{sec:data}

\begin{table*}[t!]
    \centering
    \begin{tabular}{|c|c|c|c|c|c|c|c|c|c|}
         \hline
         \hline
         sample & $z_{\rm eff}$ & $\sigma(z)$ & $f_{\rm star}$ & $b_{1E}$ & $\alpha_{\mu}$ & $10^{6}\textrm{SN}_{2D}$ & $\textrm{SN}_{3D}$ [$h^{-3}$Mpc$^3$] & $\bar{n}\, [\textrm{deg}^{-2}]$ & $\ell_{\rm max,fid}$\\
         \hline
         BGS0 & 0.229 & 0.0597 & 0.00278 & 0.99 & 1.62 & 0.463 & 90 & 627 & 134\\
         BGS1 & 0.363 & 0.0621 & 0.00216 & 1.34 & 1.60 & 0.918 & 430 & 317 & 267\\
         LRG0 & 0.469 & 0.0636 & 0.000634 & 1.72 & 1.916 & 3.89  & 2835 & 74.9 & 400\\
         LRG1 & 0.626 & 0.0715 & 0.000602 & 1.96 & 2.078 & 2.16  & 2600 & 135 & 533\\
         LRG2 & 0.794 & 0.0766 & 0.00146 & 2.73 & 1.956 & 2.03 & 3350 & 148 & 667\\
         LRG3 & 0.932 & 0.0913 & 0.00218 & 2.54 & 1.952 & 2.24 & 5295 & 136 & 767\\
         \hline 
    \end{tabular}
    \caption{Summary of quantities pertaining to the lens samples used in this analysis. $z_{\rm eff}$ is the effective redshift of the sample (see Eq.~\ref{eqn:zeff}), $\sigma(z)$ is the width of the lens redshift distributions, $f_{\rm star}$ is the stellar contamination fraction, $b_{1E}$ is the Eulerian linear bias, $\alpha_{\mu}$ is the lens magnification coefficient given by Eq.~\ref{eq:magnification_coeff}, $\textrm{SN}_{2D}$ is the Poisson angular shot noise, $\textrm{SN}_{3D}$ is the best-fit three-dimensional shot noise, allowing for deviations from the Poissonian expectation, $\bar{n}$ is the angular number density, and $\ell_{\rm max,fid}$ is the maximum $\ell$ value that we fit to for our fiducial analysis.}
    \label{tab:lens_info}
\end{table*}

Here, we summarize the data used in this analysis, as well as our angular power spectrum measurements and our covariance estimation methodology. Table \ref{tab:lens_info} contains a summary of a few quantities relevant to the lens samples used in this analysis.

\subsection{Lens galaxies}
\subsubsection{DESI LRGs}

\begin{figure*}[t!]
    \centering
    \includegraphics[width=\textwidth]{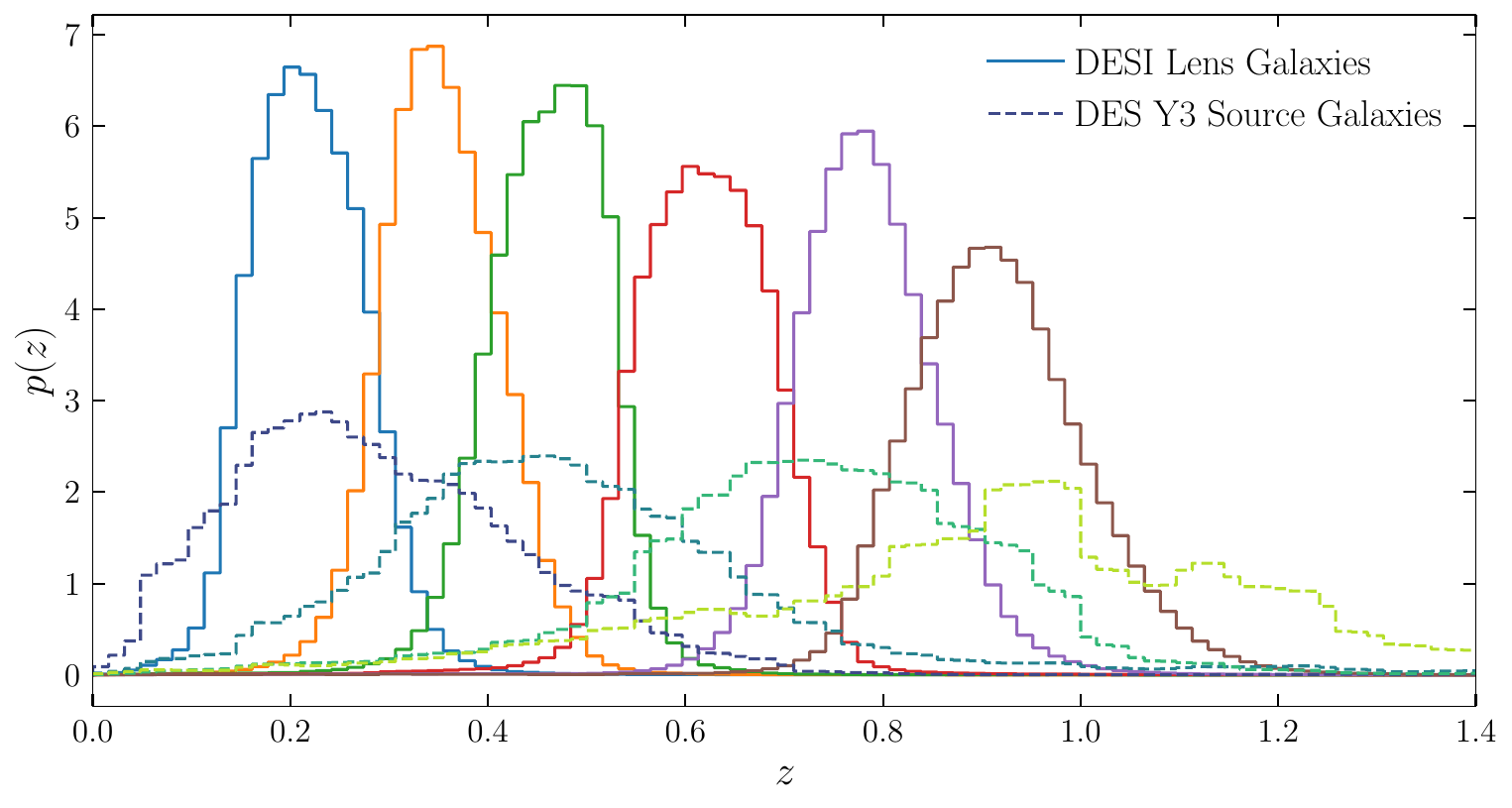}
    \caption{Redshift distributions of the DESI lens (solid) and DES Y3 source (dashed) galaxies. The two lowest redshift lens bins are comprised of the DESI Bright Galaxy Sample, and the four higher redshift bins are made up of DESI Luminous Red Galaxies.}
    \label{fig:nz_fid}
\end{figure*}

We make use of the DESI LRG target sample \cite{Zhou2023b, LRG.TS.Zhou.2023} defined over the full footprint of DR9 of the DESI Legacy Imaging Survey \cite{LS.Overview.Dey.2019}, which constitutes the parent imaging survey for this work. We briefly describe the LRG sample here, and refer the reader to \cite{Zhou2023b} for more details. This sample is selected from the parent imaging catalogs by applying cuts in extinction-corrected $g$, $r$, $z$ and WISE \cite{Wright2010} $W1$ bands. In particular, although the DESI footprint does not cover the entire DES footprint, the LRG sample that we use in this work does. Furthermore, the photometry used to select the LRG sample makes use of the full six years of DES imaging data. One major advantage of this sample is that it is one of the primary DESI target classes. With a spectroscopic success rate of greater than $99\%$, we are able to train accurate photometric redshifts, which can be used to bin the sample into four well-localized redshift bins, as shown in Figure~\ref{fig:nz_fid}. This training procedure is described in detail in \cite{Zhou2023b}, but in essence it trains a random forest regression model to produce redshift estimates given Legacy Survey photometry using the DESI Y1 redshift catalogs and the DR9 Legacy Survey imaging data. Redshift distributions and stellar contamination fractions for each of the four LRG redshift bins are estimated using the redshifts obtained for the LRG sample over the first year of DESI main survey observations. For our fiducial analysis, we specifically use the redshift distribution of these galaxies inferred from DESI spectroscopy in the overlapping DES region; we comment on the negligible effect of using the full Y1 area instead in Appendix~\ref{app:lens_calibration}.

We apply masking following \cite{Zhou2023b} to remove regions of the sky near bright stars and large galaxies included in the Sienna Galaxy Atlas (SGA)\cite{Moustakas2023}, and to avoid the Galactic plane and areas of high extinction. In addition, we apply the masking used for the DES Y3 \metacal sample \cite{Gatti2021}, as described below, in order to measure our galaxy clustering and galaxy--galaxy lensing statistics over the same area. We do not apodize our masks as they contain a large number of small holes and doing so would significantly decrease the effective area of our measurements.

Random points are sampled uniformly over the footprint and the same masking is applied to them as for the LRG catalogs. Weights are assigned to the randoms independently for each LRG redshift bin, such that the weighted random densities are correlated with imaging and foreground systematics with the same trends as measured in the galaxy catalogs in that redshift bin. These weights are constructed by performing linear regression on the correlations between LRG density and $g$, $r$, and $z$ band, extinction corrected imaging depths, PSF sizes, as well as $E(B-V)$ as estimated by \cite{Schlegel1998}. We find negligible differences when removing the weights that correct for $E(B-V)$ correlations. We use the weights computed by performing a linear regression over the full \decals region, but we have verified that our $C(\ell)$ measurements are stable to performing this regression over just the DES footprint where we measure our power spectra. The weighting methodology and null tests are presented in \cite{Zhou2023b}, and we note that these weights are necessarily different from the LRG weights used for the key DESI BAO and RSD analyses, given the differences in binning used in this analysis and the LRG BAO and RSD analyses.

\subsubsection{DESI BGS}
In addition to the DESI LRG sample described above, we make use of the DESI Bright Galaxy Sample \cite{BGS.TS.Hahn.2023} as an additional lens galaxy sample designed to trace $z \lesssim 0.4$ structure. This sample is particularly useful for galaxy--galaxy lensing science as it has minimal redshift overlap with two of the four DES Y3 source galaxy redshift bins. The BGS sample has many of the same advantages as the DESI LRG sample, with comparably high spectroscopic completeness, allowing us to bin galaxies into two narrow redshift bins using photometric redshifts, robustly calibrate the redshift distributions of these bins, and estimate systematics such as stellar contamination. The photometric redshifts that we use to bin the BGS sample are trained in a manner identical to that described for LRGs in \cite{Zhou2023b}. We briefly describe our treatment of this sample here, and refer the reader to additional systematics tests, mirroring those done in \cite{Zhou2023b} for the LRG sample, in Appendix~\ref{app:lens_calibration}. 

Similarly to the LRGs, redshift distributions for each of the two BGS redshift bins are estimated using the redshifts obtained for the BGS sample over the first year of DESI main survey observations in the overlap region with the DES Y3 footprint. These are shown in Figure~\ref{fig:nz_fid}. Unlike the LRG sample, we do not apply weights to the redshift distributions to correct for spectroscopic incompleteness, given the $>99\%$ spectroscopic completeness of this sample. While we do include these weights in the LRG sample, they have a negligible impact on the LRG redshift distributions, and so for simplicity we have omitted them for the BGS sample.

We apply the same masking as for the LRG sample, and we have checked that the SGA masking done for LRGs does not significantly impact our measured statistics, despite the fact that the redshift distribution of SGA galaxies slightly overlaps our BGS samples. The BGS samples, which are generally brighter galaxies detected at higher signal-to-noise, exhibit even less significant trends with potential contaminants than the LRG samples. Correcting for these trends in our angular power spectrum measurements has a significantly smaller impact than our statistical uncertainty and thus we do not apply weights correcting for these trends for our fiducial BGS measurements.

\subsubsection{Galaxy overdensity maps}
To construct galaxy overdensity maps, $\delta_{g,p}$, we first bin galaxies into \HEALPIX\, \cite{healpix,healpy} maps ($\texttt{NSIDE}=2048$), $N_{p} = \sum_{g\in p} v_{g}$, where $v_{g}$ is an ``effective redshift weight" assigned to galaxy $g$ that will be described in \S\ref{sec:effective_redshift}, and the sum runs over all galaxies in pixel $p$. We then compute weighted random counts, $R_{p} = \sum_{r\in p} v_{r}$, and pixel averaged random weights using our random catalog: $w_{p} = R_p/\sum_{r\in p}1$, where $v_{r}$ are weights assigned to the randoms to correct for angular systematics and the denominator in the second equation is simply counting the total number of randoms in each pixel. For each lens bin, we construct five different galaxy count maps: one with no weights applied to the galaxies, and four with galaxy weights constructed to bring the effective redshift of our clustering measurements into agreement with our lensing measurements for each of the four DES Y3 \metacal source bins. Random weights are always applied to correct for angular systematics for the LRG samples.

In terms of the above quantities, the projected galaxy density is
\begin{align}
    \rho_{g,p} = \frac{N_{p}}{R_p}, \, \delta_{g,p} = \frac{\rho_p}{\bar{\rho}_p} - 1 \, ,
\end{align}
where $\bar{\rho}_{p}$ is the mean of $\rho_p$ taken over all unmasked pixels. We then define the mask $W^{\delta_g}_p = \Theta(R_{p} - 0.2\bar{R}_{p})$, where $\bar{R}_{p}$ is the average of $R_p$ over all pixels with $R_p>0$, and $\Theta$ is the Heaviside step function, i.e., the mask is one where the average random density is greater than $20\%$ of the mean, and zero otherwise following, e.g. \cite{white2022cosmological,Chen22b}. We also compute the Poisson shot noise for each redshift bin as $1/\bar{n}$ using
\begin{equation}
    \bar{n} = \frac{(\sum_{p} N_p w_p)^2}{\Omega_{s} \sum wN_p w_p^2}\, ,
    \label{eq:nbar}
\end{equation}
\noindent where $wN_p = \sum_{g\in p} v^2_{g}$ and $\Omega_{s}$ is the survey area in steradians. 

\subsection{DES Y3 \metacal}
\label{sec:des}

\begin{table}[htb!]
    \centering
    \begin{tabular}{|c|c|c|c|c|}
         \hline
         \hline
         Source bin & $\sigma \left(\Delta z_{s}\right)$ & $\langle m \rangle$ & $\sigma (m)$ & $\sigma_{e}^2/\bar{n}$\\
         \hline
         0 & 0.018& -0.006& 0.009& 0.040\\
         1 & 0.015& -0.020& 0.008& 0.046\\
         2 & 0.011& -0.024& 0.008& 0.045\\
         3 & 0.017& -0.037& 0.008& 0.062\\
         \hline 
    \end{tabular}
    \caption{Summary of quantities pertaining to the source samples used in this analysis. The first column indexes the source bin, $\sigma \left(\Delta z_{s}\right)$ is the uncertainty marginalized over in the mean redshift of each source bin, $\langle m \rangle$ and $\sigma (m)$ are the mean and standard deviation of the prior on the multiplicative bias correction applied to each source sample, and $\sigma_{e}^2/\bar{n}$ is the shape noise divided by the average angular number density.}
    \label{tab:source_info}
\end{table}

We make use of the \metacal shape catalog constructed from the first three years of DES data \cite{Gatti2021} to measure gravitational lensing through the cross correlation between galaxy ellipticities, $e_{g,i}$, in DES and galaxy overdensities measured from our DESI samples. The catalog contains 100 million galaxies over an area of 4142 square degrees, with an effective number density of $\bar{n}=5.59\, \mathrm{arcmin}^{-2}$. The shape measurement process is known to be biased by a number of observational factors, and so the raw galaxy ellipticities, $e_{g,i}$, with $i$ indexing the two galaxy ellipticity components, must be corrected in order to obtain an unbiased measurement of the gravitational lensing signal. 

To account for this, the \metacal algorithm computes the response, $R_{g_i}$, of observed galaxy shapes to an artificial shear. By appropriately weighting $e_{g,i}$ by $R_{g_i}$, the biases to $e_{g_i}$ can be removed in estimators using these ellipticities \cite{Huff2017,Sheldon2017}. Residual biases to $e_{g,i}$ at the $\sim 2-4\%$ level, mostly sourced by blending of galaxy shapes, must be calibrated using image simulations \cite{MacCrann2021};  uncertainties in this calibration are marginalized over in our cosmological analysis. 

We make use of the fiducial DES Y3 redshift calibration, binning the \metacal sample into four coarse redshift bins, and using the ensemble $n(z)$s provided for these bins. The $n(z)$ estimates for the four bins are obtained using a combination of \texttt{SOMPZ} photometric redshifts \cite{Buchs2019, Myles2021} and clustering cross-correlations \cite{Gatti2021b}, additionally corrected for the effects of redshift dependent blending \cite{MacCrann2021}. Furthermore, the \texttt{SOMPZ} algorithm relies on a combination of wide and deep field photometry \cite{Hartley2021} which are related to each other through the synthetic source injection software \texttt{Balrog} \cite{Everett2020,Suchyta2021}, as well as catalogs of spectroscopic and high-quality photometric redshifts. These redshift distributions are shown alongside those of the DESI lens galaxies in Figure~\ref{fig:nz_fid}.

In each tomographic bin, we divide each ellipticity component by the mean \metacal response measured in that bin as in \cite{Gatti2021,Amon2021,Secco2021}, and subsequently subtract the mean ellipticity in each component. Once we have calibrated the ellipticities in this manner, we construct galaxy ellipticity maps as

\begin{equation}
    e_{p,i} = \frac{\sum_{g\in p} v_{g} e_{g,i}}{\sum_{g\in p} v_{g}} \, 
\end{equation}
\noindent where $v_g$ are the inverse variance weights provided with the \metacal catalog, and $i$ indexes over the two galaxy ellipticity components. Because our signal is weighted by the number of source galaxies per pixel divided by the ellipticity dispersion, $\sigma_e$, which can vary quite significantly over the footprint, we compute the mask for our ellipticity maps as

\begin{equation}
    W_{p}^{e} = \sum_{g\in p} v_{g}\, 
\end{equation}
where $\sigma_e$ enters through $v_{g}$, since $v_g$ are inverse variance weights. We also compute the mode-coupled noise bias, sometimes known as the noise power spectrum, which enters into our covariance calculations as

\begin{equation}
    N_{\ell > 2} = A_{p} \left \langle \sum_{g\in p} v_{g}^2 \sigma^2_{e,g} \right \rangle_{\rm pix}\, 
    \label{eq:nell_gammae}
\end{equation}
\noindent where $\sigma^2_{e,g} = 0.5 (e_{g,1}^2 + e_{g,2}^2)$ and $A_{p}$ is the area of a pixel in steradians, and the average is taken over all pixels in the map. As shown by \cite{Nicola2019}, this is equivalent to what would be measured from repeatedly rotating all galaxy ellipticities randomly and measuring power spectra, i.e., it is the contribution from uncorrelated shape noise.

\subsection{Angular Power Spectra}

\begin{figure*}[htb!]
    \centering
    \includegraphics[width=\textwidth]{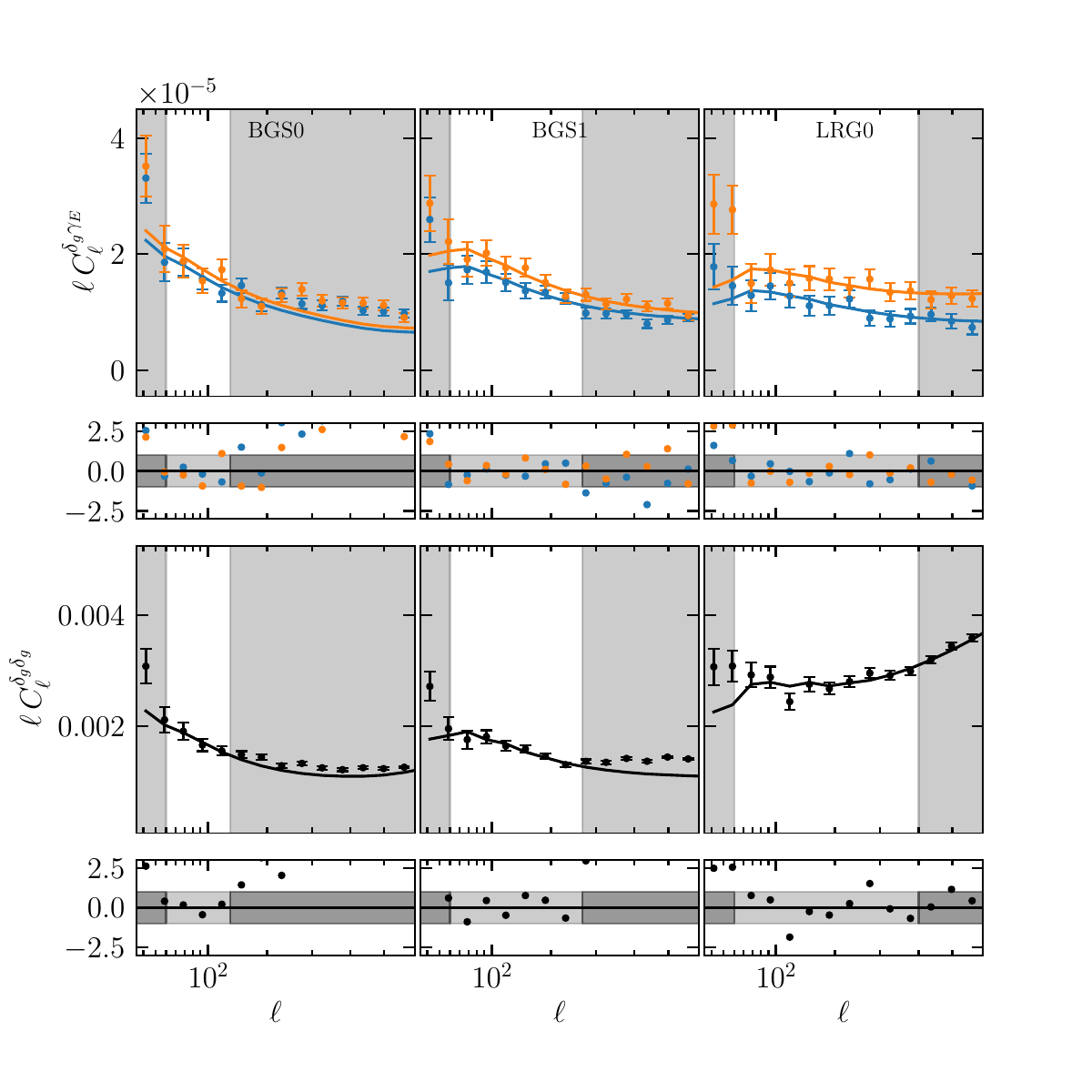} 
    \caption{Galaxy density-shear E-mode cross-spectra (top) and galaxy-density auto-spectra (bottom) compared with the best-fit model from our fiducial analysis (lines). Note that we do not use all source--lens bin combinations in this fit, as detailed in Section~\ref{sec:noiseless_sims}. A fit to the full data vector is shown in Figure~\ref{fig:bestfit_model_allbins}. Smaller sub-panels show the residuals in units of the estimated uncertainty on each data point. We do not fit to the points in the gray regions to avoid unmodeled RSD contributions on large scales, and higher order bias and baryonic contributions on smaller scales as described in Section~\ref{sec:model}. The blue and orange points in the top row are measurements using the third and fourth highest redshift DES Y3 source bins. We find $\chi^2=27.5$ for 54 data points using 21 free parameters, not counting the IA, magnification or source sample uncertainties since these are prior dominated, equivalent to $\chi^2_{\rm red} = 0.86$.}
    \label{fig:bestfit_model}
\end{figure*}

In order to extract cosmological information from our data, we measure auto and cross angular power spectra of the galaxy overdensity fields, $\delta_{g,\rm obs}^{i}$, and E-mode galaxy ellipticity fields, $\gamma_{E}^{j}$, where $i$ and $j$ index the lens and source galaxy redshift bins. As we explain in Section~\ref{sec:model}, our fiducial analysis setup uses data from only the first three lens bins, whose auto and cross correlations we shown in Figure~\ref{fig:bestfit_model} along with the error bars computed as in Section~\ref{sec:covariance} and the best-fit model. In order to compute these harmonic-space 2-point functions we use the pseudo-$C_{\ell}$ estimator implemented in the \namaster code. We briefly review this methodology here, and refer the reader to \cite{Alonso2019} for further details. 

A map $a(\hat{\bf n})$ on the unit sphere, where $\hat{\bf n}$ denotes an angular position on the sky, can be decomposed into spherical harmonics via:
\begin{equation}
    \tilde{a}_{\ell m} = \int d\Omega \, a(\hat{\bf n})\,  W^{a}(\hat{\bf n}) \, Y_{\ell m}(\hat{\bf n})\, ,
\end{equation}
\noindent
or in the case of a spin-2 field, like the galaxy ellipticity field, we have 

\begin{equation}
    \tilde{\gamma}_{E,\ell m} \pm i\tilde{\gamma}_{B,\ell m}= \int d\Omega \, (\gamma_1(\hat{\bf n}) \pm i \gamma_2(\hat{\bf n}))\,  W^{\gamma}(\hat{\bf n}) \, _{\pm 2}Y_{\ell m}(\hat{\bf n})\, ,
\end{equation}
\noindent 
where $W^{a}(\hat{\bf n})$ is the mask, and $Y_{\ell m}(\hat{\bf n})$ and $_{\pm 2}Y_{\ell m}(\hat{\bf n})$ are spherical harmonics and spin-weighted spherical harmonics \cite{Hikage2011}, respectively. Without loss of generality, we consider only the scalar field case for the rest of this section. We also use the shorthand

\begin{equation}
    \tilde{a}(\hat{\bf n}) = a(\hat{\bf n})\,  W^{a}(\hat{\bf n})\, .
\end{equation}
Given two sets of spherical harmonic coefficients, $\tilde{a}_{\ell m}$ and $\tilde{b}_{\ell m}$, we can compute the angular power spectrum of these two fields as:

\begin{equation}
    \tilde{C}^{ab}_{\ell} = \frac{1}{2\ell + 1}\sum_{m=-\ell}^{\ell} \tilde{a}_{\ell m} \tilde{b}_{\ell m}^{*}\, ,
\end{equation}
\noindent which is then related to the true unmasked angular power spectrum, $C^{ab}_{\ell}$ as
\begin{equation}
    \langle \tilde{C}^{ab}_{\ell} \rangle = \sum_{\ell^{\prime}} M_{\ell \ell^{\prime}}^{ab} C^{ab}_{\ell^{\prime}}\,
\end{equation}
\noindent 
where $M_{\ell \ell^{\prime}}^{ab}$ is the mode-coupling matrix (MCM), which can be computed analytically from the masks of the two fields, $a$ and $b$ \cite{Hivon2002}. See \cite{Alonso2019} for the expressions of $M_{\ell \ell^{\prime}}^{ab}$ given masks for spin-0 and spin-2 fields that we use in this work.  

In order to obtain unbiased angular power spectrum estimates, we must invert $M_{\ell \ell^{\prime}}^{ab}$, but in the case of masks that remove large fractions of the sky this matrix is singular. To circumvent this issue, it is necessary to bin $\tilde{C}^{ab}_{\ell}$ into bandpowers, with each bandpower $L$ containing (potentially weighted) sums over many $\ell$ values. The binned MCM, $M_{L L^{\prime}}^{ab}$, is then invertible and we have:
\begin{align}
    \langle \tilde{C}^{ab}_{L} \rangle &=  \sum_{L^{\prime}} C^{ab}_{L^{\prime}} M_{L L^{\prime}}^{ab} \\
            &= \sum_{L^{\prime}} \sum_{\ell^{\prime}} B_{\ell^{\prime} L^{\prime}} C^{ab}_{\ell^{\prime}}  \sum_{\ell} B_{\ell L} B_{\ell^{\prime} L^{\prime}} M_{\ell \ell^{\prime}}^{ab}\,
\end{align}
\noindent where $B_{\ell L}$ is the weight given to $\ell$ in bandpower $L$. $M_{L L^{\prime}}^{ab}$ can then be inverted to give an estimate of $\hat{C}^{ab}_{L}$:

\begin{equation}
    \hat{C}^{ab}_{L} = \sum_{L^{\prime}} (M^{ab})^{-1}_{L L^{\prime}} \tilde{C}^{ab}_{L}\, .
    \label{eq:decoupling}
\end{equation}
\noindent $\hat{C}^{ab}_{L}$ is an unbiased estimate of $C^{ab}_{L}$ in the limit that $C^{ab}_{\ell}$ is piecewise constant over each bandpower, $L$. In general, this is not the case, and so we must account for binning into bandpowers using a bandpower convolution matrix, $\mathcal{F}_{L \ell}^{ab}$ which connects a theory prediction for $C^{ab}_{\ell}$ to the bandpowers $C^{ab}_{L}$, 

\begin{align}
    C^{ab}_{L} &= \sum_{\ell} \mathcal{F}_{L \ell}^{ab} C_{\ell}^{ab} \\
    &= \sum_{\ell} \sum_{L^{\prime}} (M^{ab})^{-1}_{L L^{\prime}} \sum_{\ell^{\prime}} B^{ab}_{\ell^{\prime} L^{\prime}} M^{ab}_{\ell \ell^{\prime}} C^{ab}_{\ell}\, ,
    \label{eq:model_model_coupling}
\end{align}
\noindent where $\mathcal{F}_{L \ell}^{ab}$ combines the mode coupling, binning, and de-coupling procedures. Note that we could just as well have avoided deconvolving our measurements, and evaluated our model prediction by removing the inverse mode coupling matrix in Equation~\ref{eq:model_model_coupling}, but following convention we have chosen to deconvolve our measurements.

We compute our bandpowers and bandpower convolution matrices using the \namaster \texttt{compute\_full\_master} function. Figure~\ref{fig:bestfit_model} shows these angular power spectrum measurements for the first three lens (BGS0, BGS1 and LRG0) bins and two highest redshift source bins, which are the spectra used in our fiducial analysis as described in \S~\ref{sec:noiseless_sims}, as well as our best-fit model. This fit will be further described in \S~\ref{sec:results}. Unlike some other works making use of pseudo-$C_{\ell}$ estimators, we do not correct for the pixel window function, as the form of this correction depends on the number of source galaxies per pixel \cite{Nicola2019}, and because even in the limit of infinite sampling the pixel window depends on azimuthal angle due to the variation in \HEALPIX\, pixel shape with azimuth. Although algorithms exist to circumvent these issues, for example \cite{BaleatoLizancos2024}, we opt to simply take the pixel size to be small ($\texttt{NSIDE}=2048$) compared to the scales of interest in this work, such that the impact of the azimuthally averaged pixel window function on our measurements is significantly below $1\%$ even for $\ell=1200$, which is the largest $\ell$ that we use in this work for the simulated tests extending beyond our fiducial scale cuts to $k_{\rm max}=0.6\, h\, \rm Mpc^{-1}$. We note that the largest $\ell$ used in our fiducial analysis is much smaller than this, at $\ell_{max}=400$ for the first LRG bin.

\begin{figure*}[htb!]
    \centering
    \includegraphics[width=\textwidth]{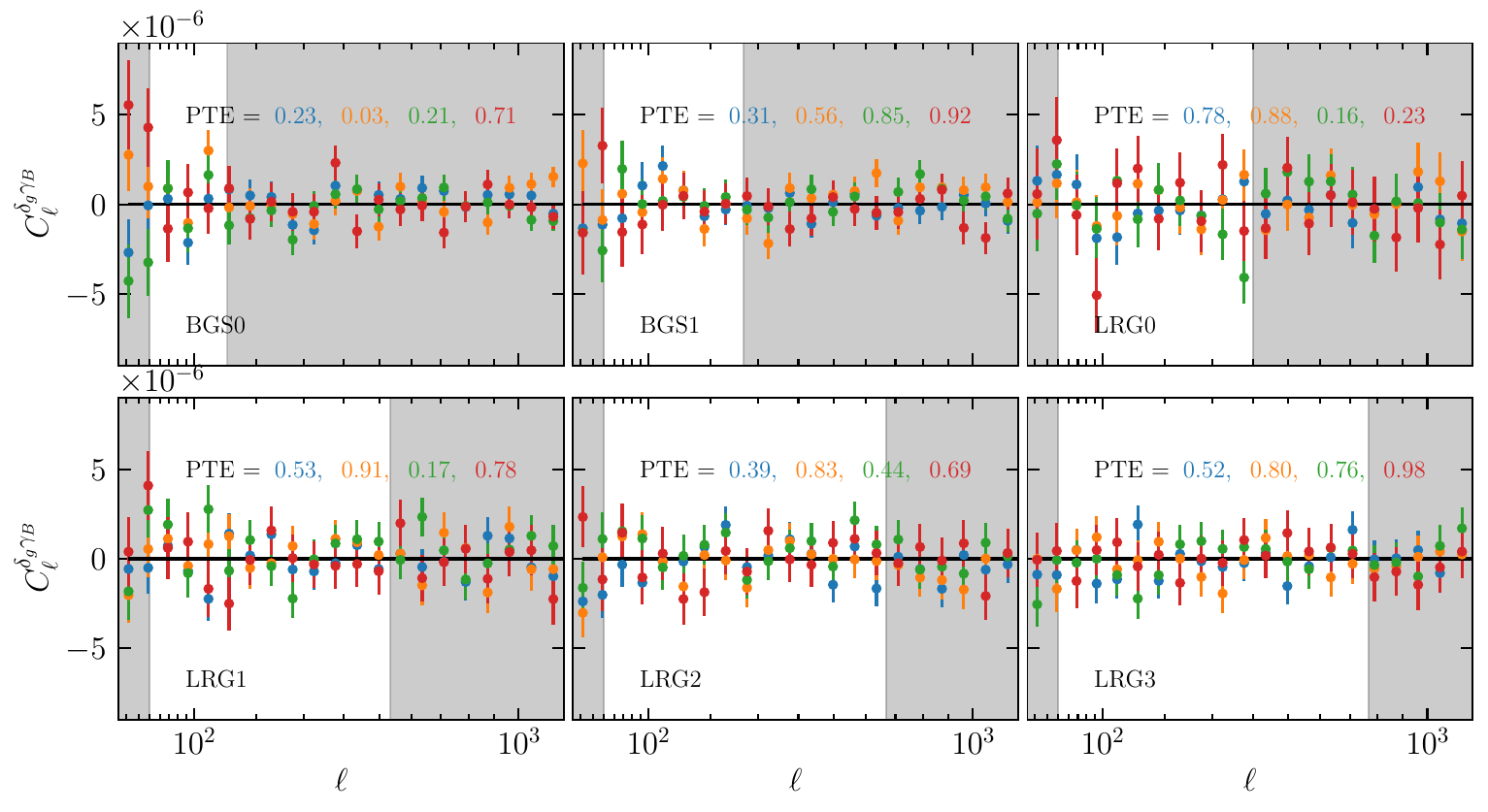}
    \caption{Measurements of galaxy density - B mode galaxy ellipticity power spectra for all lens and source bin combinations. The different colors represent the four different source bins, going from low to high redshift from blue, orange, and green to red. The covariance for all of these measurements is estimated from the Gaussian simulations described in Sec \ref{sec:covariance}. The probability that the chi-squared value measured in our Gaussian simulations with no B-modes exceeds the chi-squared for each spectrum in the data measured over the scales used in this analysis (non-greyed regions) is displayed on each panel as ``PTE''. }
    \label{fig:bmodes}
\end{figure*}

Finally, as a systematics test, we also measure the galaxy-overdensity--B-mode angular power spectra for each source--lens bin configuration, shown in Figure~\ref{fig:bmodes}. The different panels show each of the six lens bins considered in this work, and the different colored points show measurements for each source bin. Error bars are derived from the Gaussian simulations described in \S~\ref{sec:covariance}. Inset in the figure, we quote the probability that the $\chi^2$ value measured for each spectrum in our Gaussian simulations over the scales used in our analysis exceeds that measured in our data (PTE). No spectrum has a PTE of less than $3\%$, and of the spectra used in our fiducial analysis the lowest PTE value is $16\%$. As such, we conclude that B-mode contamination contributes negligibly to our analysis.

\subsection{Covariance}
\label{sec:covariance}

We make use of a Gaussian covariance matrix computed analytically with the \namaster function \texttt{gaussian\_covariance}, where we use as input the best-fit theory spectra shown in Figure~\ref{fig:bestfit_model}. In order to avoid complications in implementing an accurate model for  $C^{\gamma_E,\gamma_E}_{\ell}$, we instead use a third order B-spline fit to the measured, noise bias subtracted $C^{\gamma_E,\gamma_E}_{\ell}$s as input to our covariance calculations. A number of works \cite{friedrich2021pdf,GarciaGarcia2021} have shown that Gaussian covariance matrices are sufficient for $\Lambda$CDM analyses of very similar statistics for a comparable sky area and level of constraining power, and so we focus on validating the computation of the disconnected (Gaussian) contribution to the covariance in this section. 

It has been shown that the narrow kernel approximation (NKA) that is used to accelerate the computation of the effect of survey geometry on the Gaussian part of the covariance from an $\mo(\ell_{\rm max}^6)$ operation to a tractable $\mo(\ell_{\rm max}^3)$ is inaccurate at the $10-30\%$ level for galaxy lensing surveys, which have very complicated masks. These masks break the main assumption of the NKA, which is that the MCM is close to diagonal. Ref.~\cite{Nicola2019} showed that replacing the input theory spectra with their mode-coupled counterparts scaled by the mean of the product of their masks as:
\begin{equation}
    C_{\ell}^{ab} \rightarrow \frac{\sum_{\ell^{\prime}}M^{ab}_{\ell \ell^{\prime}} C^{ab}_{\ell}}{\langle W^a_p W^b_p \rangle_{\rm pix}}\, 
    \label{eq:inka}
\end{equation}
\noindent significantly improved the agreement of the NKA and their Gaussian simulations with realistic galaxy lensing survey geometries. Note that we include the noise terms in Equation~\ref{eq:inka} as $f_{\rm sky}/\bar{n}$, where $\bar{n}$ is given by Equation~\ref{eq:nbar} for galaxy densities and as Equation \ref{eq:nell_gammae} for $\gamma_{E}$.

The additional subtlety that we incur due to our choice to use different galaxy weights in our auto- and cross-spectrum measurements, is that the shot noise of the galaxy maps that enter these measurements are different. To account for this we simply use the geometric mean of the shot-noise values obtained for the maps that enter into the auto- and cross-spectrum measurements when computing the shot-noise contribution for $C_{\ell}^{gg}$ of galaxy samples that differ only by different effective redshift weights. 

We validate these approximations using Gaussian random field simulations, generating correlated realizations of the fields $\{\tilde{\delta}_{p}^{0},...,\tilde{\delta}_{p}^{5},\tilde{e}_{p}^{0},...,\tilde{e}_{p}^{3}\}$ with \namaster's \texttt{synfast} function. Instead of Poisson sampling a Gaussian density field to obtain the correct shot-noise values for the galaxy overdensity fields, we include the Poisson shot-noise in our input, noiseless auto spectra. In principle, we should generate five overdensity maps per lens bin, one with the shot-noise appropriate for the unweighted lens catalog, and four with shot-noise values appropriate for the lens catalogs with effective redshift weights applied for each source bin. In order to reduce the computational cost of these simulations, we have opted to generate only the field with shot noise appropriate for the unweighted catalogs. As such, we compare to a slightly modified version of our covariance, where we have used the unweighted lens catalog shot noise for all relevant spectra, and so we do not explicitly validate our treatment of the impact of effective redshift weighting on our covariance. Nevertheless, the difference between our fiducial analytic covariance and the analytic covariance we use for this comparison is at the level of $\sim2-5\%$, and so it is not important for interpreting the results presented here.

\begin{figure*}[htb!]
    \centering
    \includegraphics[width=\textwidth]{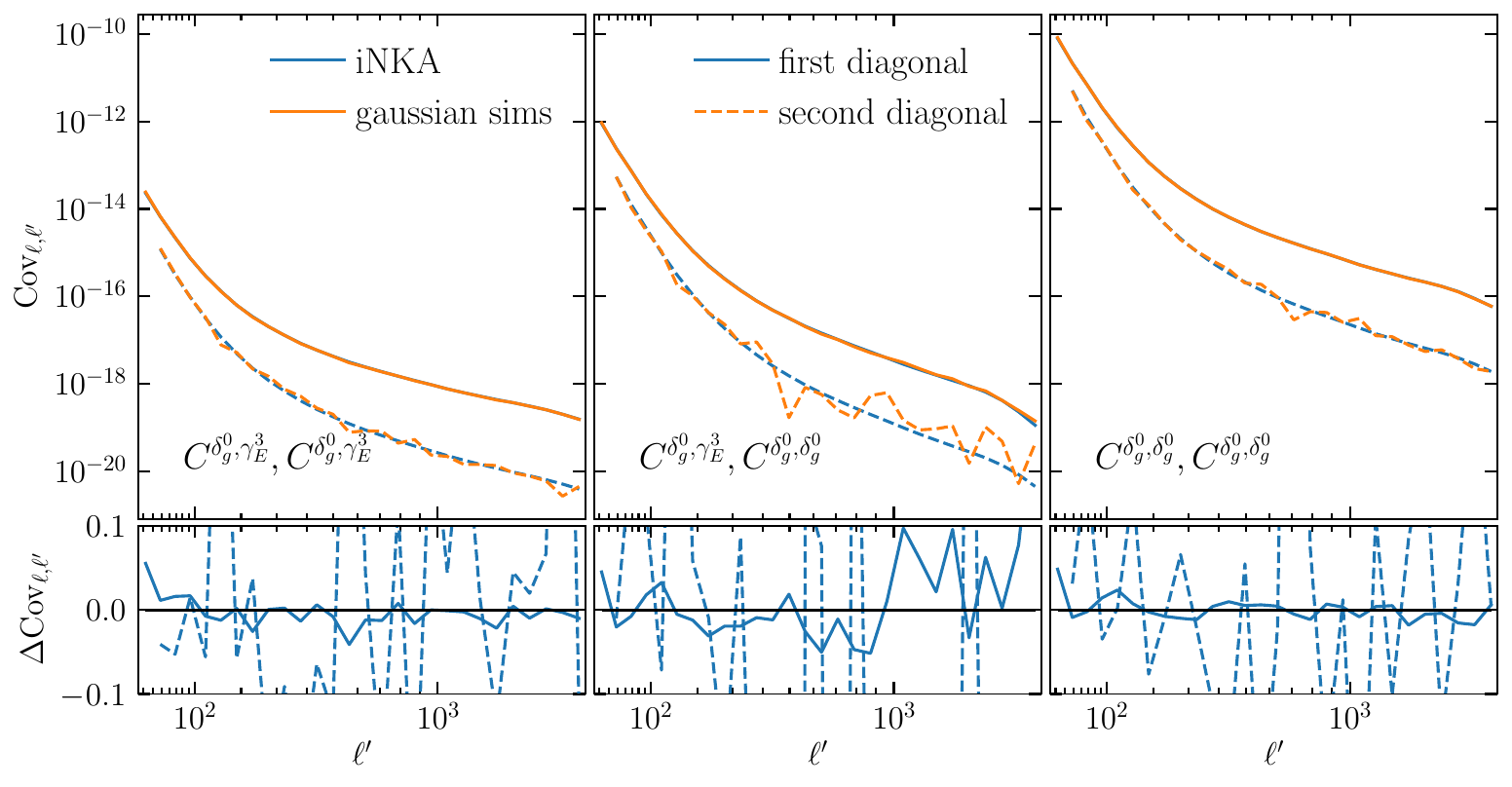}
    \caption{Comparison of three representative sub-blocks of our fiducial covariance to the covariance estimated from Gaussian simulations as described in \S\ref{sec:covariance}. The three different rows show the covariance between $C^{\delta_g,\gamma_{E}}$ and itself (left), $C^{\delta_g,\gamma_{E}}$ and $C^{\delta_g,\delta_{g}}$ (middle), and $C^{\delta_g,\delta_{g}}$ and $C^{\delta_g,\delta_{g}}$ (right). The solid lines show the first diagonal, while the dashed lines show the first off diagonal, while orange shows the estimate from Gaussian sims, while blue is our fiducial analytic covariance assuming the iNKA. The bottom panels show the fractional deviation of our fiducial covariance from our simulated covariance.}
    \label{fig:cov_comparison}
\end{figure*}

In order to simulate sheared galaxy shape fields, for each source bin we generate a noiseless convergence field, $\kappa_p$, correlated with the other source galaxy convergence and lens galaxy overdensity fields. We then transform convergence to shear, $\gamma_{p,i}$, using the inverse Kaiser--Squires algorithm \cite{Kaiser1993}. Using the actual positions and ellipticities of the DES \metacal catalog for the source bin in question, we apply a random rotation to all $e_{g,i}$, and then shear these ellipticities:
\begin{equation}
    \tilde{e}_{g}^{\prime} = \frac{e^{i\pi\theta_{g}}\tilde{e}_{g} + \tilde{\gamma}_{g}} {1 + (e^{i\pi\theta_{g}}\tilde{e}_{g} \tilde{\gamma}_{g}^{*})}\,
\end{equation}
where $\tilde{e}_g$ and $\tilde{\gamma}_g$ are the complex galaxy ellipticity and shear at the position of the galaxy, and $\theta_g$ is the random rotation generated for galaxy $g$. We then apply the relevant masks for the galaxy overdensity fields, and use the map-making procedure outlined in \S\ref{sec:des} for our source galaxy maps. We then measure the auto- and cross-spectra of all the generated fields, including both E- and B-mode components for relevant spectra. We compare the covariance computed with these simulations to our analytic Gaussian covariance in Figure \ref{fig:cov_comparison}. 

\subsection{Analysis Blinding}
In order to mitigate observer bias, we blinded the results of our constraints until we had finalized all aspects of our analysis that we believed could shift our constraints. In order to do so, after measuring the angular power spectra used in this analysis, we produced a blinded measurement given by:
\begin{equation}
    \tilde{C}_{\ell}^{\prime ab} = \tilde{C}_{\ell}^{ab} + \left(C_{\ell}^{ab}(\theta^{\prime}) - C_{\ell}^{ab}(\theta^{\rm fid})\right)\, ,
\end{equation}
where $\tilde{C}_{\ell}^{ab}$ are the unblinded measurements, while $C_{\ell}^{ab}(\theta^{\prime})$ and $C_{\ell}^{ab}(\theta^{\rm fid})$ are model predictions at a randomly chosen and a fiducial set of parameters. The fiducial parameters, $\theta^{\rm fid}$ are chosen to be the best fit from \cite{Aghanim:2018eyx}, while the randomly chosen cosmology, $\theta^{\prime}$, is generated by applying a hashing algorithm to a known string, and using this to seed a random sample, $\theta^{\prime}$, from the initial proposal distribution used in our MCMC analyses. The standard deviation of shifts in $\sigma_{8}$ that we expected by performing this procedure was $\sim 0.1$, i.e., about two times larger than our expected constraining power.

We chose to perform this blinding operation in a reduced parameter space from that of our fiducial model described in \S\ref{sec:model} in order to limit the size of the change in our data vector that was allowed to be $\lesssim 20\%$. In particular, we applied the blinding shift using a linear galaxy bias model, as well as a nonlinear alignment IA model (NLA) \cite{Joachimi2010} with linear redshift evolution (\S\ref{sec:model}). We also did not allow the shot noise, source redshift or shear multiplicative bias parameters to vary, as these were well known from previous analyses. All other parameters in our fiducial analysis set up were allowed to vary.

Before un-blinding we performed a series of tests in order to ensure the robustness of our results. The tests that were passed before unblinding were:
\begin{enumerate}
    \item Recovered the input cosmology within noise ($\sim 0.25\sigma$) on the \texttt{Buzzard} simulations with the fiducial modeling pipeline.
    \item Posterior projection effects were well understood on noiseless simulations.
    \item No significant detection of B-modes for measurements used in fiducial analysis ($p>0.05$).
    \item No significant detection of cross-correlation between systematics maps and galaxy density maps.
    \item Galaxy density cross-power spectra consistent with predictions given by $p^{\delta_g^{i}}(z)$ overlap and magnification.
    \item Acceptable goodness of fit to blinded data ($\chi^2=27.7$ for 54 data points)
    \item Insensitivity of blinded results to changing footprint used for $p^{\delta_g^{i}}(z)$ estimation from overlap region with DES Y3 footprint to full DESI Y1 footprint.
    \item Insensitivity of results to inclusion of $\ell\le 50$ for \cdk.
    \item No preference for nuisance parameters at edges of priors.
    
    \end{enumerate}

After unblinding, we updated our covariance to use the best-fit model predictions from an analysis of all lens and source bins with our fiducial model.


\section{Model}
\label{sec:model}
Having validated our measurement and covariance methodologies, we now discuss our forward model. The following sections aim to provide a high-level overview of all of the components entering into our model predictions. See Table \ref{tab:params} for a list of all free parameters and their priors.

\begin{table*}
\caption{Parameters and priors}
\begin{center}
\begin{tabular}{| c  c  c  |}
\hline
\hline
Parameter & Prior & Reference\\  
\hline 
\multicolumn{3}{|c|}{{\bf Cosmology}} \\
$\omega_c$  &  $\mathcal{U}$($0.08, 0.16$) &  \\ 
$A_s$ &  $\mathcal{U}$($1.1\times 10^{-9},3.1\times 10^{-9}$) & \\
$n_s$ & $\mathcal{U}$($0.93, 1.01$) & Sec. \ref{sec:sampling} \\
$\omega_b$ & $\mathcal{U}$($0.0173, 0.0272$) &  \\
$h$  & $\mathcal{U}$($0.52, 0.82$) & \\
$\log_{10}\sum m_{\nu}$  & $\mathcal{U}$($-2.0, -0.301$) & \\

\hline
\multicolumn{3}{|c|}{{\bf Lens Galaxy Bias}} \\
$\frac{\sigma_8(z)}{\sigma_{8,\rm fid}}(1 + b_{1}^{i}) $  & $\mathcal{U}$ ($0.5, 3.5$) & Eq. \ref{eq:bias_exp}-\ref{eq:pgm}\\
$(\frac{\sigma_8(z)}{\sigma_{8,\rm fid}})^2b_{2}^{i} $  & $\mathcal{N}$ ($0, 1$) & Eq. \ref{eq:bias_exp}-\ref{eq:pgm}\\
$(\frac{\sigma_8(z)}{\sigma_{8,\rm fid}})^2b_{s}^{i} $  & $\mathcal{N}$ ($0, 1$) & Eq. \ref{eq:bias_exp}-\ref{eq:pgm} \\
$\frac{2k_{\rm max}^2}{1+b^{i}_1} b_{\nabla^2 a}^{i} $ & $\mathcal{N}$ ($0, 0.2$) & Eq. \ref{eq:bias_exp},\ref{eq:pgg},\ref{eq:counterterm}\\
$\frac{2k_{\rm max}^2}{1+b^{i}_1}b_{\nabla^2 \times}^{i} $ & $\mathcal{N}$ ($0, 0.2$) & Eq. \ref{eq:bias_exp},\ref{eq:pgm},\ref{eq:counterterm}\\
$b_{\nabla^2 m}$ & $\mathcal{N}$ ($0, 0.5$) & Eq. \ref{eq:bias_exp},\ref{eq:pmm}\\
$\textrm{SN}^{i}$ & $\mathcal{N}(\textrm{Table \ref{tab:lens_info}}, 30\%)$ & Eq. \ref{eq:bias_exp},\ref{eq:pgg}\\

\hline
\multicolumn{3}{|c|}{{\bf Intrinsic Alignment}} \\
$(\frac{\sigma_8(z)}{\sigma_{8,\rm fid}})c_{1}^{ij}$   & $\mathcal{N}$ ($-1,5$) & Eq. \ref{eq:shape_exp}\\
$(\frac{\sigma_8(z)}{\sigma_{8,\rm fid}})^2c_{2}^{ij}$   & $\mathcal{N}$ ($0,5$) & Eq. \ref{eq:shape_exp}\\
$(\frac{\sigma_8(z)}{\sigma_{8,\rm fid}})^2c_{\delta 1}^{ij}$   & $\mathcal{N}$ ($0,5$) & Eq. \ref{eq:shape_exp}\\
$(\frac{\sigma_8(z)}{\sigma_{8,\rm fid}})^2c_{t}^{ij}$   & $\mathcal{N}$ ($0,5$) & Eq. \ref{eq:shape_exp}\\
$(\frac{\sigma_8(z)}{\sigma_{8,\rm fid}})^2\alpha_{s}^{ij}$   & $\mathcal{N}$ ($0,45$) & Eq. \ref{eq:shape_exp}\\
\hline 
\multicolumn{3}{|c|}{{\bf Magnification}} \\
$\alpha^{i}_{\rm \mu}$ & $\mathcal{U}(\textrm{Table \ref{tab:lens_info}}\pm 0.1)$ & Eq. \ref{eq:magnification_coeff}\\
$c_{\mu, \rm UV}$ & $\mathcal{N}$($0,0.4$) & Eq. \ref{eq:alpha_uv_corr} \\
\hline
\multicolumn{3}{|c|}{{\bf Source \photoz\ }} \\
$\Delta z^{i}_{\rm s}$  & $\mathcal{N}$ ($0.000$, \text{Table \ref{tab:source_info}}) & Eq. \ref{eq:delta_z_source}  \\
\hline
\multicolumn{3}{|c|}{{\bf Shear calibration}} \\
$m^{i}$ & $\mathcal{N} \text{(Table \ref{tab:source_info})}$ & Eq. \ref{eq:shear_calib} \\
\hline
\end{tabular}
\end{center}
\label{tab:params}
\end{table*}

\subsection{Field level description}
\label{sec:field_level}
We make use of two types of fields in this analysis: the projected galaxy density field, $\delta_{g,\rm obs}(\hat{\bf n})$, and the projected E-mode galaxy ellipticity field, $\gamma_{E}(\hat{\bf n})$. We do not treat B-modes in our model, as a cross correlation between a scalar field and B-modes can only be generated by a parity violating process. We can express $\gamma_{E}(\hat{\bf n})$ as \cite{Vlah21}
\begin{align}
    \gamma_{E}^{i}(\hat{\bf n}) = \gamma_{E,I}^{i}(\hat{\bf n}) + \kappa^{i}(\hat{\bf n})\, ,
\end{align}
\noindent 
where $i$ indexes the source galaxy bin in question. The first term on the right-hand side is the intrinsic alignment contribution to galaxy ellipticity while the second term is the contribution due to gravitational lensing. We neglect higher-order terms related to source magnification and reduced shear, as these are insignificant at the scales used in this analysis \cite{Krause2021, Lange2024}. We verify this assumption on $N$-body simulations that include these effects in \S\ref{sec:simval}. The intrinsic alignment contribution can be expressed as 
\begin{align}
    \gamma_{E,I}^{i}(\hat{\bf n}) =  \int d\chi\ w^{\gamma_{E,I}^{i}}(\chi) \gamma_{E,I}^{i}(\hat{\bf n}\chi, z(\chi)),
\end{align}
\noindent 
where $w^{\gamma_{E,I}^{i}}(\chi) = p^{\gamma_{E}^{i}}(z(\chi)) E(z(\chi))$ and $p^{\gamma_{E}^{i}}$ is the source galaxy selection function, i.e., the galaxy redshift distribution for the $i$th source bin normalized to integrate to one, and $E(z)$ is the Hubble parameter at $z$. The gravitational lensing term is given by
\begin{align}
    \kappa^{i}(\hat{\bf n}) =  \int d\chi\ w^{\kappa^{i}}(\chi) \delta_{m}(\hat{\bf n}\chi, z(\chi))\, ,
\end{align}
\noindent 
where
\begin{align}
    w^{\kappa^{i}}(\chi) &= \frac{3}{2}\Omega_{m,0}H_{0}^2(1+z(\chi))\int_{z(\chi)}^{\infty} dz^{\prime} g^{i}(z(\chi), z^{\prime}) \nonumber \\
    g^{i}(z, z^{\prime}) &= \frac{\chi(z)(\chi(z^{\prime}) - \chi(z))}{\chi(z^{\prime})} p^{\gamma_{E}^{i}}(z^{\prime}).
\end{align}

Similarly, the observed galaxy density field for the $i$th lens bin can be expressed as 
\begin{align}
    \delta_{g,\rm obs}^{i}(\hat{\bf n}) = \delta_{g}^{i}(\hat{\bf n}) + \delta_{g,\mu}^{i}(\hat{\bf n}) \, ,
\end{align}
where $\delta_{g}^{i}(\hat{\bf n})$ is the projected intrinsic real-space galaxy density field, and the second term on the right-hand-side is the lens magnification contribution. In order to neglect the impact of redshift-space distortions, we fit only to $\ell > 50$, where the beyond-Limber and redshift-space-distortion effects impact our observables at the $<1\%$ level \cite{Sailer24}. 

We can express the projected intrinsic real-space galaxy density field as:
\begin{align}
    \delta_{g}^{i}(\hat{\bf n}) = \int d\chi\ w^{\delta_{g}^{i}}(\chi) \delta_{g,3D}(\hat{\bf n}\chi, z(\chi))\, ,
\end{align}
\noindent
and $w^{\delta_{g}}(\chi) = p^{\delta_{g}^{i}}(\chi) E(\chi(z))$, and $p^{\delta_{g}^{i}}(\chi)$ is the lens galaxy selection function. The magnification contribution is
\begin{align}
    \delta_{g,\mu}^{i}(\hat{\bf n}) = 2(\alpha_{\mu}^{i} - 1) \int d\chi\ w^{\delta_{g,\mu}^{i}}(\chi) \delta_{m}(\hat{\bf n}\chi, z(\chi))\, ,
\end{align}
\noindent where $w^{\delta_{g,\mu}} = w^{\kappa}$, and $\alpha_{\mu}^{i}$ is the response of the galaxy angular number density, $n^{i}$, to a change in convergence:
\begin{align}
\label{eq:magnification_coeff}
    \frac{1}{n^{i}}\frac{d n^{i}}{d\kappa} = 2 (\alpha_{\mu} - 1)\, .
\end{align}

\subsection{Angular power spectra and effective redshifts}
\label{sec:effective_redshift}

In order to predict the angular power spectra of the projected field discussed above, we use the Limber approximation\cite{Limber53,LoVerde08}
\begin{align}
\label{eq:limber}
    C^{ab}_{\ell} &= \int d\chi\ \frac{w^{a}(\chi)w^{b}(\chi)}{\chi^2} \nonumber \\
    &P_{ab}\left( k_{\perp}=\frac{\ell + \frac{1}{2}}{\chi}, k_{\parallel}=0 ; z(\chi)\right) + \mathcal{O}(\ell^{-2}) ,
\end{align}
where $w^{a}(\chi)$ and $w^{b}(\chi)$ are the projection kernels appropriate for fields $a$ and $b$, and $P_{ab}$ is the cross-power spectrum between these fields evaluated at wave-vectors $\bk = (k_\perp,k_\parallel)$ perpendicular to the line of sight. This is an excellent approximation for angular scales $\ell$ that we fit in this work. Given the field level description presented above, we can express the two main spectra of interest:
\begin{align}
C^{\delta_{g, \rm obs}^{i},\delta_{g, \rm obs}^{j}}_{\ell} &= C^{\delta_{g}^{i},\delta_{g}^{j}}_{\ell} + C^{\delta_{g}^{i},\delta_{g, \mu}^{j}}_{\ell} + C^{\delta_{g}^{j},\delta_{g, \mu}^{i}}_{\ell} + C^{\delta_{g, \mu}^{i},\delta_{g, \mu}^{j}}_{\ell} \nonumber \\
C^{\delta_{g, \rm obs}^{i}, \gamma_{E}^{j}} &= C^{\delta_{g}^{i},\kappa^{j}}_{\ell} + C^{\delta_{g}^{i},\gamma_{E,I}^{j}}_{\ell} + C^{\delta_{g, \mu}^{i},\kappa^{j}}_{\ell}.
\label{eq:cell_components}
\end{align}

The spectra in Equation~\ref{eq:cell_components} with at least one power of $\delta_g^i$ have Limber integrals that are highly localized due to the narrowness of the lens galaxy redshift distributions $p^{\delta^i_g}(z)$. This implies that we can make an additional approximation and subsitute $z(\chi) \rightarrow z^{ab}_{\rm eff}$ in Equation~\ref{eq:limber}, where the effective redshift is given by
\begin{align}
    z_{\rm eff}^{ab} = \int d\chi\ \frac{w^{a}(\chi)w^b(\chi)}{\chi^2}z(\chi)
    \label{eqn:zeff}.
\end{align}
This choice cancels corrections to the evolution of clustering at linear order, with corrections coming in at quadratic order in the width of $p^{\delta_g^i}$\edit{, $\Delta z_i^2$ \cite{Chen22b}, removing the need to marginalize over the redshift evolution of galaxy clustering within each lens sample. While the precise size of corrections to the effective redshift approximation depends on the steepness of this evolution, theoretical calculations show that they should be sub-percent for the widths of our lens distributions \cite{Vlah_2016}, which we further validate against simulations with realistic redshift evolution in \S\ref{sec:simval}.}

An immediate consequence of Equation~\ref{eqn:zeff} is that $C^{\delta_g^{i}\delta_g^{i}}_{\ell}$ is sensitive to a different effective redshift than $C^{\delta_g^{i}\gamma_{E}^{j}}_{\ell}$, and the latter is sensitive to a different effective redshift for each $j$. Similar to \cite{Chen22b}, we remedy this by applying additional weights to our the $i$th lens galaxy sample when constructing galaxy overdensity maps for the purpose of measuring $C^{\delta_g^{i}\gamma_{E}^{j}}_{\ell}$:
\begin{equation}
    v_{g}^{j} = \frac{w^{\delta_{g}^{i}}(\chi(z_g))}{w^{\kappa^{j}}(\chi(z_g))}\, ,
\end{equation}
\noindent where $z_g$ is the photo-z estimate of each lens galaxy. In doing so we make the effective redshifts of $C^{\delta_g^{i}\delta_g^{i}}_{\ell}$ and $C^{\delta_g^{i}\kappa^{j}}_{\ell}$, which are the most significant terms in the spectra that dominate our cosmological constraining power, equal to each other. In addition to constructing four additional galaxy overdensity maps, we must also compute four new lens galaxy selection functions, $p^{\delta_{g}^{i}}_j(z)$ taking into account the weights defined above for each source galaxy sample. These are then used to compute the model predictions for $C^{\delta_g^{i}\gamma_{E}^{j}}_{\ell}$. Adopting this additional weight in the cross correlation insulates our measurements against the redshift evolution of $P^{\delta_g\delta_g}$ and $P^{\delta_g\kappa}$, so that the galaxy auto and lensing cross correlations are probed at precisely the same epoch. Since the effective redshift is fixed to that of the lens auto correlation, there is no additional dependence on which source bin the lensing is measured from in these contributions. This implies that the two cosmological correlations from which we derive our constraining power can be modeled at equal times using a consistent set of parameters.

However, because we have chosen to construct weights to make the effective redshifts of $C^{\delta_g^{i}\delta_g^{i}}_{\ell}$ and $C^{\delta_g^{i}\kappa^{j}}_{\ell}$ equal, we must resign ourselves to the fact that the cross correlation of galaxy densities with intrinsic alignments\edit{, whose redshift distribution follow that of the source samples,} are sensitive to clustering at effective redshifts distinct from the effective redshift of the \edit{autocorrelation of galaxy densities $\delta_g^i$}, and will in addition also be dependent upon the source bin. However, we can use the fact that the lens distributions $p^{\delta_{g}^{i}}(z)$ are rather narrow to approximate the galaxy clustering sampled by these cross correlations to be the same as that for their auto-correlation. Since, however, $p^{\gamma_{e}^{i}}(z)$ is quite broad for all of our source galaxy bins due to photometric redshift uncertainties inherent to the much fainter source galaxy samples, the parameters describing the intrinsic alignments of the source galaxies cannot be treated as constant over the source bins. Rather, we must describe the intrinsic alignments of each source sample narrowly localized at each lens bin---this naturally leads to a proliferation of the possible degrees of freedom in our model, since each intrinsic alignment parameter must be described per source and per lens redshifts, i.e., $N_{\rm source} \times N_{\rm lens}$ times. We describe various ways to describe this freedom in \S\ref{sec:bias_evolution}. While the effective redshifts of $C^{\delta_g^{i}\gamma_{E,I}^{j}}_{\ell}$ are slightly different than those of the lens auto-correlations and are different for each source bin, since we are interested in IA primarily as a contamination to the main signal \edit{and analyze only cross correlations where IA constitute a few percent of the lensing signal, we expect the impact on our model predictions to be quite small, with most of the differences further soaked up by the definition of the IA bias parameters.}

Similarly, in the case of magnification, we expect that model predictions due to variations of $\alpha$ over our lens redshift bins are relatively small and thus we only leave one magnification coefficient, $\alpha_{i}$, free per lens bin. Ref.~\cite{Wenzl24} investigated the effect of redshift evolution of the magnification coefficient in the BOSS survey and found that ignoring it incurred systematic errors in the predicted clustering roughly comparable with $\pm 0.1$ errors in the magnification coefficient, though this error is again tied to the width of the redshift distribution and could be removed by accurately measuring this evolution for spectroscopically calibrated samples. Rather than include this effect in our modeling, since the measurements in our fiducial setup (\S\ref{sec:projection_effects}) are relatively insensitive to magnification, we simply include this error in the width of our priors on $\alpha_i$.

Finally, we note that we have omitted the cross-term between lens magnification and intrinsic alignments, $C^{\delta_{g, \mu}^{i},\gamma_{E,I}^{j}}_{\ell}$. This is because our fiducial modeling choices allow for one set of IA parameters per lens-source bin combination, as discussed in Sec. \ref{sec:bias_evolution}. Under this assumption, there is no unique way to interpolate and extrapolate the IA parameters as a function of redshift in order to model $C^{\delta_{g, \mu}^{i},\gamma_{E,I}^{j}}_{\ell}$ over the very broad redshift range required, due to the width of the source bin redshift distributions. Evaluating the impact of this term using our fiducial cosmology and nuisance parameters, and a constant value of $c_s=-1$, we find its impact to be very small, contributing a $\Delta \chi^2<0.7$. As such, we neglect this cross-term in the analysis presented here.

\subsection{Lagrangian Perturbation Theory and Hybrid Effective Field Theory}
\label{sec:lpt}

The only remaining ingredients required to specify our models for the angular power spectra above are the power spectra, $P_{ab}(k)$, to be used in Eq. \ref{eq:limber}. In this work, we adopt the formalism of Lagrangian perturbation theory (LPT) and hybrid effective field theory (HEFT), which model the formation of large scale structure by predicting the displacements, $\Psi(\bq, \tau)$ of fluid elements originating at Lagrangian positions $\bq$, mapping to final positions $\bx = \bq + \Psi(\bq, \tau)$. These fluid elements follow Newtonian gravity in an expanding space-time such that $\ddot \Psi + \mathcal{H} \dot \Psi = -\nabla_{\bx}\Phi$, where the dots denote derivatives with respect to conformal time. The potential, $\Phi(\bx,\tau)$ is sourced by the matter density $\delta_{m}(\bx,\tau)$, which is given by number conservation as
\begin{equation}
    1 + \delta_{m}(\bx,\tau) = \int d^{3} \bq\ \delta_{D}(\bx - \bq - \Psi(\bq, \tau)). 
\end{equation}
Within LPT these displacements are computed perturbatively order-by-order, and the first order solution is often referred to as the Zeldovich approximation. In HEFT, these displacements are computed non-perturbatively using $N$-body simulations.

LPT makes predictions for the large-scale statistics of galaxy properties, such as the galaxy overdensity field $\delta_g$ or density-weighted galaxy ellipticity field $M_{ij}(\bx)=(1+\delta_{g}(\bx)) I_{ij}(\bx)$, where $I_{ij}$ is the galaxy shape, by enumerating their responses to their local initial conditions order-by-order in a bias functional 
\begin{align}
    F(\bq) = \sum_{a} b_{\mo_a} \mo_{a}(\bq)\, ,
\end{align}
and advecting this field to the late-time coordinates following
\begin{equation}
\label{eq:advection}
    1 + \delta(\bx,\tau) = \int d^{3}\bq\, (1 + F(\bq)) \delta_{D}(\bx - \bq - \Psi(\bq, \tau))\, .
\end{equation}
The operators $\mo_a$ can either be scalars or tensors for densities and ellipticities, respectively.
For convenience we can also define the advected operators
\begin{align}
    \mo_a(\bx,\tau) &= \int d^3\bq \ \mo_a(\bq)\ \delta_D(\bx - \bq - \Psi(\bq,\tau)) \nonumber \\
    \mo_a(\bk,\tau) &= \int d^3\bq \ e^{-i\bk\cdot(\bq+\Psi(\bq))}  \ \mo_a(\bq)\,.
    \label{eq:zel_ops}
\end{align}
Both the perturbative dynamics and bias expansion described above are properly thought of as effective theories, and the inclusion of additional operators, or counterterms, to tame the dependence on small-scale physics will require additional free parameters in the model. On the other hand, we emphasize that the bias expansion is a systematic one, by which we mean that any physical effect on perturbative scales can necessarily be expressed as a bias contribution at some order in the theory, without needing to individually account for such effects (see e.g. ref.~\cite{Kokron_2021} for the case of assembly bias). We now describe LPT as applied to densities and ellipticities in turn.


\subsubsection{Matter and galaxy density}
\label{sec:density}

In the case of galaxy densities, we have up to one-loop order \cite{McDonald_2009,Desjacques:2016bnm}:
\begin{align}
 \label{eq:bias_exp}
    \delta_g[\delta(\bq)] &\approx  1 + b_1 \delta_0 + b_2 (\delta^2_0 - \langle \delta^2_0\rangle) + \\
 & b_s (s^2_0 - \langle s^2_0 \rangle) + b_3 \mo_3(\bq) +  b_{\nabla^2}\nabla^2 \delta_0(\bq) + \epsilon(\bq) \nonumber \, ,
\end{align}
where the subscript ``$0$'' denotes that all quantities are computed according to the linear initial field, $s^2_0$ is the square of the traceless tidal tensor, and we have suppressed the $\bq$ dependencies on the RHS of this equation. The contribution from $b_{\nabla^2}$ is an effective theory term that captures both short-range non-localities in galaxy formation and other small-scale effects in the dynamics of galaxies, while $\epsilon$ stands for uncorrelated stochastic modes that have a white spectrum. The operator $\mo_3 = \frac13 st$ is a stand-in cubic operator, since all cubic operators contribute identically to the power spectrum at one-loop order. Since the contributions to our galaxy samples are expected to be small, and $b_3$ is rather degenerate with $b_{\nabla^2}$, we do not vary it here.
Finally, we make the ansatz that all of these quantities are computed from the CDM+baryon field, rather than the total matter field. This is motivated by the fact that neutrinos do not cluster on the typical scale of dark matter halos, and thus we expect galaxies to trace the CDM+baryon field rather than the total matter field. This ansatz was shown to be in excellent agreement with CDM+neutrino simulation predictions of dark matter halo clustering by \cite{Castorina15,Bayer2020}.

In addition to the analytic one-loop effective Lagrangian perturbation theory, also known as convolutional Lagrangian effective field theory (CLEFT), in this work we also use a simulation-enhanced extension called Hybrid Effective Field Theory (HEFT). HEFT assumes the Lagrangian bias expansion in Equation \ref{eq:bias_exp}, but uses nonlinear displacements computed exactly from $N$-body simulations in Equation \ref{eq:advection}, rather than perturbatively computing $\Psi$, as is done in CLEFT. In doing so, it has been shown that real-space galaxy--galaxy and galaxy--matter power spectra can be jointly fit to $k=0.6\, h\, \rm Mpc^{-1}$, well beyond the scales where perturbation theory models are traditionally used \cite{modichenwhite19,Kokron_2021,zennaro2023bacco,Nicola24}. This is because for sufficiently low mass and low bias tracers, the dynamical nonlinear scale $\Psi$ is larger than the halo scale controlling the convergence of the bias expansion. For highly biased tracers, it is possible that this no longer holds, but in \S\ref{sec:simval} we show that for simulations that match the bias and number densities of the samples used here, we can obtain unbiased cosmological constraints fitting to $k=0.6\, h\, \rm Mpc^{-1}$. Thus, here we adopt HEFT as our fiducial model.

Our angular power spectra require as inputs the real-space galaxy--galaxy, galaxy--matter and matter--matter power spectra. The former two can be expressed in both CLEFT and HEFT as quadratic and linear polynomials in the bias parameters:
\begin{align}
    \label{eq:pgg}
    P_{\delta_{g}\delta{g}}(k) &= \sum_{\mo_i,\mo_j\in \delta_g} b_{\mo_i}b_{\mo_j}P_{\mo_i\mo_j}(k) + P_\epsilon \\ 
    \label{eq:pgm}
    P_{\delta_g\delta_{m}}(k) &= \sum_{\mo_i\in \delta_g} b_{\mo_j}P_{1\mo_i}(k)
\end{align}
where again, $b_{\mo_i}$ are free bias coefficients that we marginalize over. These are shared between $P_{\delta_{g}\delta{g}}$ and $P_{\delta_{g}\delta_{m}}$, except for the counter-terms, $b_{\nabla^2 a}$ and  $b_{\nabla^2 x}$, that contribute to $P_{\delta_{g}\delta{g}}(k)$, $P_{\delta_g\delta_{m}}(k)$ and $P_{\delta_{m}\delta_{m}}(k)$ respectively. The stochastic spectrum $P_{\epsilon}$ is given at leading order by a single constant ``SN'' that varies for each tracer---we ignore any additional scale dependence in this work. 
Finally, we model the matter power spectrum as
\begin{equation}
    \label{eq:pmm}
    P_{\delta_{m}\delta_{m}}(k) = P^{\rm cdm}_{mm}(k) \left (1 - \frac{b_{\nabla^2 m} k^2}{1 + (kR)^2}\right) \,
\end{equation}
where $P^{\rm cdm}_{mm}$ is the matter power spectrum in the absence of baryonic feedback, equal to $P_{11}$ in HEFT. Here $b_{\nabla^2 m}$ is a counterterm that accounts for the leading-order effect of feedback and we have included a Padé factor $R = 2\, h^{-1} \Mpc$ to tame the large-$k$ behavior similar to that used in \cite{Sullivan21}. We discuss the efficacy of this parametrization further in \S\ref{sec:scale_cuts}, but we emphasize that it only contributes to the magnification terms in our model and as such we are quite insensitive to the impact of baryonic feedback. As an extreme example of this insensitivity, we can replace the nonlinear matter power spectrum in Equation $\ref{eq:pmm}$ with the one-loop matter power spectrum and our results are unchanged.

\subsubsection{Intrinsic galaxy ellipticity}
\label{sec:ias}

We can similarly expand \edit{the galaxy shape field} $M_{ij}(\bq)$ perturbatively using a bias expansion. Following ref.~\cite{Chen24}, this bias expansion can be expressed in terms of the Lagrangian shear tensor $L_{ij} = \del_{i}\Psi_{j}$ (see also refs.~\cite{Vlah20,Taruya21}). It will be useful to decompose $M_{ij}$ into its scalar trace and trace- free components:
\begin{equation}
    \mathrm{Tr}\{ M_{ij} \} = M  \, , \mathrm{TF} \{ M_{ij} \} =  M_{ij} - \frac{1}{3} \delta_{ij} M\,
\end{equation}
\noindent 
\edit{such that we can write $M_{ij} = \bar{M} ((1 + \delta_{M}) \delta_{ij} / 3 + g_{ij, I})$, where $\delta_{ij}$ is the Kronecker delta and $\bar{M}$ is the mean galaxy density weighted by size. The trace-free component with $\bar{M}$ normalized out} $g_{ij}$ is the three-dimensional intrinsic shape overdensity field, which is what we require in order to make contact with the quantities reported in the DES Y3 \metacal catalog.

To one-loop order in perturbation theory, we can write 
\begin{align}
    g_{ij}[ & L_{ij}(\bq)] \approx  \, A_{1} s_{ij} + A_{\delta1}\delta s_{ij} + A_{t} t_{ij}  + A_{2} \mathrm{TF}\{s^2\}_{ij} \nonumber \\
 & + A_{\delta t} \delta t_{ij} + A_{3} \mathrm{TF}\{L^{(3)}\}_{ij} + \alpha_{s}\nabla^2 s_{ij} + \epsilon_{ij},
  \label{eq:shape_exp}
\end{align}
\noindent 
where $A_{i}$ are free galaxy shape bias coefficients and we have kept only two cubic operators as the rest are degenerate at one-loop order. Where it is possible in the above, we have rewritten contributions from the Lagrangian shear tensor in terms of quantities more familiar to the IA literature. For example, the linear Lagrangian shear has the density and tidal field as its trace and trace-free components:
\begin{equation}
    L_{ij}^{(1)} = -\frac{1}{3} \delta(\bq) \delta_{ij} - s_{ij}(\bq)\, ,
\end{equation}
and the Lagrangian
\begin{align}
    t_{ij} 
    = \frac{4}{3}\mathrm{TF}\left \{\mathbf{L}^{(2)}\right \}_{ij} \nonumber\, ,
\end{align}
\noindent is equal at leading order to the difference between the second-order matter overdensity and velocity divergence in Eulerian perturbation theory. 

The Lagrangian IA model, as defined by the above bias expansion, reflects a full accounting of all possible contributions to the galaxy shape at one-loop order. Previous analyses of cosmic shear and GGL have also employed perturbative models such as the nonlinear alignment (NLA) \cite{Joachimi2010} or the tidal alignment and tidal torquing models (TATT) \cite{Blazek19}. These models represent subsets of the space spanned by the six bias parameters above with one and three degrees of freedom, respectively: roughly, the NLA corresponds to a model with only $A_1$, while the TATT model also frees the equivalent of $A_{\delta 1}$ and $A_2$ in Eulerian space. However, we note that since the Lagrangian bias model includes nonlinear contributions from dynamical nonlinearities through the displacements $\Psi$ the predictions cannot be matched simply by setting the bias coefficients equal in both models \cite{Schmitz18} and that, at least for halos, the leading nonlinearities are qualitatively close to low-order \textit{Lagrangian} bias coupled with the nonlinear dynamics of the displacements\edit{, including when compared to the subset of nonlinearities included in the NLA model} \cite{Maion22,Akitsu23,Chen24}. In addition, the effective theory model includes corrections $\alpha_s$ and $\epsilon_{ij}$ which, while not included in previous models, is essential to account for the dependence on small scales beyond the reach of perturbation theory including baryonic effects and galaxy formation.

We can express the galaxy shape fields in Fourier space through the helicity basis \cite{Vlah20}
\begin{equation}
    M_{ij}(\bk) = \sum_{\ell=0}^2 \sum_{m=-\ell}^\ell M_{\ell,m} Y^{(m)}_{\ell,ij}(\hat{k})\, .
\end{equation}
where the basis tensors satisfy $\hat{z}_i \hat{z}_j Y^{(m)}_{\ell, ij} = Y^m_\ell(\Omega_{\bk})$. The trace-free and symmetric component of the shape field, in particular, is described by the five components with spin $\ell = 2$, while the galaxy density can be equivalently thought of as a one-component spin $\ell = 0$ field with $Y^{m}_{\ell,ij} = \delta_{ij}$. In this basis, the angular structure of tensor correlators in can be greatly simplified by symmetry arguments. In particular, rotational symmetry about $\mathbf{k}$ means that non-zero correlations can only exist between components of the same helicity independently of spin, e.g. only the $m=0$ component of the shape field correlates with the galaxy density \cite{Vlah20}. There is thus only one non-zero component of the density-shape cross-power spectrum:
\begin{equation}
    \langle \delta(\bk) g_{ij}(\bk^{\prime}) \rangle = \frac{3}{2}\left (\hat{k}_i\hat{k}_j - \frac{\delta_{ij}}{3}\right) P_{\delta I}(k) \delta_D(\bk - \bk^{\prime})\, ,
\end{equation}
\noindent where $P_{\delta I}$ is a scalar power spectrum. We can then write 
\begin{align}
\label{eq:ia_basis}
    P_{\delta_g I}(k) &= \sum_{\mo_{i} \in \delta_g, \mo_{j} \in g_{0}} b_{\mo_i}A_{\mo_j}P_{\mo_{i}\mo_{j}}(k) \\
    P_{\delta_m I}(k) &= \sum_{\mo_{j} \in g_{0}} A_{\mo_j}P_{1\mo_{j}}(k)\, ,
\end{align}
\noindent 
where $P_{\mo_{i}\mo_{j}}(k)$ are cross-spectra between advected operators $\mo_{i}(\bk)$ and $\mo_{j}(\bk)$ contributing to $\delta_{g}(\bk)$ and $g_{0}(\bk)$ respectively. The galaxy density-shape power spectrum can in addition receive a stochastic contribution proportional to $k^2$ due to the cross correlation of $\langle \epsilon \epsilon_{ij} \rangle$ but it is expected to be small for low-mass halos so we neglect it in this paper \cite{Chen24}. The shape--shape auto-spectra are similarly described by the three helicity auto-spectra for $m=0,1,2$, with helicities of different sign described by the same spectra due to parity symmetry \cite{Vlah20}.

Ref.~\cite{Chen24} showed that this model can fit three-dimensional shape--shape auto-spectra to $k\sim0.3\, h\, \rm Mpc^{-1}$ at a volume and statistical precision well beyond what is required in this work, while \cite{Bakx23} showed that a similar model \cite{Vlah20} is able to fit projected density-shape cross-spectra to the same scale similarly well. We fit slightly beyond this scale for our fiducial analysis, but because the spectra where we obtain most of our constraining power have relatively small IA contributions, and taking into account the stringent nature of the tests in the aforementioned works, we believe that this is not an issue.

Galaxy lensing surveys measure the projected, rather than  three-dimensional, shapes of galaxies. These two-dimensional shape fields are conventionally decomposed into E and B modes, with the weak lensing signal captured by the former. The angular power spectra of the shape fields and their cross correlations with galaxy densities can be expressed in terms of the three-dimensional helicity spectra above. For the density E-mode cross spectrum with the density we are interested in this work we have \cite{Vlah21,Bakx23}
\begin{equation}
    P^{\delta \gamma_{E,I}}(k) = \frac{3}{4} (1 - \mu^2) P_{\delta I}(k),
\end{equation}
where $\mu = k_{\parallel}/k$, which we can then plug into Equation~\ref{eq:limber}. The same logic dictates that the E- and B-mode auto-spectra are given by linear combinations of the helicity spectra, with the former given by the $m=0,2$ spectra and the latter by $m=1$ in the plane of the sky ($k_\parallel = 0$), while parity dictates that the cross-correlations of B modes with the density and E modes must be zero.\footnote{Ref.~\cite{Bakx23} pointed out that the definition of galaxy shape used in conventional weak lensing surveys is normalized by the \textit{projected} shape of galaxies, itself a line-of-sight dependent quantity, and therefore it breaks many of the symmetry properties discussed above. However, these symmetry-breaking effects seem to be tolerable for the purpose of galaxy--galaxy lensing analyses and suppressed at leading order in perturbation theory, so we leave the proper definition of galaxy shapes for future work.}

\subsection{Bias priors and redshift evolution}
\label{sec:bias_evolution}

Given \S\ref{sec:effective_redshift} and \S\ref{sec:lpt}, the available dynamical degrees of freedom in our model are therefore the bias and effective-theory parameters describing matter, galaxy, and intrinsic alignments clustering at each lens redshift and for each independent (source or lens) galaxy sample. Our fiducial choice will be to sample combinations of bias parameters and the matter clustering amplitude $\sigma_8$ that roughly correspond to the same physical galaxy clustering. For example, for the linear bias, we sample the combination
\begin{equation}
    \tilde{b}_1^E = (1 + b_1) (\sigma_8(z)/\sigma_{8,\rm fid}),
\end{equation}
which denotes the linear clustering of galaxies on $8\ h^{-1}\ \Mpc$ scales. Similarly for each higher-order bias parameter our fiducial choice will be to sample them in the combination
\begin{equation}
    \tilde{b}_{O^{(n>1)}} = b_{O^{(n>1)}} \left( \sigma_8(z)/\sigma_{8,\rm fid} \right)^n,
    \label{eqn:tilde_bias}
\end{equation}
where $n$ is the order of the bias operator, such that the clustering due to each operator $O^{(n)}$ is roughly constant when the sampling parameter is fixed. We explore the consequence of this choice, particularly in the case of intrinsic alignments, in \S\ref{sec:noiseless_sims}. \edit{Similar bias parameter scalings when setting priors have been used in a number of previous works (e.g. 
 \cite{Pandey2020,Chen22b,Maus24b}), but this is the first such application in the case of intrinsic alignments.}

For the bias counterterms we choose to sample over their contribution quoted as a fraction of the linear contribution at $k_{\rm max}$, i.e.,
\begin{equation}
\label{eq:counterterm}
    \tilde{b}_{\nabla^2,a/x} =
    \left( 2 k_{\rm max, fid}^2/(1 + b_1)\right) b_{\nabla^2,a/x}
\end{equation}
where $k_{\rm max, fid}=0.4\, h\, \rm Mpc^{-1}$. Should the data push $\tilde{b}_{\nabla^2,a/x}$ to the edge of its prior, it would directly indicate that this correction is not perturbative at $k_{\rm max}$, requiring us to relax the analysis scale cut. Note also that had we used a prior independent of $b_1$, we would need to use significantly different counter-term priors for each lens bin in order to obtain reasonable priors on these terms' contributions as a fraction of linear theory for all bins given the very different biases of the BGS and LRG samples. For our HEFT analyses, we set priors centered at zero such that our counterterms contribute $20\%$ of the linear bias contribution at $k_{\rm max, fid}$ at $1\sigma$, while for CLEFT analyses we relax this to $50\%$ to account for additional dynamical uncertainty.

In the case of the intrinsic alignment parameters we additionally use the normalization convention
\begin{equation}
    A_{O^{(n)}} = c_{O^{(n)}} \bar{C}_1 \rho_c \Omega_{m,\rm fid}
    \label{eq:ia_normalization}
\end{equation}
in order to make contact with constraints from existing surveys. Here $\rho_c$ is the comoving critical density, $\Omega_{m,\rm fid}=0.31$ and $\bar{C}_1$ is a constant conventionally fixed to $\bar{C}_1 = 5 \times 10^{-14}h^2 M_{\odot}{\rm Mpc}^{3}$ \cite{Brown02}. We use a fiducial value of $\Omega_{m, \rm fid} = 0.31$ in the pre-factor to avoid unmotivated cosmological dependence in our prior, which can additionally lead to projection effects in our marginalized posteriors. In the literature, this normalization often also includes a factor of $D(z)^{-n}$, where $D(z)$ is the growth factor; in our case this additional factor is implicitly included by sampling $\tilde{A}$ instead. It is useful to note that that the constant normalization factors in front of each IA coefficient are equal to 0.0043.

Let us turn to the redshift evolution of the bias parameters. For the galaxy density, the effective redshift approximation implies that we only need to sample the bias parameters at the effective redshift for each lens bin without worrying about the redshift evolution in each sample. This is the choice adopted by most galaxy clustering analyses, including this one, and also spans the full physical degrees of freedom allowed.

For galaxy--IA cross correlations, the same logic implies that we need to sample the value of each IA parameter at each of the $N_{\rm source} \times N_{\rm lens}$ effective redshifts in our problem. This product accounts for the fact that (a) each source bin is an independent sample that (b) is spread over a significant redshift range $p^{\gamma^i_e}(z)$ such that significant redshift evolution can occur between each lens bin. This \textit{maximally agnostic intrinsic alignment redshift dependence} (\texttt{MAIAR}) will be our fiducial choice, and results in a large multiplication in the number of IA parameters. These parameters enter linearly into our model predictions for $C^{\delta_g^{i}\gamma_{E}^{j}}_{\ell}$, and so can be analytically marginalized over making our analyses computationally tractable. \edit{The number of parameters that are included in the \texttt{MAIAR} model, particularly non-linear combinations of IA and other nuisance parameters, lead to significant projection effects if care is not taken to mitigate them through careful choices of priors and removal of unconstraining data as discussed in Section~\ref{sec:noiseless_sims}. Note that similar effects have been observed in previous work \cite{Joachimi2020,Amon2021,Krause2021}.}

For the purposes of comparison with the IA parameterizations made in past works, we also investigate models where each IA parameter has a straightforward redshift dependence $c_{O}(z)$, independent of the source sample. A common choice (e.g. \cite{desy3}) is to assume a power-law redshift dependence
\begin{equation}
\label{eq:ia_power_law}
    c_{O}(z) = c_O(z_{\rm fid})  \left(\frac{1 + z}{1 + z_{\rm fid}}\right)^{\eta_{O}},
\end{equation}
where $z_{\rm fid}=0.62$ is the pivot redshift and the free parameters are then the normalization $c_O$ and slope $\eta_O$. As an alternative choice we can use a spline basis \cite{KP4s2-Chen}
\begin{equation}
    c_O(z) = \sum_{m=0}^{N} c_{O,m} W\left( \frac{z - z_{\rm min}}{\Delta} - m \right),
    \label{eqn:spline}
\end{equation}
where $\Delta$ is a pre-set redshift spacing defining the smoothness of the redshift dependence and the spline covers points between $z_{\rm min}$ and $z_{\rm max} = z_{\rm min} + N \Delta$. For simplicity we choose a linear spline basis such that $W(x) = \text{max}(0,1-|x|)$. In the limit of two points $m=0,1$ this is equivalent to a linear $c_O(z)$ with the two coefficients being the value of the bias parameters at the bracketing redshifts. The advantage of this basis, in addition to being more flexible, is that the free parameters $c_{O,m}$ enter linearly into $C^{\kappa,g}$ and so can be analytically marginalized. For both of the above parameterizations we scale the amplitudes as above.

Finally, let us briefly describe our specific choices of priors for the ($\sigma_8$-normalized) density and shape bias parameters, as listed in Table~\ref{tab:params}. For the density biases, in addition to the counterterm priors discussed above, we sample the linear term with an uninformative, uniform prior and the rest with normal distributions $\mathcal{N}(0,1)$.  The latter choice is substantially wider than those found in simulations for galaxy samples like our own \cite{Abidi:2018eyd,Kokron:2021faa, zennaro2021priors,Ivanov24,Cabass24}. The stochastic contribution to the density is rather degenerate with the counterterm contribution for galaxy densities, and as such we choose an (informative) Gaussian prior allowing for up to $30\%$ deviations from Poissonian shot noise based on results in simulations \cite{Kokron:2021faa}.

For the intrinsic alignment priors, we choose $c_{\mo_{i}}$ such that our priors on $A_{\mo_i}$ cover the values that we expect of the halos hosting the DES Y3 source galaxies \cite{Bakx23,Akitsu23}, with the assumption that the shapes of halos carry higher degrees of IAs than do those of galaxies. We further assume that the priors on the linear alignment contribution are centered at negative values. Our priors further generously cover the values of IAs found in direct measurements of LRGs from spectroscopic sample, which are expected to be less stochastic and more aligned than the DES source galaxies \cite{Mandelbaum06,Singh2015, Samuroff2023, Kurita23}. Our normalization convention further allows the priors to widen with redshift beyond $z=0$ roughly as have been observed in simulated halos  \cite{Akitsu23}. \edit{We note that our priors are somewhat wider than the constraints obtained from analyses of cosmic shear data. This is because, as we will discuss in Section~\ref{sssec:iaz}, the IA constraints from these analyses rely on more rigid parametrizations of the redshift evolution of IAs, such that they mainly probe IAs close the the peak of source galaxy distributions which contribute negligibly to our cosmological constraints.}

\subsection{Scale cuts}
\label{sec:scale_cuts}
\begin{figure*}[htb!]
    \centering
    \includegraphics[width=\textwidth]{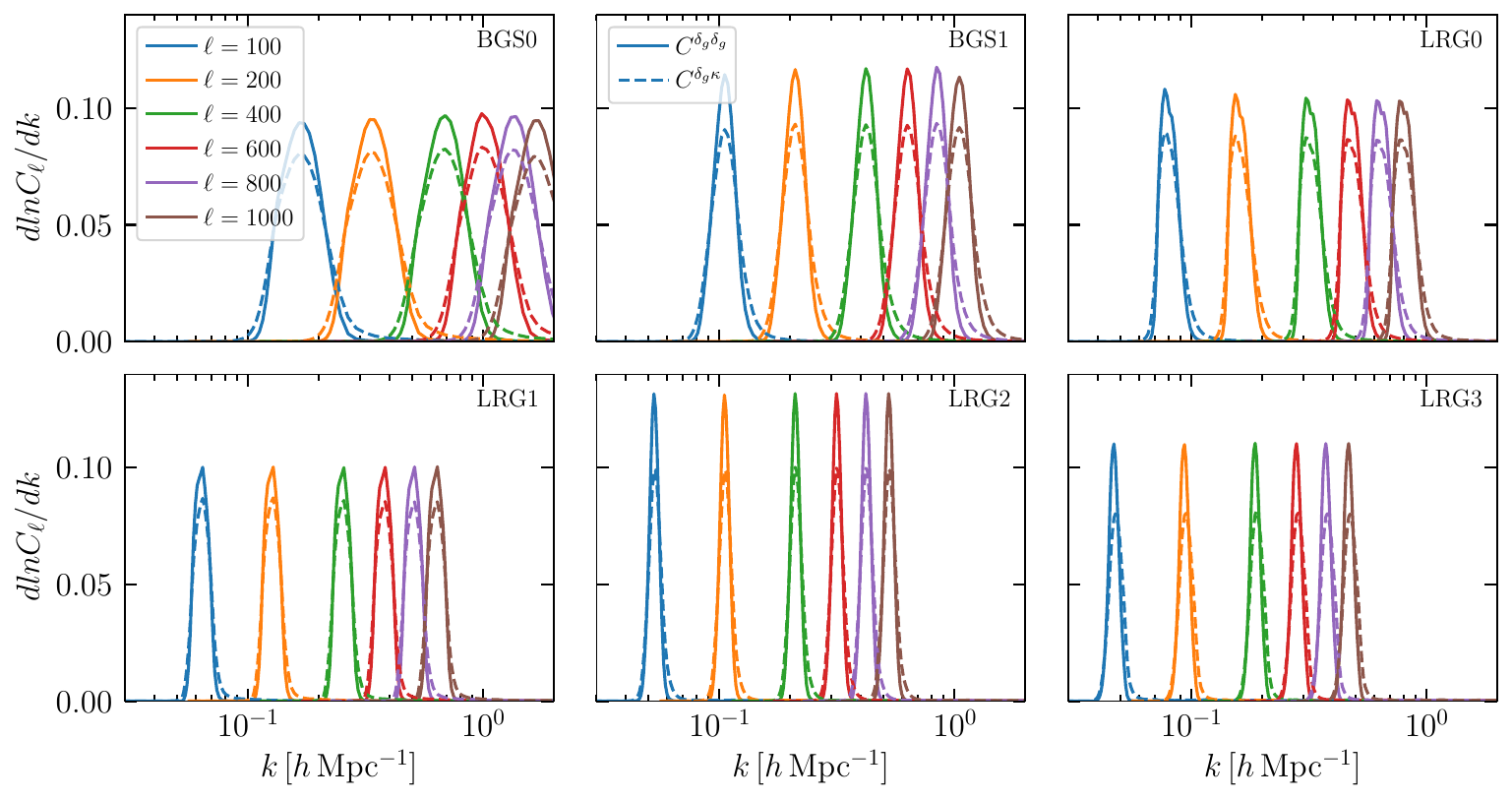}
    \caption{Fractional contributions to $C^{\delta_{g}\delta_{g}}_{\ell}$ (solid) and $C^{\delta_{g}\kappa}_{\ell}$ (dashed) as a function of $k$ in the Limber approximation. We make scale cuts to remove any $\ell$ bins that receive more than a $10\%$ contribution from $k>k_{\rm max}$, taking $k_{\rm max}=0.4\, h\, \rm Mpc^{-1}$ in our fiducial analysis.}
    \label{fig:k_support}
\end{figure*}

Now that we have specified our models for the angular power spectra of interest, we describe how we determined which scales to use in our likelihood analysis. In order to mitigate any theoretical systematics, we wish to remove data points from our analysis that receive contributions from scales where we believe our model is inapplicable. In order to determine this, we compute the response
\begin{align}
    \frac{1}{\ell + \frac{1}{2}}\frac{\delta \ln C^{ab}_{\ell}}{\delta \ln P^{ab}(k)} = \frac{1}{C^{ab}_{\ell}}w^{a}\left(\frac{\ell + \frac{1}{2}}{k}\right) w^{b}\left(\frac{\ell + \frac{1}{2}}{k}\right)P^{ab}(k)\, 
\end{align}
\noindent of our projected observables to the three-dimensional power spectrum in order to determine the fractional contribution of each $k$ to a given angular scale $\ell$. The results are shown in Figure~\ref{fig:k_support}. Importantly, both $C^{\delta_g \delta_g}_\ell$ and $C^{\delta_g \kappa}_\ell$ have rather narrow support in $k$-space, allowing us to cleanly separate perturbative and very nonlinear scales in our analysis. We note the same would not be true for the lensing auto-correlation due to the width of the lensing kernel. We can thus make scale cuts such that the total contribution from $k>k_{\rm max}$ is less than $10\%$ in both $C^{\delta_{g}\delta_{g}}$ and $C^{\delta_{g}\kappa}$. For our fiducial analysis, we use $k_{\rm max}=0.4\, h\,\rm Mpc^{-1}$, and the scales used with this scale cut are shown in the non-greyed out regions of Figure~\ref{fig:bestfit_model}. The impact of intrinsic alignment contributions in this context is negligible, as for a given $\ell$ the IA contribution almost always comes from equal or lower values of $k$ than the $C^{\delta_{g}\kappa}$ contribution to $C^{\delta_{g,\rm obs}\gamma_{E}}$. 

\begin{figure}[htb!]
    \centering
    \includegraphics[width=\linewidth]{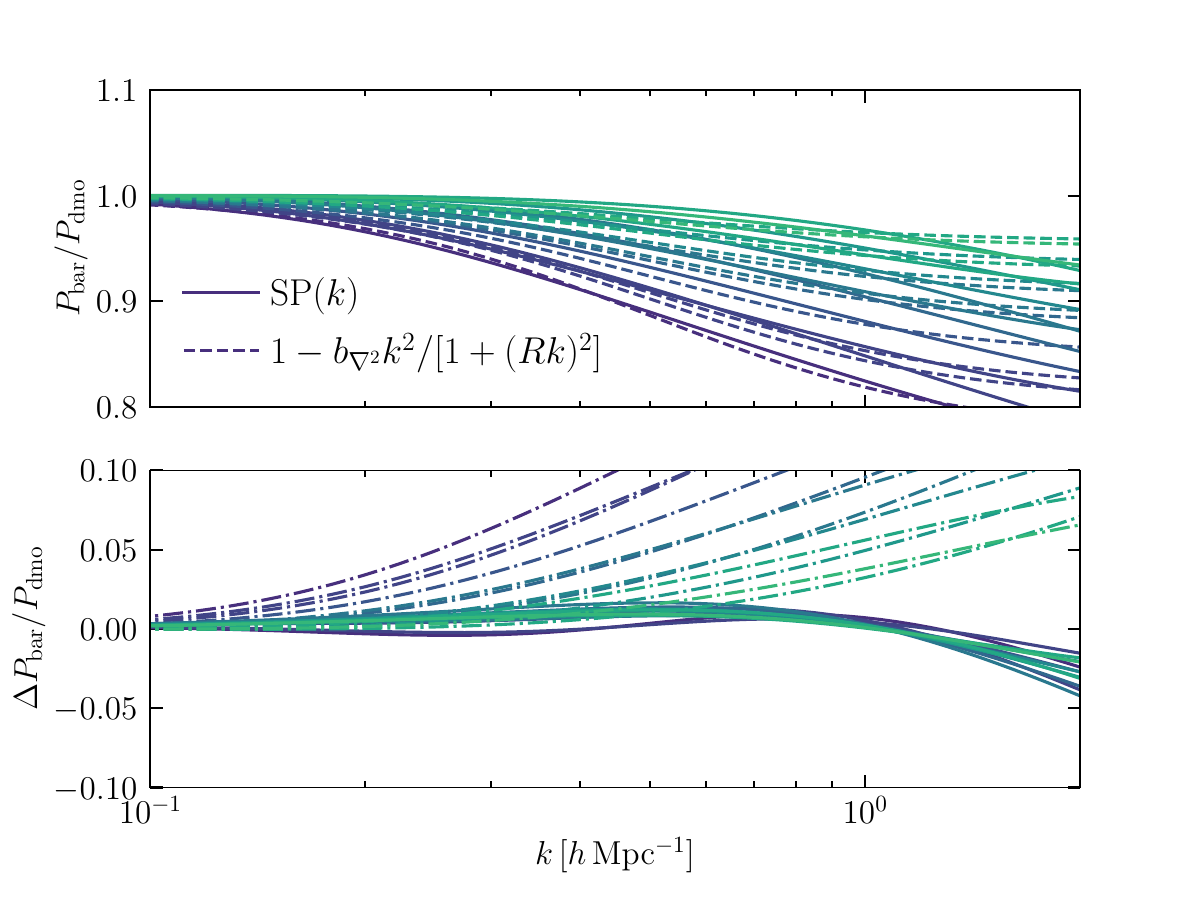}
    \caption{Comparison of our counter-term model for the impact of baryons on the matter power spectrum with SP($k$), a model fit to a broad range of hydrodynamic simulations \cite{Salcido23}. The top panel shows a comparison of the two models, where we have fit our counter-term model (dashed) to each of a number of different points in the SP($k$) parameter space (solid). The bottom panel shows the fractional error of our model (solid) as well as the error one would make by neglecting the impact of baryons entirely and just using our dark-matter-only $P_{mm}(k)$ model (\edit{dot-}dashed).}
    \label{fig:baryon_counterterm}
\end{figure}

On the other hand, the lens magnification contribution to $C^{\delta_{g,\rm obs}\gamma_{E}}$ is sensitive to significantly more nonlinear scales than the $C^{\delta_{g}\kappa}$, due to the significant support of the lensing kernels at low redshift. This issue is partially mitigated by the fact that this term only requires knowledge of the matter power spectrum, and does not rely on a perturbation theory and so our modeling of it is limited mainly by our ability to model the effects of baryonic feedback on $P_{mm}(k)$. As shown in Figure~\ref{fig:baryon_counterterm}, the counter-term that we include in our model for $P_{mm}(k)$ is capable of fitting a broad range of baryonic feedback scenarios as modeled by SP($k$) \cite{Salcido23} at the $\sim 1-3\%$ level to $k=2\, h\, \rm Mpc^{-1}$. Here we evaluate SP($k$) at our fiducial cosmology, varying the power-law parameterization of the baryon fraction as a function of halo mass at $z=0.125$, the minimum redshift that SP($k$) is reliable to, although we do not find that the performance of our counter-term model is significantly sensitive to the redshift that we perform this test at.

Since we believe we can sufficiently model the matter power spectrum to $k=2\, h\, \rm Mpc^{-1}$, we just need a method for marginalizing over the residual magnification contributions from scales smaller than $k_{\rm UV} = 2\, h\, \rm Mpc^{-1}$. Making the observation that $w^{\kappa}\sim \chi \sim \ell / k$ for small $\chi$, we can approximate this contribution to $C^{\delta_{g,\mu}\kappa}$ and $C^{\delta_{g,\mu}\delta_{g,\mu}}$
noting that
\begin{align}
    C^{\kappa,\kappa}_{\ell,\rm UV} &\rightarrow \left( \frac{3}{2}  \Omega_{m,0} H_0^2 \right)^2\int_0^{\chi_{\rm min}} d\chi\ P_{mm}(\ell/\chi) \nonumber \\
    &=  \left( \frac{3 }{2} \Omega_{m,0} H_0^2 \right)^2 \ell \int_{k_{\rm UV}}^{\infty} dk\, k^{-2}P_{mm}(k) \nonumber \\
    &\equiv \left( \frac{3 }{2} \Omega_{m,0} H_0^2 \right)^2 \ell\ \alpha_{\mu, \rm UV}
    \label{eq:alpha_uv}
\end{align}
i.e., the unknown effect of short-wavelength modes on magnification, and lensing in general, can be approximated through a counterterm proportional to $\ell$. Note that the integrand is highly suppressed at small scales due to the $k^{-2}$ factor, with the largest contributions coming from modes with $k \gtrsim k_{\rm max}$---an estimate in a Planck $\Lambda$CDM cosmology using {\tt halofit} \cite{Smith2003,Takahashi2012} gives $\alpha_{\mu, \rm UV} \approx 30\ h^{-4} \text{Mpc}^4$. This counterterm is a \textit{universal} counterterm having to do with the small-scale matter density at $z\approx 0$ and not tracer dependent; its contribution to e.g.\ $C^{\delta^i_{g,\mu} \kappa}_\ell$  simply comes with an additional factor of the magnification bias $(2\alpha_\mu^i - 1)$. 

Higher $\ell$ corrections to the magnification contribution will be tracer dependent\footnote{They still depend upon universal integrals of the matter power spectrum, but with coefficients dependent on the galaxy redshift distribution.} but also significantly smaller on the scales we are interested in, and are in addition less UV sensitive. Roughly speaking the UV contributions as a function of angular scale can be written as a series $(H_0 \ell / k_{\rm UV})^n$. Note that this is also the small parameter that controls the size of the correction due to the redshift evolution of $P_{mm}$ neglected in Equation~\ref{eq:alpha_uv}, since they come about from Taylor expanding at low redshifts where $z \approx H_0 \chi \sim H_0 \ell / k_{\rm UV}$. 

In our analysis we keep the leading term with a prior width set by $40\%$ of the N-body only contribution. Specifically, since the emulator we use extends only to $k_{\rm emu} = 4\, h \text{Mpc}^{-1}$, we perform our integrals up to $k_{\rm UV}$ and compute the expected size of the correction from N-body modes up to $k_{\rm emu}$. A rough estimate using {\tt halofit} shows that this correction alone captures more than $80\%$ of the UV contribution in a dark-matter only universe, so we define
\begin{equation}
    \alpha_{\mu, \rm UV} = (1 + c_{\mu, \rm UV}) \int_{k_{\rm max}}^{k_{\rm emu}} dk\ k^{-2} P_{mm}^{\rm cdm}(k)
    \label{eq:alpha_uv_corr}
\end{equation}
and set a $40\%$ prior on $c_\mu$ such that it captures both the effect of modes missed by the emulator while also marginalizing up to a $20\%$ effect of baryons close to $k_{\rm UV}$.


\subsection{Source redshift and shear calibration uncertainty}
\label{sec:obs_sys_models}
Extensive work calibrating all sources of bias in the estimation of the source galaxy redshift distributions \cite{Myles2021} and multiplicative shear biases \cite{Gatti2021,MacCrann2021} was performed by the DES collaboration. Nevertheless there is still residual uncertainty in each of these that we must marginalize over. Following \cite{desy3}, we marginalize over a shift in the mean redshift, $\Delta z_i$, and a constant multiplicative bias, $m_i$, per source galaxy bin. 

To marginalize over $\Delta z_i$, we perform the following operation on the source galaxy selection functions:
\begin{align}
\label{eq:delta_z_source}
    p^{\gamma_{E}^{i}\prime}(z) = p^{\gamma_{E}^{i}}(z + \Delta z_i)\, ,
\end{align}
\noindent 
and to marginalize over shear multiplicative biases, we simply perform:
\begin{align}
\label{eq:shear_calib}
    C^{\delta_{g,obs}^{i}\gamma_{E}^{j}\prime}_{\ell} = (1 + m^{j})C^{\delta_{g,obs}^{i}\gamma_{E}^{j}}_{\ell}\, ,
\end{align}
where we use the same priors on these parameters as used in \cite{desy3}. Because we have spectroscopically determined the redshift distributions of the lens galaxies, we do not marginalize over any nuisance parameters related to their calibration. Nevertheless we do perform a test of the robustness of this assumption in \S\ref{sec:results}.

\subsection{\texttt{Aemulus} $\nu$ and perturbation theory codes}
\label{sec:numerics}

As described in the previous subsections, in this work we adopt Hybrid Effective Field Theory (HEFT) for galaxy and matter densities and Lagrangian perturbation theory (LPT), also known as Convolutional Lagrangian Effective Field Theory (CLEFT), for intrinsic alignments as our fiducial dynamical models. For the latter we use the publicly available code \texttt{velocileptors}\footnote{\href{https://github.com/sfschen/velocileptors}{https://github.com/sfschen/velocileptors/tree/master}} \cite{Chen_2020,Chen_2021}. For the former, we use the \texttt{Aemulus} $\nu$ emulator \footnote{\href{https://github.com/AemulusProject/aemulus_heft}{https://github.com/AemulusProject/aemulus\_heft}} \cite{DeRose2023} to generate HEFT predictions of $P_{\mo_i\mo_j}(k)$. This emulator is trained on a suite of 150 $N$-body simulations run over a seven parameter $w$CDM parameter space, including massive neutrinos. These simulations were run with \texttt{Gadget-3}\cite{Springel05}, initialized at $z=12$ using third-order LPT. The ICs were computed using an extended version of \texttt{Monofonic} \cite{Michaux2021} in order to include the effect of massive neutrinos on the initial CDM+baryon distribution \cite{Elbers2021} and \texttt{FASTDF} for the neutrino distribution \cite{Elbers2022}. These ICs properly account for the Newtonian nature of our simulations, i.e., lacking radiation and GR effects \cite{Zennaro2017} and are intentionally initialized at as low of a redshift as possible in order to mitigate discreteness effects \cite{Marcos2006, Garrison2016,Michaux2021}.

The emulator, which uses a combination of principle component analysis and polynomial chaos expansions, is trained on measurements from these simulations that have their statistical errors drastically reduced by means of Zeldovich control variates \cite{Kokron22,DeRose2022b}. \cite{DeRose2023} showed that the error on $P_{\mo_i\mo_j}(k)$ is significantly below the $1\%$ level at $k\le 1\, h\, \rm Mpc^{-1}$ and $z\le3$ for the dominant basis spectra, and we further validate this model's accuracy in \S\ref{sec:simval}.

The intrinsic alignment power spectra in Lagrangian perturbation theory were derived in ref.~\cite{Chen24} who also released the public Python-based \texttt{spinosaurus}\footnote{\href{https://github.com/sfschen/spinosaurus}{https://github.com/sfschen/spinosaurus}} code. \texttt{spinosaurus} computes these intrinsic alignment spectra using FFTs and includes a full resummation of long-wavelenght linear modes (CLEFT), as well as options to compute the unresummed and resummed Eulerian perturbation theory spectra. We use \texttt{spinosaurus} for all of our intrinsic alignment calculations in this work. Both \texttt{velocileptors} and \texttt{spinosaurus} use the same conventions for bias parameters, and we run both using the infrared resummation cutoff $k_{\rm IR}=0.1\, h\,\rm Mpc^{-1}$ and using the linear CDM+baryon power spectrum predictions from \textsc{CAMB} as input. 

\section{Likelihood, Sampling and Analytic Marginalization}
\label{sec:sampling}
The main results of this paper take the form of posterior probability distributions of parameters of interest, marginalized over a large number of nuisance parameters. In order to compute posteriors, we assume that the likelihood of our data given a set of parameters is Gaussian with a covariance given as described in \S\ref{sec:covariance}, and priors on the parameters of our model given in Table~\ref{tab:params}. We analytically marginalize over all parameters that enter into our model linearly, i.e.,~all intrinsic alignment parameters, as well as the stochastic terms and counter-terms in the bias expansion, $\textrm{SN}^{i}$, $b^{i}_{\nabla^2a}$, $b^{i}_{\nabla^2\times}$, and $b_{\nabla^2m}$. In our fiducial constraints we vary all $\Lambda$CDM parameters over the range of values spanned by the \texttt{Aemulus} $\nu$ simulations. We also investigate combining our likelihood with galaxy BAO data in \S\ref{sec:results_w_bao}.

In order to speed up the likelihood evaluation we train fully-connected neural network emulators to predict the cosmology-dependent ingredients that enter into the bias and IA expansions, i.e., the basis spectra described in Eqns. \ref{eq:pgg} and \ref{eq:ia_basis}. We largely follow the methodology presented in \cite{derose2021}, using a combination of principle component analysis and neural networks to reduce the number of required parameters in our neural networks. The main difference between the emulators used in this work and those presented in \cite{derose2021} is that we build emulators for individual basis spectra rather than the galaxy and IA power spectra that enter directly into the projection integrals, e.g. $P_{gg}(k)$. We also build an emulator for $\sigma_8(z)$ as a function of cosmological parameters so that we can bypass using a Boltzmann code to compute relevant transfer functions. 

For both bias and IA emulators, we use four fully connected layers with 150 neurons each making use of the specialized activation function presented in \cite{alsing2019} and taking the \texttt{arcsinh} of the inputs to reduce the dynamic range, keeping the first 104 and 93 principle components for the bias and IA models respectively. For $\sigma_8(z)$ we use two 150 neuron fully connected layers and 104 principle components, and do not use an \texttt{arcsinh} scaling, since the dynamic range of $\sigma_8(z)$ over the range of redshifts that we consider in this work is small. We train these emulators over the range of cosmologies spanned by the \texttt{Aemulus} $\nu$ simulations, and achieve a $1\sigma$ error of approximately $0.1\%$ between redshifts $z=0$ and $z=3$ and wave-numbers $k=10^{-2}\, h\, \Mpc^{-1}$ and $k=1\, h\, \Mpc^{-1}$ for all basis spectra other than the matter power spectrum, where we build emulators to $k=4\, h\, \Mpc^{-1}$. 

Finally, we use the Metropolis--Hastings sampler \cite{Lewis2002,Lewis2013} implemented in Cobaya \cite{torrado2020cobaya} to compute posterior distributions, running 16 independent chains simultaneously, and halting our sampling when $R-1=0.02$, where $R$ is the Gelman--Rubin statistic \cite{Gelman1992}. We plot all posterior distributions using \texttt{GetDist} \cite{lewis2019getdist}.

\section{Simulations and Model Validation} 
\label{sec:simval}
In order to demonstrate the robustness of our results to various choices and approximations we have made in our modeling, we have run a series of validation tests against two different types of simulations. First, we analyze data vectors produced using our model in order to test the robustness our model to our choice of priors and IA redshift evolution prescription. We then turn to fitting the \texttt{Buzzard v2.0} simulations \cite{DeRose2019, DeRose2022}, a suite of full $N$-body lightcones that contain realism beyond that implemented in our model, in order to test the scales on which our model based on the Limber approximation and LPT and HEFT, neglecting higher-order lensing contributions, can reliably constrain the true cosmology.

\subsection{Noiseless simulations}
\label{sec:noiseless_sims}

\subsubsection{IAs, Priors and Projection Effects}
\label{sec:projection_effects}
We begin by generating a noiseless data vector using the maximum likelihood nuisance parameter values \textit{without IA contamination} obtained from fitting the \texttt{Buzzard} simulations described in the next section, and the values of the cosmological parameters used to generate the \texttt{Buzzard} simulations. We convolve these predictions with the window functions measured from the data, and proceed to fit them assuming the window functions and covariance matrix estimated from the data. 

\begin{figure}[htb!]
    \centering
    \includegraphics[width=\columnwidth]{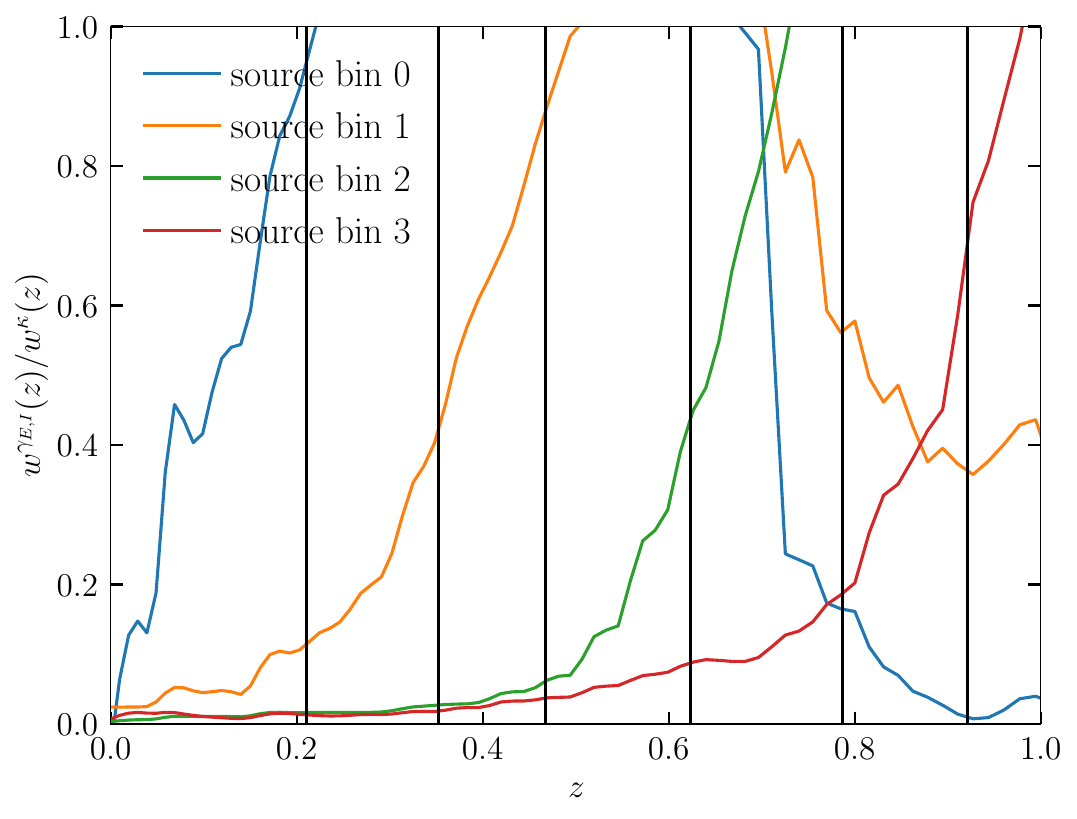}
    \caption{Fractional contribution of IAs to the galaxy--galaxy lensing signal for a lens at redshift $z$ for each source galaxy bin. For simplicity we assume only a linear IA contribution, i.e., the NLA model, and $A_1=1$. The black vertical lines represent the effective redshifts of each of our lens bins. It is apparent that the IA unacceptably high, greatly exceeding the statistical error on the measurements, for all but the first three lens bins, cross-correlated with the two highest redshift source bins. See further discussion around the balance between statistical and IA related theory error in Figure~\ref{fig:varying_cs_prior_S8}.}
    \label{fig:ia_contamination}
\end{figure}

Let us first consider the theoretical error on $S_8$ due to the unknown amplitude of intrinsic alignments. We can understand this dilution of $S_8$ information from the GGL signal by computing the relative contributions from lensing and IAs to $C^{\delta_g \gamma_E}_\ell$ in  Equation~\ref{eq:limber}. Since the lens galaxy densities $p^{\delta^i_g}(z)$ are very narrow we can approximate them as $\delta$ functions centered at $z^i_{\rm eff}$, in which case the ratio of the lensing and IA contributions is simply given by
\begin{align}
    \frac{C^{\delta_g^i \gamma_{E,I}^j}_\ell}{C^{\delta_g^i \kappa^j}_\ell} &\approx \frac{w^{\gamma_{E,I}^j}(z_{\rm eff}) P_{\delta_g^i I^j}(\ell/\chi) }{w^{\kappa^j}(z_{\rm eff}) P_{\delta_g^i m}(\ell/\chi)} \approx   \frac{ A_1(z_{\rm eff})\ w^{ \gamma_{E,I}^j}(z_{\rm eff})}{2 w^{\kappa^j}(z_{\rm eff})} 
    \label{eq:ia_cont_ratio}
\end{align}
where we have used linear theory to arrive at the final expression. 

The above calculation shows that the lensing amplitude and IA contribution are fully degenerate at leading order, such that neither can be independently determined without an informative prior on the other. However, it is important to observe that the size of the IA contribution is bounded by the size of $w^{\gamma_{E,I}}$ or, more generally, the size of the overlap integral between the source and lens distribution, and by the conservative bounds on the linear IA amplitude from simulations. The ratio of their product---which controls the size of the associated theoretical covariance induced by IA---with the lensing kernel itself for each of the DES source samples is shown in Figure~\ref{fig:ia_contamination}. Evidently, the lower redshift sources overlap sufficiently with the DESI lens samples that the IA contribution is always unacceptably high but the last two source bins (\texttt{S23}) and the first three lens bins (\texttt{L012}) are sufficiently separated in redshift that the IA contributions are limited to a few percent and subdominant to the lensing contribution. We thus expect these cross correlations (\texttt{L012xS23}) to supply essentially all of the $S_8$ signal when the theoretical error is accounted for, such that we can limit ourselves to this subset of our full data for our fiducial analysis.

\begin{figure}
    \centering
    \includegraphics[width=\columnwidth]{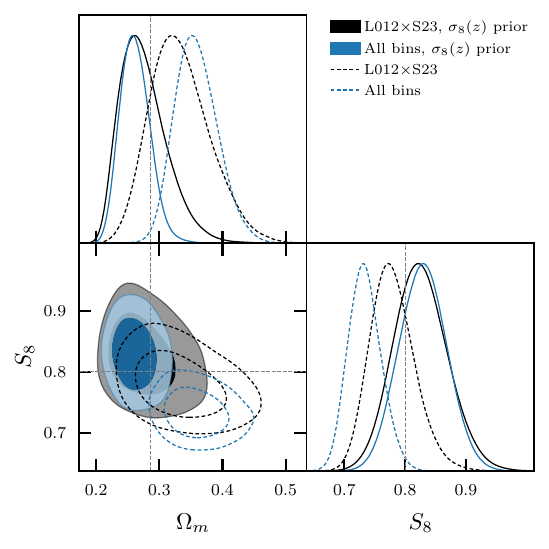}
    \caption{Sensitivity of our results to which data is included in our fits, as well as the form of the priors on the galaxy bias and IA parameters, when fitting to a noiseless data vector without intrinsic alignments but using the pairwise IA redshift evolution model (\texttt{MAIAR}). The different colored contours show how our results change as we include bins that have more IA contamination. The black contours use only the cross correlations of the first three lens bins (\texttt{L012}) and last two source bins (\texttt{S23}), as these result in less than $10\%$ contamination assuming a fiducial value of $c_s=-1$. The blue contours relax this to include all bin combinations. The solid contours use our fiducial priors on the product of bias and IA parameters and $\sigma_8(z)$. The dashed contours set priors on the bias and IA parameters without the extra $\sigma_8(z)$ dependence. With our fiducial priors, our results are stable to the inclusion of unconstraining data with large IA contamination, whereas the dashed contours show large projection effects when this additional data is included. We note that the improvement in constraining power on $\Omega_m$ when including extra lens and source bins is largely due to extra information in the galaxy density auto-power spectra.}
    \label{fig:ia_model_selection}
\end{figure}

The above intuition can be tested using the noiseless data vectors generated by our model. Figure \ref{fig:ia_model_selection} shows fits to these data varying which source--lens bin combinations are included in the fits, investigating two different choices for priors on our bias and IA parameters. We adopt our fiducial \texttt{MAIAR} redshift parametrization of IA evolution in these constraints, though we revisit this choice in the next section. The different colored contours represent fits that make use of data combinations with varying amounts of IA contamination, with the black contours showing fits to the \texttt{L012xS23} combination above that minimizes intrinsic alignment contamination. The blue contours show fits to all source--lens bin combinations. The constraints on $\Omega_m$ are significantly improved by including all redshift bins, since the full set of galaxy auto-correlations is sensitive to this parameter through the shape of the power spectrum. On the other hand, the restricted data set gives essentially identical $S_8$ constraints as the full data set, validating our heuristic argument that the theoretical covariance from unknown intrinsic alignment contamination dominates the cross-correlation pairs not included in \texttt{L012xS23}, diluting away their constraining power.  Since our aim is mainly to measure $S_8$ from these data, and we expect more stringent and robust constraints on $\Omega_m$ from external data sets, we use \texttt{L012xS23} as our fiducial data set, which in addition has the advantage of involving significantly fewer nuisance parameters and integrals required during sampling.

\edit{One frequently encountered problem when marginalizing over large numbers of nonlinear bias parameters, as we do in this work to consistently model galaxy densities, IA and nonlinear matter at 1-loop order, is that the proliferation of degrees of freedom results in large projection effects in the posteriors of cosmological parameters \cite{Joachimi2020,Amon2021,Krause2021}. As discussed in \S~\ref{sec:bias_evolution}, we can understand these projection effects by noting that the nonlinear galaxy bias and IA parameters are better constrained at high values of $\sigma_8$, where the effect of nonlinearities is more pronounced, leaving significantly more posterior volume at low $\sigma_8$. When marginalizing over our nuisance parameters, this large amount of posterior volume dominates over the likelihood term, thus shifting the marginalized posteriors away from the maximum likelihood value. Our strategy in this paper is to marginalize over the combination of bias and IA parameters scaled by $(\sigma_8(z)/\sigma_{8,\rm fid})^{n}$ described in \S\ref{sec:bias_evolution}(Eqn.~\ref{eqn:tilde_bias}), thereby removing this asymmetry in posterior volume by marginalizing over the parameter combination that governs the absolute contribution of each nuisance parameter.}

\edit{In order to test these $\sigma_8(z)$-dependent prior choices, we show results both using our fiducial priors (solid lines), which marginalize over the combination of bias and IA parameters multiplied by $(\sigma_8(z)/\sigma_{8,\rm fid})^{n}$, where $n$ is the order at which the bias or IA parameter enters into the perturbative expansion, and priors without this scaling (dashed), in Figure~\ref{fig:ia_model_selection}. Both solid and dashed results use our \texttt{MAIAR} IA model, allowing for a free set of IA parameters per source--lens bin pair. In the case of our fiducial priors, our results are stable to including bins that have potentially large IA contamination, neither significantly increasing our constraining power, nor significantly shifting our marginalized posteriors on $S_8$ and $\Omega_m$ away from the values used to generate the synthetic data. On the other hand, when marginalizing over the bare bias and IA parameters, we see that the constraints shift significantly when including bins with potentially large IA contamination. Even the black contours, with minimal IA contamination, are shifted away from the true parameter values, and this becomes even more drastic as more source and lens bin combinations are included.\footnote{We note that the improvement in constraining power on $\Omega_m$ when including all bins with our fiducial priors is real, in the sense that it is not due solely to projection effects, but rather almost entirely by extra information in the galaxy-density auto power spectra.} This symmetrization by $\sigma_8$ scaling thus largely removes the projection effects observed when sampling over the bare bias and IA parameters both in our fiducial setup and when analyzing the full set of cross correlations, as shown in the solid contours of Figure~\ref{fig:ia_model_selection}. As demonstrated here, our results are sensitive to our parameterization of the bias and IA priors, and so we report maximum likelihood points, which are less sensitive to these effects, along with all of our results.}

\subsubsection{Intrinsic Alignments and Bias Evolution}
\label{sssec:iaz}

As discussed in \S\ref{sec:bias_evolution}, the physical degrees of freedom in bias evolution sampled by the cross correlation of lensing with galaxies are, to leading order, the values of the bias parameters for each source sample at each effective redshift, with additional contributions suppressed as long as the lens samples are narrowly distributed in redshift. Setting these as the free parameters of our \texttt{MAIAR} model is therefore the most general and agnostic choice for both galaxy densities and intrinsic alignments. However, particularly in the case of the latter, previous surveys have typically selected more informative models of intrinsic alignment evolution wherein the intrinsic alignment parameters are either constant over the whole survey, or source-sample independent functions of redshift. In the case of the latter the redshift dependence is usually fit as a simple functional form like a power law with a free amplitude and slope. Our goal in this subsection is to test the robustness of both of these choices. 

We would like to investigate a scenario in which the IA parameters have non-trivial redshift dependence. As a simple example, we consider a scenario in which the ``true'' IA redshift evolution is given by
\begin{equation}
\label{eq:ia_pz_cont}
    c_s^{i}(z) = c_{s,0}\left(1 - \frac{p^{\gamma_{E}^i}(z)}{\textrm{max}(p^{\gamma_{E}^i}(z))}\right)\, 
\end{equation}
\noindent where $c_s^{i}(z)$ is the linear IA amplitude for the $i$-th source bin as defined in Equation~\ref{eq:ia_normalization}, and $c_{s,0}$ is an overall normalization constant that we vary in order to modulate the significance of the IA contamination, with other nonlinear parameters set to zero. This \edit{toy model} allows the IA amplitude to be different between the peak of the source-galaxy selection function and its tails, as we expect from realistic galaxy samples, in this case increasing from zero at the peak to $c_{s,0}$ far away from it.
We substitute these values into our pipeline in order to generate noise-free simulated data vectors with this IA redshift dependence.

\begin{figure*}[htb!]
    \centering
    \includegraphics[width=0.9\columnwidth]{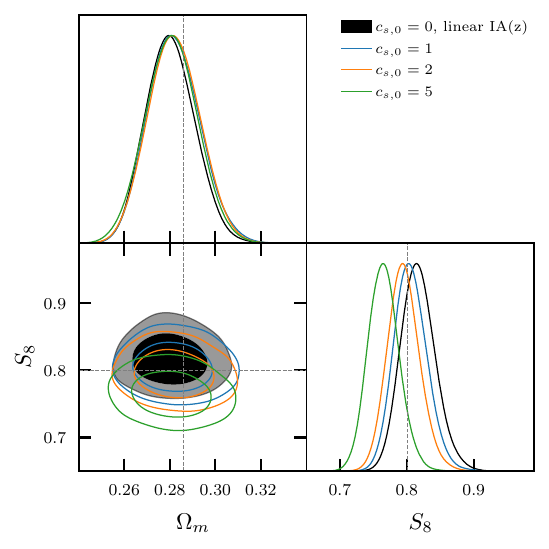}
    \includegraphics[width=0.9\columnwidth]{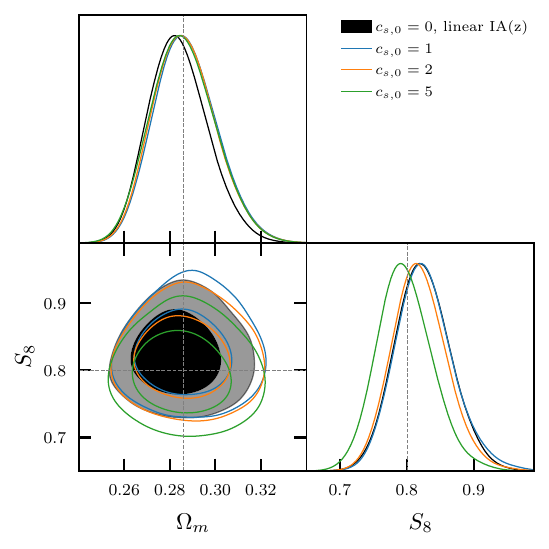}    
    \caption{Sensitivity of our results to non-trivial IA redshift evolution as defined in Equation \ref{eq:ia_pz_cont}, with different colors representing different assumptions for $c_{s,0}$ as indicated in the legend. (Left) A linear model for the redshift evolution of the IA parameters fixed with respect to different source bins, fitted to all source--lens bin combinations. This is comparable to the assumptions typically made when analyzing galaxy lensing data. (Right) Our pairwise IA redshift evolution model (\texttt{MAIAR}), that allows for an independent set of IA parameters for each source--lens bin combination. Here we only include the first three lens bins and last two source bins, i.e., our fiducial analysis choice. We observe that our fiducial model is significantly less biased than the linear model, with the observe shifts due to the fact that we have chosen to use Gaussian priors on the pairwise IA amplitudes. }
    \label{fig:ia_cont_test}
\end{figure*}

The left and right panels of Figure~\ref{fig:ia_cont_test} show the simulated constraints using a source-independent linear function in redshift for the the IA parameters and using our \texttt{MAIAR} prescription, respectively. The solid black contours show the results in both cases when $c_{s,0}=0$, i.e., there is no IA contamination. In this case both models of IA evolution are unbiased, with the source-independent linear model yielding significantly tighter $S_8$ constraints. This is expected: since all of the galaxy-lensing cross correlations have IA contributions described by two parameters in the former model, these parameters can be constrained by the cross-spectra where there the source and lens galaxies are not well separated and are thus IA dominated. When they are well constrained, the IA simply present a well-measured offset to the lensing amplitude, and indeed the $S_8$ constraints in this case are very similar to those in tests where IA are not marginalized over. On the other hand, the $S_8$ constraints in the \texttt{MAIAR} case are noticeably wider. Indeed, this would be the case even if more cross-correlation pairs were included, since in this model the measurement of IAs in one bin does not inform IA constraints in other bins.

However, it is important to note that the intrinsic alignment amplitudes measured in the linear model are those in the most intrinsic-alignment dominated source--lens bin pairs rather than those in which the lensing dominates, i.e., where our constraining power derives from. The colored contours in Figure~\ref{fig:ia_cont_test} show the constraints from each model in cases where $c_{s,0} \neq 0$. In the linear case we see that this results in systematic shifts in the contours proportional to $c_{s,0}$, leading to $0.43\sigma$, $0.84\sigma$ and $2.0\sigma$ biases in $S_8$ for $c_{s,0}=1,2,5$ respectively, well within the realm of measured IAs in simulations. On the other hand, the corresponding colored contours in the right panel show significantly less variation in the $S_8$ constraints when varying the strength of the IA contamination. This is as expected since the IA amplitude in each cross-spectrum is independent and not well-constrained. The residual shifts shown are due to the fact that the IA bias priors in these tests are centered at zero, while the cross-spectra have IA amplitudes that are small (and negative) but nonzero, such that $c_{s,0}=5$ pushes against the Gaussian prior.

\begin{figure}
    \centering
    \includegraphics[width=\columnwidth]{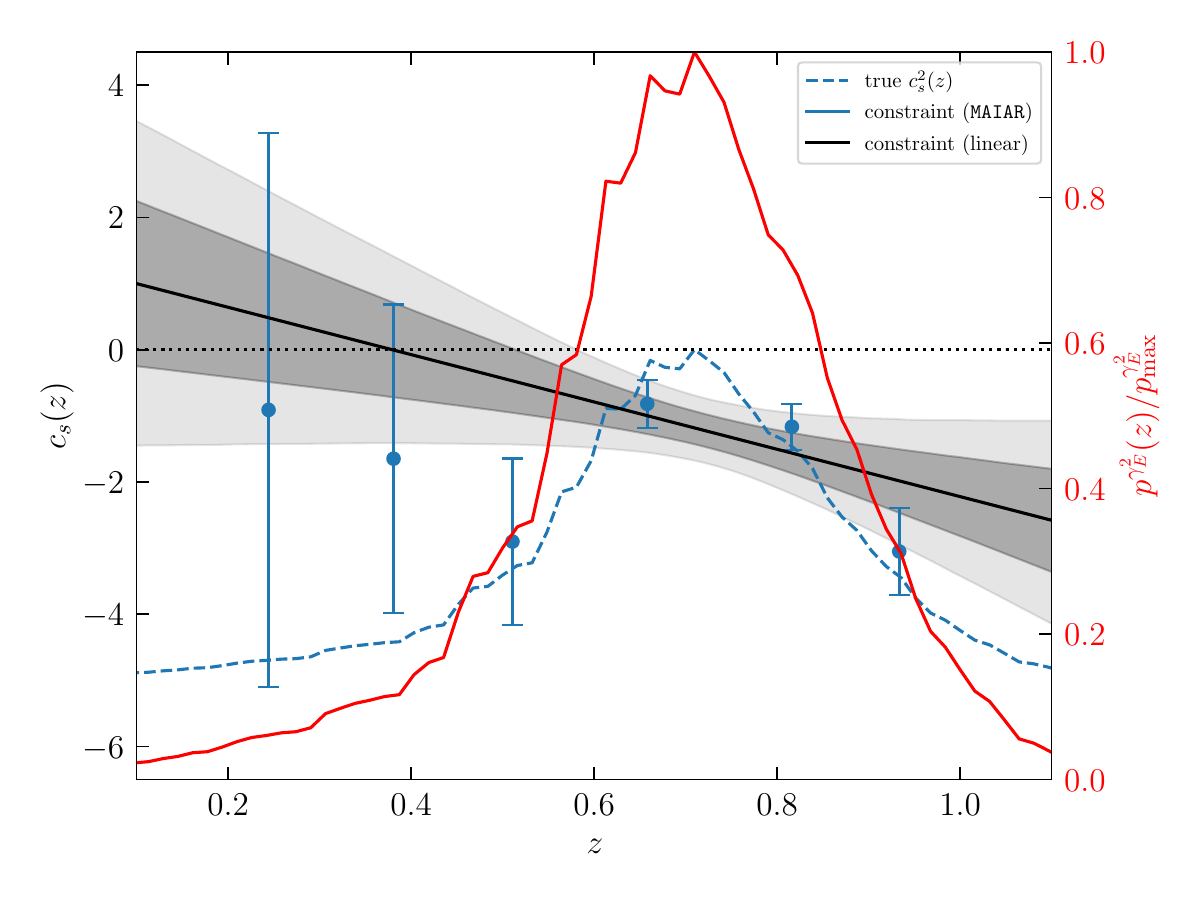}
    \caption{\edit{Constraints on the IA amplitude $c^i_s(z)$ (blue dashed) of the $i = 2$ source sample at fixed cosmology when there is nontrivial redshift evolution of intrinsic alignments away from the peak of the source redshift distribution $p^{\gamma^i_E}(z)$ (red). The \texttt{MAIAR} scheme successfully recovers the correct redshift evolution (blue points), with tighter constraints when there is more lens-source overlap, implying stronger $S_8$ constraints when there is less overlap. Gray bands show the ($1$ and $2\sigma$) constraints on IA evolution assuming a two-parameter (linear) model---in this case, the rigid functional form strictly limits the explored values of $c_s(z)$ even when the source-lens overlap is small, effectively extrapolating constraints from the peak of $p^{\gamma^i_E}(z)$ (i.e.~constraints from cross-correlating with LRG1 and LRG2) to other redshifts to erroneous values several $\sigma$ away from the true (blue-dashed) value. This discrepancy leads to biased $S_8$ constraints when the cosmology is allowed to vary.}}
    \label{fig:ia_schematic}
\end{figure}

\edit{In order to illustrate the cause of the biases in $S_8$ caused by insufficiently flexible models of IA sample and especially redshift dependence, Figure~\ref{fig:ia_schematic} shows constraints on the intrinsic alignment amplitude $c_s(z)$ for a single source bin ($i = 2$) at \textit{fixed} cosmology when the ``truth'' is given by Equation~\ref{eq:ia_pz_cont}. This is a useful test since the lensing amplitude and $c_s(z)$ are exactly degenerate at linear order in any given cross-correlation pair, such that we can infer the errors in $S_8$ due to each pair from the error in $c_s(z)$. In the \texttt{MAIAR} scheme, $c_s(z)$ is most tightly constrained at the lens redshifts that overlap most with the source distribution $p^{\gamma^i_E}(z)$, with relatively weak constraints in the lowest redshift lens bins. This latter observation explains why these source-lens combinations are able to constrain $S_8$, since they imply that the impact of IAs is small in these pairs. Nonetheless, the constraints on $c_s(z)$ are centered around its ``true'' redshift evolution, as expected.}

\edit{On the other hand, the black line and shaded region in Figure~\ref{fig:ia_schematic} show the constraints on IA evolution assuming a linear $c_s(z)$. While the constraints are again tightest around the support of $p^{\gamma^i_E}(z)$, the rigidity of the linear functional form means that the extrapolated constraints on on $c_s(z)$ at low redshifts is still substantially tighter than in the \texttt{MAIAR} case. In particular, the two parameters of the linear fit can be extremely well determined from the IA-dominated source-lens pairs, roughly by the value and slope of $c_s(z)$ about the central redshift. However, as this example shows, it can be quite dangerous to extrapolate the IA amplitude in this way, leading to inferred IA amplitudes several $\sigma$ away from truth, especially since we have no \textit{a priori} knowledge of the redshift evolution of intrinsic alignments in (complicated, photo-z selected) source galaxies.}

\edit{The above examples underscore the importance of setting sufficiently wide priors, with sufficient redshift flexibility, for} galaxy--galaxy lensing analyses: IA contamination can be marginalized over as a ``theoretical error'', and these errors are directly proportional to size of intrinsic alignments allowed by the priors. Since we (so far) can only bound the expected magnitude of IAs through measurements of halos in N-body simulations, the only direct method to reduce the size of these theoretical errors---themselves quite comparable to the size of the statistical errors of our data---is to reduce the overlap between lens and source samples. On the other hand, if trends with between IA parameters and galaxy luminosity or color can be measured in the future, e.g. with deep spectroscopy, then it may be possible to use these to set informative priors on a galaxy-formation informed model for the redshift evolution of IA, and recover some of the lost constraining power \cite{Krause2016}.

\subsection{\texttt{Buzzard} $N$-body simulations}
In order to test our measurement and modeling codes in a  realistic setting, we use the \texttt{Buzzard v2.0} suite of simulations \cite{DeRose2019, DeRose2022}. These are a set of synthetic galaxy catalogs populated in lightcone outputs of $N$-body simulations run using \texttt{Gadget-2}, and initialized using \texttt{2LPTIC}, with a linear power spectrum produced using \textsc{CAMB} \cite{Lewis2002}. Halos were identified using the \texttt{ROCKSTAR} halo finder \cite{Behroozi_2013}, and galaxies were assigned using the \texttt{ADDGALS} method \cite{wechsler2021addgals} fit to the subhalo abundance matching models presented in \cite{derose2021modeling}. 

DESI BGS and LRGs are selected out of the simulations using color-magnitude cuts based on those used in the DESI data, but modified slightly to match the simulated angular number densities with those measured during DESI science verification as described in \cite{DeRose2024}. DES-like source galaxy samples are selected to reproduce the redshift distributions and number densities of the DES Y3 \metacal catalog. We make measurements without shape noise in order to reduce the statistical errors on our simulated measurements. We must exclude the second BGS bin from our simulated analysis due to the fact that its redshift distribution significantly overlaps with the transition between two different simulations in our lightcones, and thus has a resulting cross-correlation coefficient between galaxies and matter of $r\sim 0.95$ on linear scales (see \cite{DeRose2024}), which would result in a bias to $\sigma_8$ of about $5\%$ in this bin.

We run the same measurement pipeline that is applied to the data on these simulated galaxies, with the only difference being that systematic weights for the lens sample and \metacal responses for the source sample are set to one. Averaging over 7 quarter-sky simulations, our measurements have a resulting statistical error that is approximately four times smaller than that of our measurements in the data, not accounting for the lack of shape noise in our simulations. Thus we should be able to detect systematic errors at the $< 0.25\sigma$ level of our constraining power in the data at $1\sigma$ confidence with these simulations.

We fit all source and lens bin combinations, excluding the second BGS bin for reasons mentioned above, with our fiducial model and fix all IA parameters to zero in order to further increase the precision with which we can measure systematic biases in our model. To include BAO priors in these mock tests we simply adjust the central values of the BAO measurements in the BOSS and eBOSS data to match the truth in the \texttt{Buzzard} cosmology. In the case of the Ly$\alpha$ constraints, the individual Ly$\alpha$ auto-spectrum and cross-correlations with QSOs, the posteriors are marked by a degree of non-Gaussianity---for simplicity we simply combine them and move the resulting likelihoods, which are then well-described by Gaussians, to the central values expected in \texttt{Buzzard{.}}

\begin{figure}[htb!]
    \centering
    \includegraphics[width=\columnwidth]{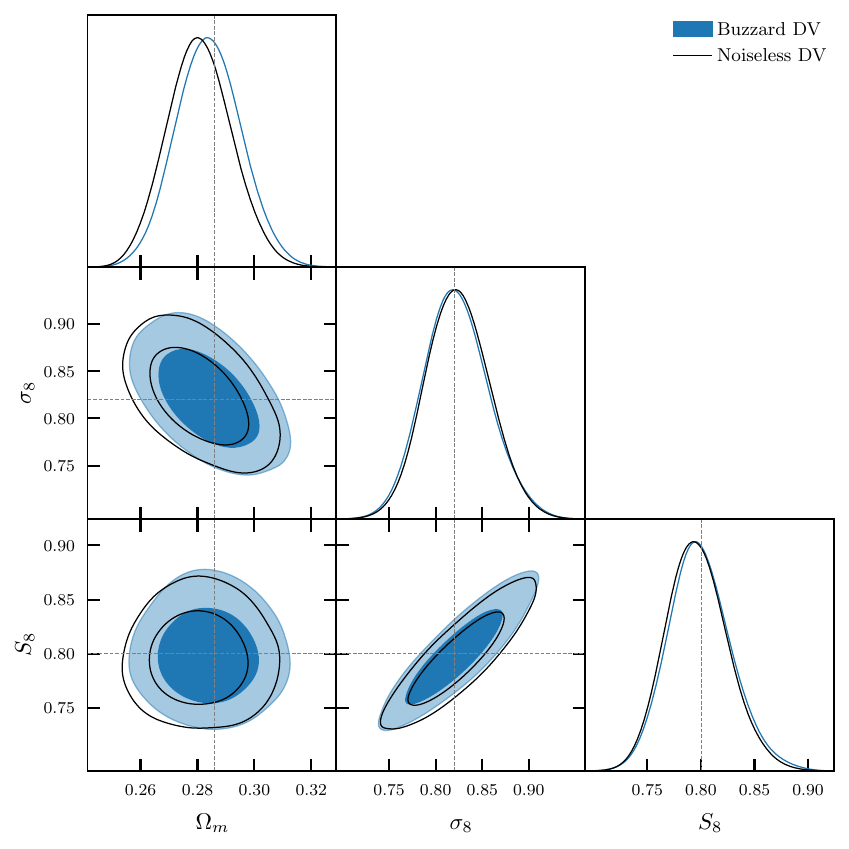}    
    \caption{A comparison of a fit to the  \texttt{Buzzard} data vector (blue) using our fiducial model without marginalizing over IAs and including a BAO prior, to a fit to a noiseless data vector generated with our fiducial model (black). The near perfect agreement, even in a much more constraining scenario than the setup used in the data, demonstrates the ability of our fiducial model to describe the \texttt{Buzzard} simulations at high precision. We observe a nearly identical level of agreement without including a BAO prior.}
    \label{fig:simcomp_noia}
\end{figure}

The resulting constraints are shown in Figure~\ref{fig:simcomp_noia}. The blue filled contours are the constraints that we obtain fitting to the \texttt{Buzzard} simulations, and the black constraints are obtained by fitting to noiseless data generated at the best-fit parameter values taken from the blue contours. The fact that these two contours agree nearly perfectly is a validation that our fiducial model is unbiased when fit to this suite of simulations.

\begin{figure*}[htb!]
    \centering
    \includegraphics[width=0.75\textwidth]{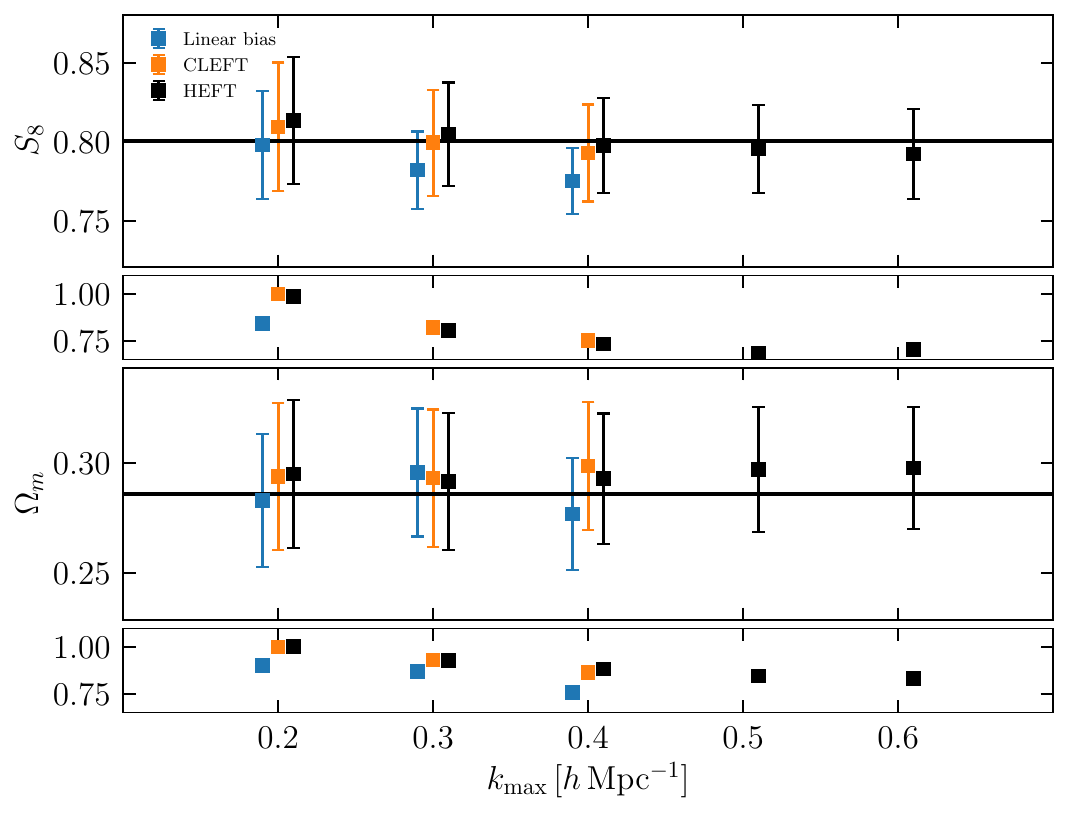}
    \caption{A comparison of the constraints obtained from fitting the \texttt{Buzzard} simulations varying the bias model and scales fit. In all cases, we do not marginalize over IAs, but otherwise keep the same modeling choices as used in the data, modulo changing the bias model. The smaller panels show the improvement in constraining power measured with respect to the \texttt{HEFT} model fitting to $k_{\rm max}=0.2\, h\,\rm Mpc^{-1}$. We observe that both the HEFT models are unbiased up to $k_{\rm max}=0.4\, h\,\rm Mpc^{-1}$, while the linear bias model becomes biased at $k_{\rm max} > 0.2\, h\,\rm Mpc^{-1}$, and the error bar is significantly smaller than both HEFT and CLEFT at $k_{\rm max}=0.2\, h\,\rm Mpc^{-1}$, indicating that the model is incomplete even at these scales. We only plot the HEFT model constraints past $k_{\rm max}=0.4\,h\rm Mpc^{-1}$, since linear theory and CLEFT do not apply on such small scales. The HEFT constraints are unbiased up to $k_{\rm max}  = 0.6\, h\,\rm Mpc^{-1}$.}
    \label{fig:simcomp_noia_bias_model_comp}
\end{figure*}

In order to gauge the sensitivity of our constraints to the assumed galaxy bias model and scale cuts, we have also run analyses on these simulations allowing for $k_{\rm max}=0.2-0.4\,h\, \rm Mpc^{-1}$ for our fiducial HEFT model, as well as a models using CLEFT and the commonly used combination of linear bias with a nonlinear matter power spectrum. We also show HEFT results using $k_{\rm max}=0.5-0.6\,h\, \rm Mpc^{-1}$, although we do not apply these scale cuts when analyzing data, as our IA model likely breaks down on these scales. The results of these variations are summarized in Figure~\ref{fig:simcomp_noia_bias_model_comp}. For these tests, we again fix the intrinsic alignment contributions to our model to zero in order to perform as precise a test as possible, although we do not include BAO here as we wish to investigate the improvement in constraining power of this data on $\Omega_m$ as a function of $k_{\rm max}$.  We do not show CLEFT or linear bias results for $k_{\rm max}>0.4\,h\, \rm Mpc^{-1}$ as we found that the acceptance rates in the MCMCs for these models were low, making these analyses very expensive. Given our expectation that these models are insufficient at these scales, we do not show these results here. 

For the $k_{\rm max}=0.2\, h\rm Mpc^{-1}$ case, we find excellent agreement between HEFT and CLEFT, both in terms of the posterior means for $S_8$ and $\Omega_m$, as well as the $1\sigma$ errors. At these scales, we already see that assuming a linear bias model results in approximately $15\%$ smaller errors on $S_8$ than HEFT and CLEFT. This indicates that although the linear bias constraints are unbiased on this set of simulations for $k_{\rm max}=0.2\, h\rm Mpc^{-1}$, there are values of the nonlinear bias parameters that are within our assumed priors that would cause detectable biases, had it so happened that our simulations preferred those bias values. 

For $k_{\rm max}>0.2\, h\rm Mpc^{-1}$ we find detectable biases when assuming a linear bias model, illustrating a clear breakdown of this assumption. The HEFT and CLEFT constraints are very similar to the maximum scale that we fit the CLEFT model, $k_{\rm max}=0.4\, h\rm Mpc^{-1}$. We thus conclude that both HEFT and CLEFT are unbiased at this scale, and that differences between the two at two-loop order are subdominant in our constraints at $k_{\rm max} = 0.4\, h\rm Mpc^{-1}$. There is a shift to lower values of $S_8$ for the HEFT model above $k_{\rm max}=0.2\, h\rm Mpc^{-1}$, but analyzing a noiseless HEFT data vector to the same scales shows nearly identical, small shifts that must therefore be due to projection effects. As such we infer that the HEFT model is unbiased in these simulations to $k_{\rm max}=0.6\, h\rm Mpc^{-1}$.

We observe an additional $10\%$ improvement in constraining power on $S_8$ going from $k_{\rm max} = 0.4\, h\rm Mpc^{-1}$ to $k_{\rm max} = 0.6\, h\rm Mpc^{-1}$, while the HEFT model remains unbiased at high significance. Given that we are using CLEFT to model intrinsic alignments, which has been shown to be accurate to $k\sim 0.4\, h\rm Mpc^{-1}$ \cite{Chen24}, we choose $k_{\rm max} = 0.4\, h\rm Mpc^{-1}$ for this analysis. Furthermore, it is unclear that we would still see the quoted $10\%$ gain in constraining power had we not neglected the connected trispectrum terms in our covariance matrix, given the significantly nonlinear nature of these scales. We plan on improving our covariance matrix treatment and investigating HEFT modeling for intrinsic alignments similar to \cite{Maion2024} in upcoming analyses.

\section{Results}
\label{sec:results}

\begin{figure*}[htb!]
    \centering
    \includegraphics[width=0.75\textwidth]{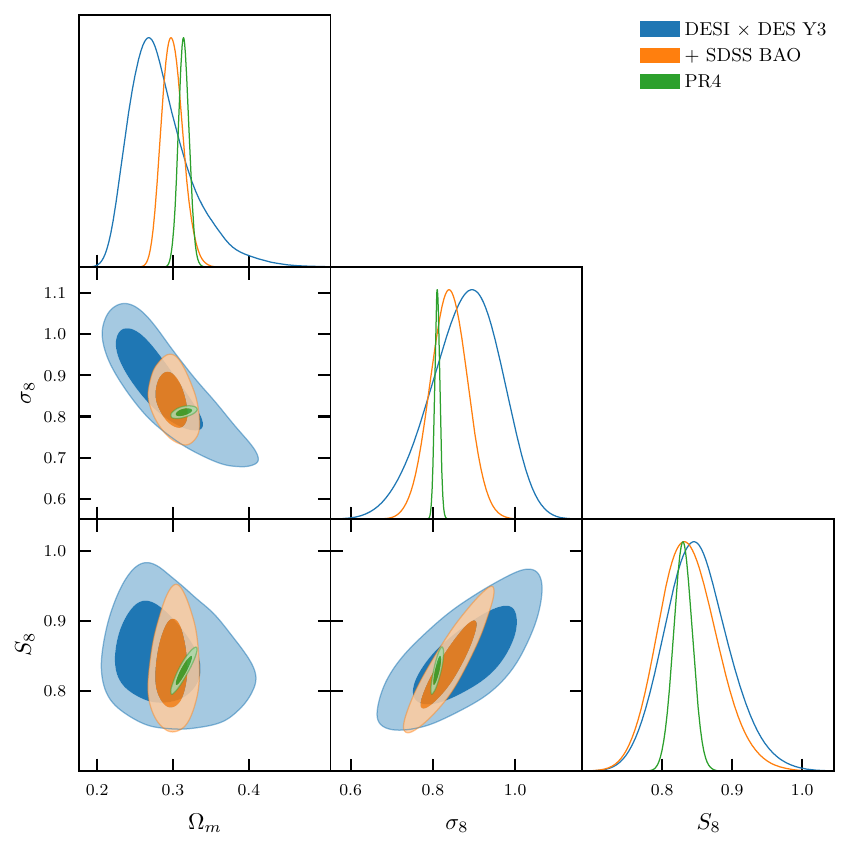}
    \caption{Cosmological constraints from our ``$2\times2$-point'' DESIxDES analysis with and without external priors from SDSS BAO measurements. Adding the BAO data significantly tighten constraints on $\Omega_m$ and lead to slight improvement in the overall $S_8$ constraint. Both constraints are in excellent agreement with those from the primary CMB.}
    \label{fig:w_bao}
\end{figure*}

We now proceed to describe the main results of this work. Figure \ref{fig:w_bao} shows the results of our fiducial $2\times2$-point analysis. In particular we find that the BGS0, BGS1 and LRG0 bins yield consistent constraints, and when combined we obtain
\begin{align}
    S_8 &= 0.850^{+0.040}_{-0.051}\, (0.834) \nonumber \\ 
    \Omega_m &= 0.286^{+0.025}_{-0.049}\, (0.294) \quad \text{(Fid.)} \\ 
    \sigma_8 &= 0.878^{+0.089}_{-0.080}\, (0.842)\, . \nonumber 
\end{align}
The best-fit model and measurements used in this fit are shown in Figure~\ref{fig:bestfit_model}. Our model has a $\chi^2=27.5$ for 54 data points using 21 free parameters, not counting the IA, magnification or source sample uncertainties since these are prior dominated, equivalent to $\chi^2_{\rm red} = 0.86$. A full accounting of the number of degrees of freedom in our fits requires a more sophisticated statistical framework (e.g. \cite{Doux2021}), that we may pursue in future work.

\begin{figure}[htb!]
    \includegraphics[width=\columnwidth]{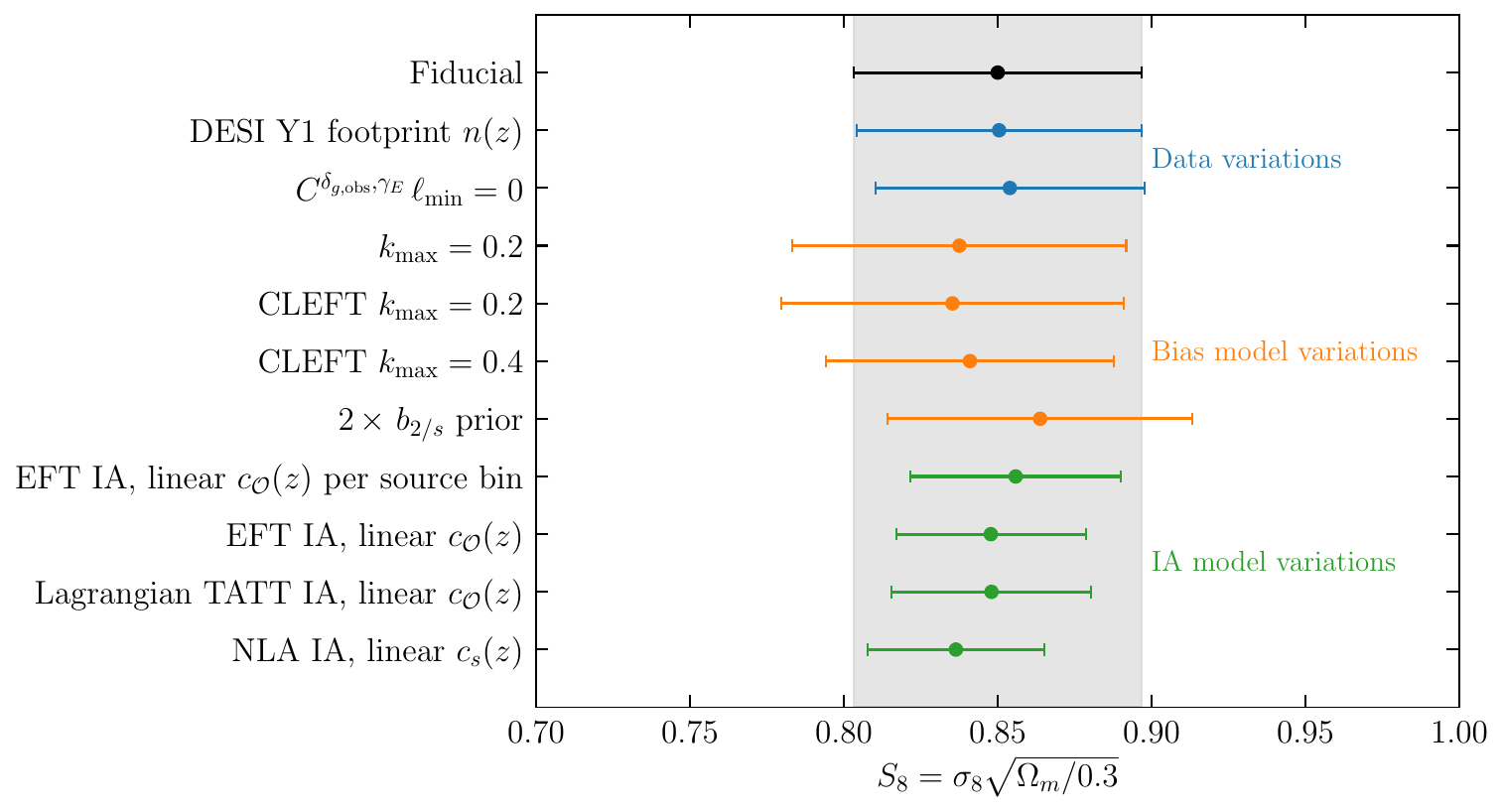}
    \caption{Summary of the one-dimensional $S_8$ posteriors for all of the analysis variations considered here. Blue points show variations in the data systematics treatments, orange show changes in bias modeling and scale cuts, and green points show variations in IA model assumptions. See \S\ref{sec:results} for details.}
    \label{fig:s8_consistency}
\end{figure}

Of particular note is that our fiducial $S_8$ constraint is consistent with measurements from the primary CMB, e.g. Planck PR4 gives $S_8=0.830\pm0.014$ \cite{Tristam2023} making this one of the few galaxy lensing analyses that has yielded a value of $S_8$ that is within $1\sigma$ of the primary CMB constraints. We perform comparisons with other analyses in the literature below, but for now we investigate whether there is part of our model or data that is driving this preference for higher $S_8$ than usually reported from galaxy lensing studies. A summary of these investigations is shown in Figure~\ref{fig:s8_consistency}.

As a direct check we can consider whether our constraints are robust to our choice of dynamical model. Fitting to the same scales but using CLEFT---that is, employing a fully perturbative model for all physical quantities---we obtain 
\begin{align}
    S_8 &= 0.841^{+0.038}_{-0.051}\, (0.831) \nonumber \\ 
    \Omega_m &= 0.296^{+0.028}_{-0.052}\, (0.243) \quad \text{(CLEFT)} \\ 
    \sigma_8 &= 0.854 \pm 0.082\, (0.924)\, ,\nonumber 
\end{align}
with a best-fit chi-squared value of $\chi^2=28.1$. This constraint is in excellent agreement with our fiducial one, both with respect to the posterior means and errors, as expected from our analyses on the \texttt{Buzzard} simulations. 

\begin{figure}[t!]
    \includegraphics[width=\columnwidth]{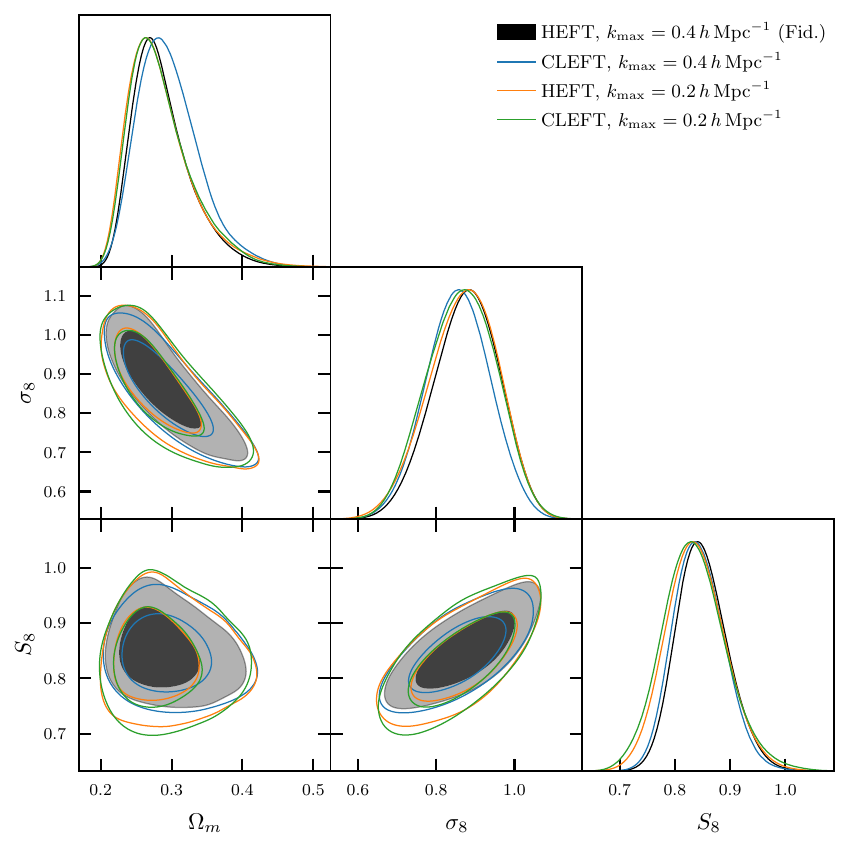}
    \caption{Comparison of constraints varying the maximum scale included in the analysis, as well as the galaxy bias model employed. Our fiducial constraints using HEFT with $k_{\rm max}=0.4\, h\, \rm Mpc^{-1}$ are shown in black. A purely perturbative analysis, i.e., using CLEFT rather than HEFT, but with all other settings kept the same is shown in blue. When using CLEFT we adopt broader counterterm priors, as described in the text. The orange and green contours show similar analyses using $k_{\rm max}=0.2\, h\, \rm Mpc^{-1}$ with HEFT and CLEFT respectively. Our fiducial analysis is $20\%$ more constraining than fitting to $k_{\rm max}=0.2\, h\, \rm Mpc^{-1}$, but otherwise the differences that we observe are consistent with tests done with noiseless simulations.}
    \label{fig:scalecut_sens}
\end{figure}

We also consider what happens to our constraints when varying the scale cuts in both modeling approaches, e.g. when we restrict the range of scales that we fit to $k_{\rm max}=0.2\, h\, \rm Mpc^{-1}$. Figure \ref{fig:scalecut_sens} shows these variations, with the black contours again showing our fiducial constraints using HEFT to fit to $k_{\rm max}=0.4\, h\, \rm Mpc^{-1}$ and the blue contours showing the CLEFT version of this analysis. The orange and green contours show the same analyses as black and blue respectively, but restricting the scales modeled to $k_{\rm max}=0.2\, h\, \rm Mpc^{-1}$. Our $S_8$ constraints are expectedly stable to these changes, with the main difference being that the fiducial chains result in $17\%$ more constraining power than $k_{\rm max}=0.2\, h\, \rm Mpc^{-1}$. The fact that our constraints are stable when moving to larger scales, or using the fully perturbative model, suggests that our analysis is under excellent perturbative control: any differences in the modeling between CLEFT and HEFT, which must exist beyond one-loop order, are not important at the accuracy level of our data, or indeed for the more stringent requirements of our mock tests.

\begin{figure}[t!]
    \includegraphics[width=\columnwidth]{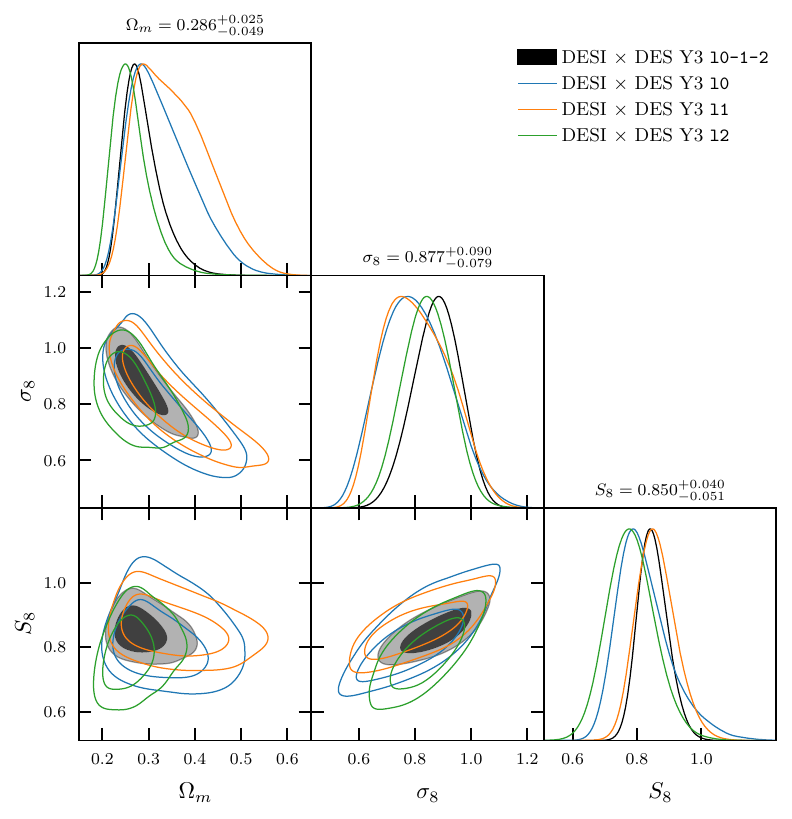}
    \caption{Comparison of constraints from individual lens bins (colored contours) and the fiducial combined constraints (black contours). It is apparent that the second BGS bin (orange, labeled $l1$) is the most constraining bin on $S_8$, while the first LRG bin (green, labeled $l2$) drives our $\Omega_m$ constraining power. Both of these finding are consistent with tests we have performed on noiseless simulated data.}
    \label{fig:lensbin_cons}
\end{figure}

We can also consider the $S_8$ constraints from individual lens bins to check the internal consistency of the data. As shown in Figure \ref{fig:lensbin_cons}, all of our lens bins yield consistent results, although BGS1 shows a $\sim 1\sigma$ preference for higher $S_8$ than the other two bins. The preference for lower values of $S_8$ in the first and third lens bin is largely driven by projection effects that become more substantial when fitting to individual bins. As expected, the \texttt{BGS1} bin yields the tightest $S_8$ constraints, as it has comparably low IA contamination compared with \texttt{BGS0} but is at higher redshift and thus we are able to model to higher $\ell$ given our scale cuts of $k_{\rm max} = 0.4\, h\, \rm Mpc^{-1}$, while the \texttt{LRG1} bin drives our constraints on $\Omega_m$.

\subsection{Including BAO}
\label{sec:results_w_bao}

The line-of-sight projection in two-dimensional data like those used in this paper washes out much of the shape information in the three-dimensional power spectrum that can be used to constrain cosmological parameters beyond the lensing amplitude $S_8$ such as $\Omega_m$ and $H_0$.\footnote{While we note that there is still \textit{some} information, particularly in $C^{gg}_\ell$ (see e.g. Figure~\ref{fig:ia_sens}), in order to be conservative in our modeling choices our fiducial analysis setup is not optimized to extract this information since (a) we only use the auto-spectrum of the lens galaxies within the DES footprint to ensure sample homogeneity, thus reducing the area over which $C^{gg}_\ell$ is measured and (b) we do not include $C^{gg}_\ell$ from lenses where the cross-spectrum is IA-dominated to avoid sampling over an excessive number of free parameters. We have made these choices considering that their effect on $S_8$ is minimal and that the constraints on other cosmological parameters from these data alone are not competitive.} While this does not significantly affect our ability to constrain $S_8$, we can slightly tighten our constraints and avoid sampling ruled-out cosmological parameter space by including additional large-scale structure data. To avoid including additional amplitude information we include constraints on the expansion history from baryon acoustic oscillations (BAO) as measured in the Sloan Digital Sky Survey---specifically, we use the BAO data measured in the LRGs from BOSS (DR12) \cite{Alam17} and LRGs, QSOs and Ly$\alpha$ in eBOSS (DR16) \cite{alam2021}. These data were deemed sufficiently statistically independent to be combined in ref.~\cite{alam2021} and together (when combined with a BBN prior on $\omega_b$) significantly constrain the available parameter space in $\Omega_m$ and $H_0$.

The orange contour in Figure~\ref{fig:w_bao} shows the cosmological constraints from our data when combined with these BAO data, which yield
\begin{align}
    S_8 &= 0.840^{+0.038}_{-0.045}\, (0.843) \nonumber \\ 
    \Omega_m &= 0.300^{+0.012}_{-0.015} \, (0.295) \quad \text{(w. BAO)} \\ 
    \sigma_8 &= 0.840\pm 0.044\, (0.850)\, . \nonumber 
\end{align} 
The BAO allow us to break the $\sigma_8$-$\Omega_m$ degeneracy to measure $\sigma_8$ and, in addition, since the degeneracy is not perfect, slightly tighten the $S_8$ constraint by $9\%$. The resulting constraints are also in excellent agreement with the cosmological constraints from the Planck CMB data, which is not surprising since the BAO and lensing constraints are independently consistent with them.

\subsection{Dependence on IA Parametrization and Priors}

\begin{figure}[t!]
    \includegraphics[width=\columnwidth]{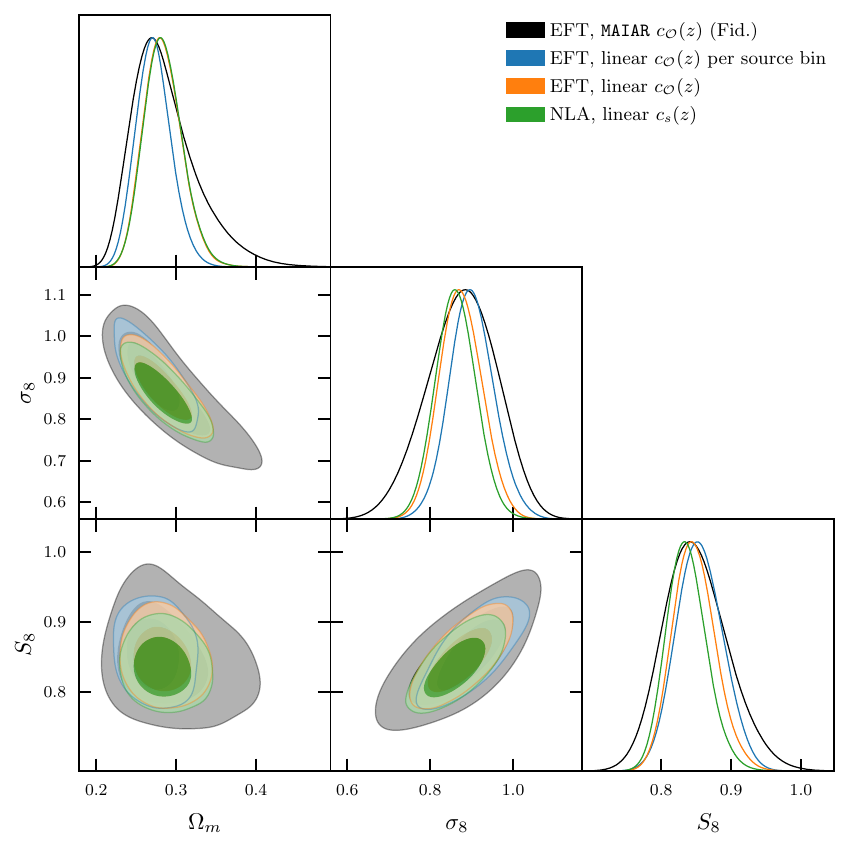}
    \caption{Comparison of constraints varying the IA model parameterization employed in our analysis. The black contours show our fiducial analysis, allowing one set of IA parameters per cross-spectrum, and restricting the source--lens bin combinations used to only those with relatively small potential IA contamination, as described in \S\ref{sec:projection_effects}. The blue, orange and green contours show analyses analyzing all source and lens bin combinations assuming that the IA parameters evolve linearly in redshift. The blue contours assume our fiducial second order EFT IA model and one linear function $c_{\mathcal{O}}(z)$ for each source bin. The orange contours assume the same linear function $c_{\mathcal{O}}(z)$ for all source bins, while the green contours assume the same but use the nonlinear alignment (NLA) model. These more restrictive IA models result in $35\%$, $46\%$ and $59\%$ more constraining power respectively, but make strong assumptions about the redshift and scale dependence of the IA signal beyond the evidence provided by current IA measurements. Until better IA measurements from the data or IA simulations are available, we favor the constraints provided by our fiducial model and quote these for the duration of this work.}
    \label{fig:ia_sens}
\end{figure}

\begin{figure}
    \centering
    \includegraphics[width=\columnwidth]{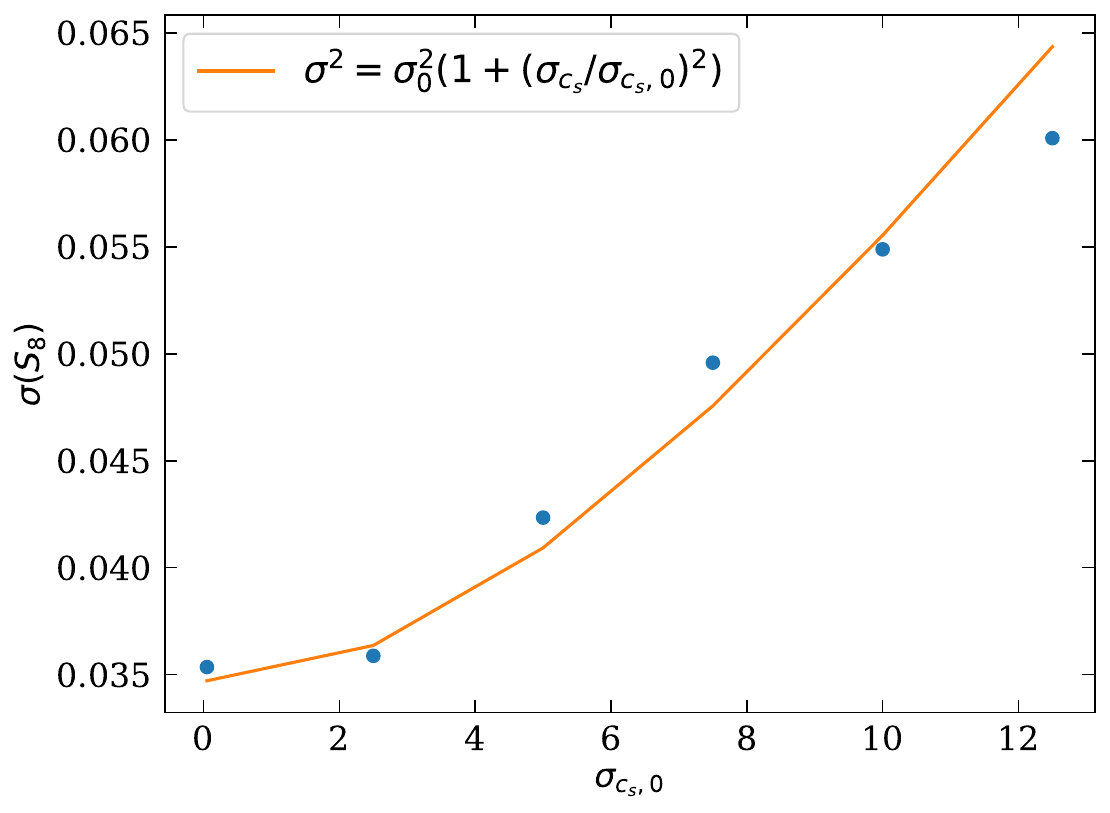}
    \caption{The width of our constraint on $S_8$ as a function of the width of the $c_s$ prior in each cross-correlation pair $\Delta c_s$. The trend is roughly fit assuming that the theoretical IA error adds to the width of the constraint in quadrature, with the theoretical error overtaking the sample variance around $\Delta c_s = 8$.}
    \label{fig:varying_cs_prior_S8}
\end{figure}

Given the novelty of the fiducial IA model that we apply in this analysis, we now present how our constraints change when making IA assumptions more in line with those used in previous galaxy lensing analyses. Figure \ref{fig:ia_sens} shows these variations, where the black contours again represent our fiducial constraints that use our EFT model for IA, keeping operators up to second order, and allowing for one set of IA parameters per source--lens bin pair (i.e., our \texttt{MAIAR} parameterization). The blue, orange and green contours fit to all bin combinations with varying levels of IA model complexity. The blue contours make the same assumptions as our fiducial analysis, except there we use a linear spline model for the redshift evolution of the IA parameters, allowing for one linear function $c_{\mathcal{O}}^{i}(z)$ per source bin. The orange contours make the additional assumption that the linear IA functions $c_{\mathcal{O}}^{i}(z)$ are shared between source bins, i.e., $c_{\mathcal{O}}^{i}(z)=c_{\mathcal{O}}^{0}(z)$ for all $i$. Finally, the green contours show an analysis assuming the nonlinear alignment (NLA) model, although we use the one-loop prediction for $P_{\delta,s_{ij}}(k)$ rather than the fully nonlinear prediction sometimes used when applying the NLA model.  These models all yield $S_8$ posterior means that are consistent with each other, but result in significantly different levels of constraining power. Indeed, adopting the two-parameter NLA model artificially improves our constraining power by $63\%$ over our fiducial results. The majority of the improvement comes from constraining the redshift evolution of the IA parameters to be linear, even when allowing separate linear functions per source bin. Doing so improves our $S_8$ constraining power by $36\%$. Further restricting to one linear function, $c_{\mathcal{O}}(z)$, for all source bins shrinks our $S_8$ errors by $51\%$ from our fiducial model.

As a further comparison to IA models used in the literature, we conducted an analysis using a Lagrangian version of the common tidal alignment and tidal torquing model (TATT), where we set $\alpha_s$ and $c_t$ to zero while assuming a linear redshift evolution of the other IA parameters. Given the additional terms generated by the advection of shapes in LPT, this is not entirely equivalent to the standard TATT model, but it is reasonably close with the same number of degrees of freedom. The resulting constraint shown in Figure~\ref{fig:s8_consistency} is very similar to using our full IA model while assuming linear redshift evolution.

We can also explore the dependence of our constraints on the width of our IA priors, i.e., the extent to which the $S_8$ constraint is weakened by conservative assumptions about the ranges that IA parameters can assume for weak-lensing source samples. Figure~\ref{fig:varying_cs_prior_S8} shows our $S_8$ constraint as a function of the prior width $\Delta c_s$, where we also scale the nonlinear IA parameter priors by an equal factor for consistency. The theoretical error induced on our measurement by unknown IA bias parameters adds to the $S_8$ posterior width roughly in quadrature
\begin{equation}
    \sigma^2 = \sigma_0^2 \Big( 1 + \Big(\frac{\sigma_{c_s}}{\sigma_{c_s,0}}\Big)^2 \Big) \ ,
\end{equation}
becoming dominant at $\sigma_{c_{s,0}} = 8$. This number is large, even relative to (conservative) expectations from simulations, because we have purposefully picked only those bins where the IA contamination is small in our fiducial setup. Had we chosen to utilize all the cross-correlation pairs in our measurements, the IA-free error $\sigma_0$ would be reduced, but $\sigma_{c_s,0}$ would decrease, making the total constraint relatively unchanged for reasonable choices of prior width as we saw in Figure~\ref{fig:ia_model_selection}.

\subsection{Fixed-Cosmology IA Measurement}

\begin{figure}
    \centering
    \includegraphics[width=\columnwidth]{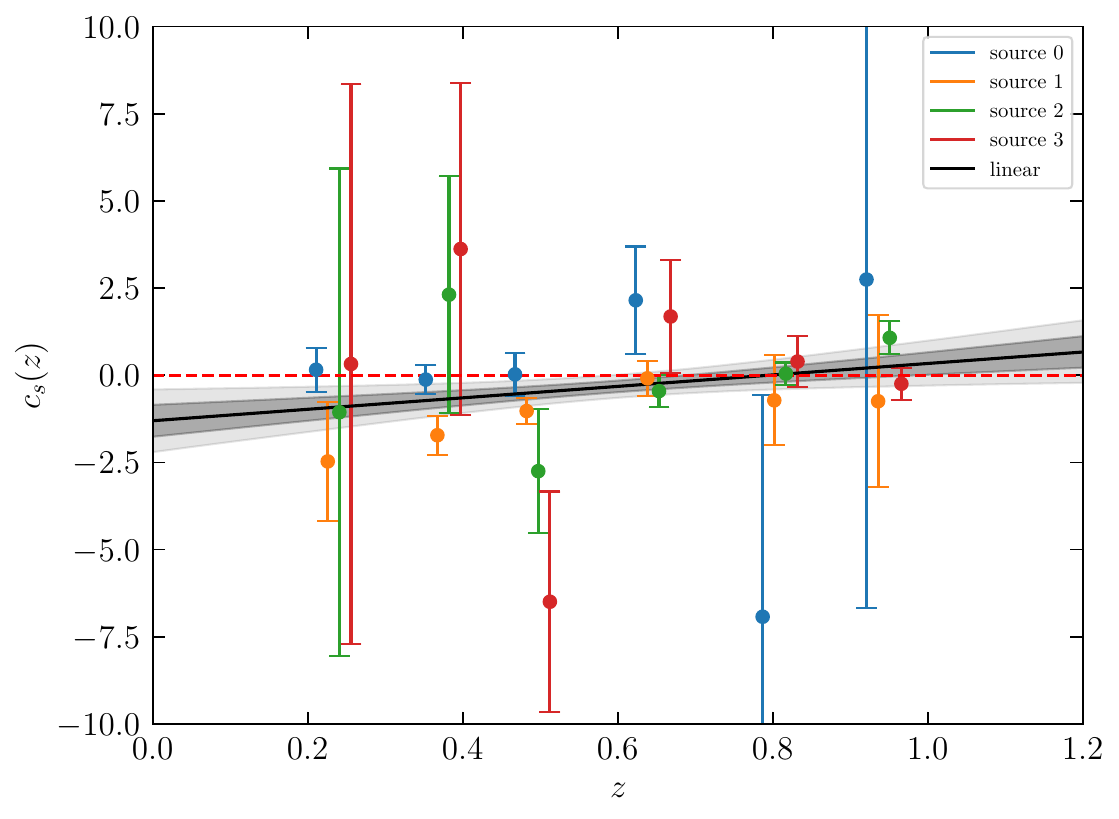}
    \caption{Tomographic constraints on the intrinsic alignment amplitude of each source sample. The black line and shaded bands show the mean and 1- and 2 sigma regions when fitting $c_s(z)$ independently of source sample in the linear model.}
    \label{fig:cs_z_constraints}
\end{figure}

As we have seen, allowing wide IA-parameter priors substantially degrade our ability to constrain the matter clustering amplitude $S_8$. The flip side of this observation is that our measurements allow us to tomographically constrain the intrinsic alignments of each source sample $c_s(z)$ at the lens redshifts. We show these constraints, where the cosmology is fixed to the mean Planck PR4 cosmology, and we analyze all source and lens bin combinations, in Figure~\ref{fig:cs_z_constraints}. Since our aim is now to measure $c_s(z)$ we  widen the priors on $c_s^{ij}$ to $\mathcal N(0,10)$ so as to not bias the resulting constraints. Intrinsic alignments, or deviations from the Planck prediction for lensing amplitudes, are detected to be nonzero at more than $2\sigma$ in several bins and, importantly, also deviate by similar amounts for a given lens sample redshift. The tightest constraints come from lens-source pairs with substantial redshift overlap, such that the signal becomes IA dominated and, conversely, the pairs we have opted to include in our fiducial analysis show rather weak constraints, with $\sigma( c_s )$ wider than our fiducial priors in many cases. This indicates that the statistical error in the lensing cross-correlation amplitude $C^{\delta_{g}\gamma_{E}}_{\ell}$ is not subdominant to the theoretical error in these bins. Our IA constraining power also degrades slightly when going to higher redshift, as uncertainties in the magnification contributions to these spectra become a confounding factor.

In order to illustrate the differences of these tomographic measurements of IA with the two-parameter source-independent IA parameterizations common in the literature we also perform a fixed-cosmology fit to a model that assumes a linear redshift evolution for the IA parameters (Eqn.~\ref{eqn:spline}). The results are shown in the black shaded regions---since all IA amplitudes are forced to be the same at a given redshift in this model the constraints are driven by the most tightly-constrained bins, which as noted above do not always have the same implied clustering amplitudes as the lensing-dominated bins constraining $S_8$. For example, for the cross-correlations between LRG0 and the first two source bins, whose shapes we omit in our fiducial analysis, prefer IA amplitudes closer to zero than the last two. Even the more tightly constrained bins, such as \texttt{BGS0xS0} and \texttt{BGS0xS1}, differ by several $\sigma$ in their preferred value of $c_s$. Taking into account the covariance between the individual bins, the pairwise IA posteriors in Figure~\ref{fig:cs_z_constraints} deviate from the source-independent 2-parameter model in black by $\chi^2 = 29.0$, with nonzero linear IA amplitudes being detected at $\chi^2 = 34.8$. 

It is worth noting that since these differences exist at the same redshift they cannot be due to a mismatch with the true and assumed cosmologies. However, if we assume that IAs can only be anti-aligned with the tidal field, the fact that several bins prefer a positive IA signal, rather than the negative radial alignment signal predicted by simulations, may be taken as evidence that either $S_8$ is higher than Planck, as our cosmology fits indicate, or the presence of a systematic or statistical fluctuation. \edit{Indeed, from Equation~\ref{eq:ia_cont_ratio} we can see that an exact physical degeneracy exists between the lensing kernel ampltiude $w^{\kappa_j}(z_{\rm eff})$ and the IA bias paramter $A_1(z_{\rm eff})$---if, for example, the amplitude of the lensing kernel were altered due to uncertainties in the DES photo-z's beyond those reported by the collaboration, such changes in $w^{\kappa_j}(z_{\rm eff})$ would be reflected in our fixed-cosmology fits as source-dependent shifts in the IA amplitude. Since this exact degeneracy with photo-z's also exists with $S_8$ itself when we vary cosmological parameters, it is a fundamental premise of galaxy-galaxing lensing analyses like this one that photometric redshifts are well-understood from the outset.}

\section{Conclusions}
\label{sec:conclusions}

\begin{figure*}[htb!]
    \includegraphics[width=0.75\textwidth]{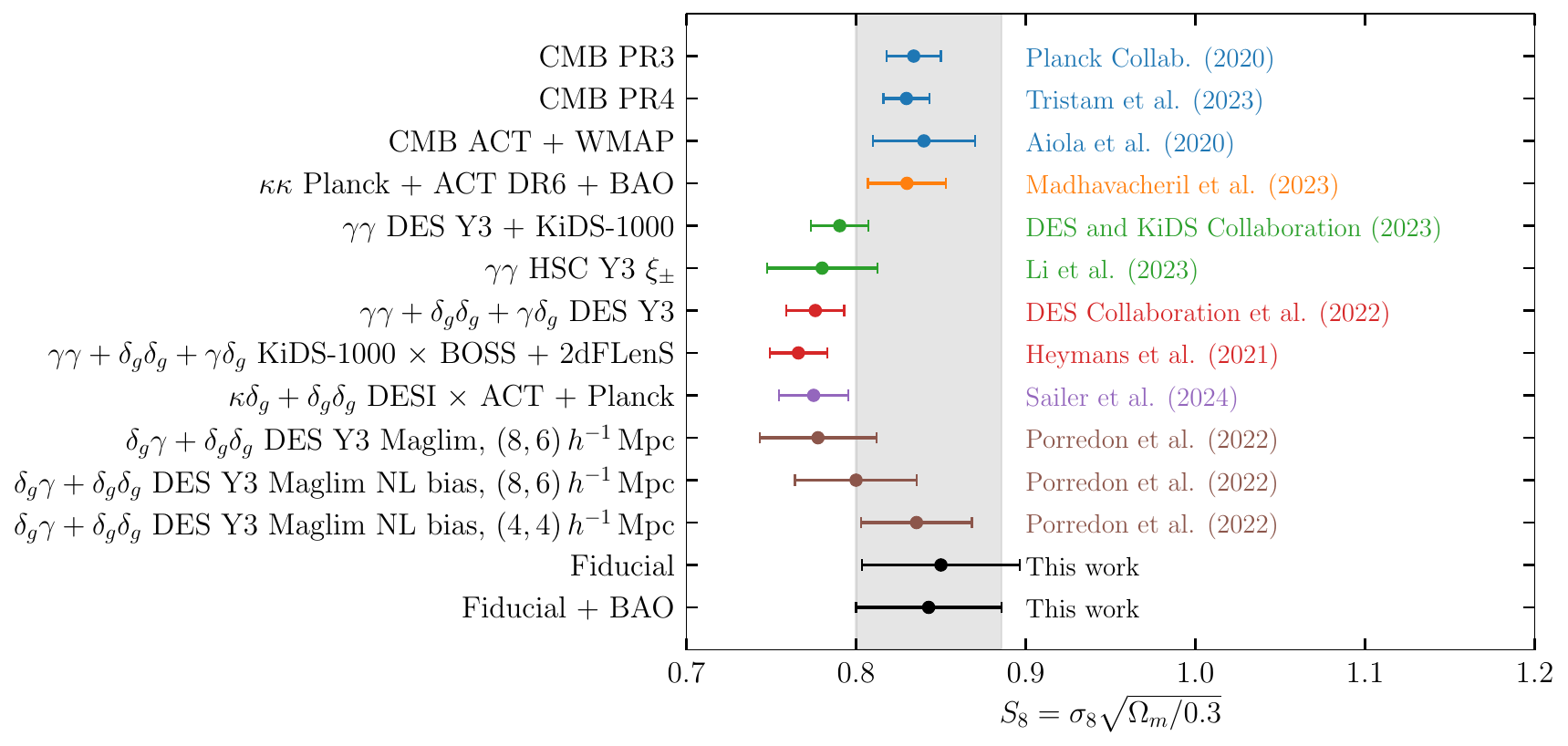}
    \caption{A comparison of our $S_8$ constraints with a number of results in the literature. Blue points denote constraints from the primary CMB, while orange show CMB lensing auto-spectrum constraints. Green, red, purple and brown points show cosmic shear, $3\times2$-point, CMB-galaxy lensing and galaxy--galaxy lensing constraints respectively. Our results with and without BAO are shown at the bottom in black. See \S\ref{sec:conclusions} for more discussion.} 
    \label{fig:s8_summary}
\end{figure*}

\begin{figure}[htb!]
    \includegraphics[width=\columnwidth]{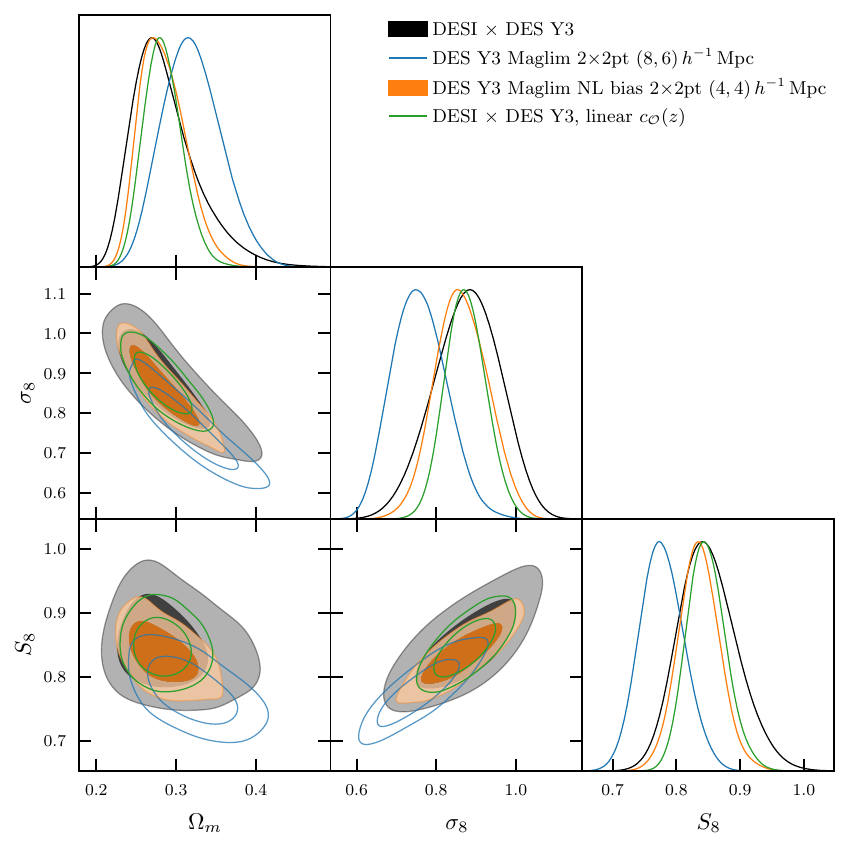}
    \caption{Comparison of the constraints derived in this work to those of \cite{Porredon2021}, using the \texttt{Maglim} lens sample and the \metacal source sample. The black contours here show our fiducial results, while the blue and orange contours show the linear and nonlinear bias constraints from \cite{Porredon2021}, although note that these constraints also use different minimum scales as described in \S \ref{sec:conclusions}. Of particular note is that our fiducial constraints agree quite well with the nonlinear bias constraints derived using the \texttt{Maglim} sample, although the latter use a very different nonlinear bias model. These constraints should be very correlated, given that they share the same source catalog and sky area, so this finding is reassuring. Furthermore, when we restrict to a similar IA treatment, assuming a single linear function of redshift for each IA parameter (green), we obtain similar uncertainties on $S_8$, although the analysis presented here is slightly more constraining likely due to our inclusion of higher redshift lens bins.}
    \label{fig:des_2x2}
\end{figure}

The cross correlations of the projected densities and shapes of galaxies, as measured in imaging surveys, directly probe the clustering of matter on cosmological scales, providing a precision test of the growth of structure in the standard model of cosmology. In this work we present measurements of these cross correlations using the target imaging samples for the DESI BGS and LRG samples and the \metacal galaxy shape catalog derived from the first three years of the Dark Energy Survey. In addition to having comparable signal-to-noise to other state-of-the-art galaxy--galaxy lensing measurements, the spectroscopic calibration of the DESI samples means that our measurements do not suffer from the photometric uncertainties typical of imaging surveys. Firstly, these samples have well measured and extremely small contamination from stars and systematic variations in galaxy photometry. Additionally, due to the accurate photometric redshift estimates available for these samples derived from the abundant spectroscopy that has been gathered for them, each lens sample can be well-localized along the line of sight. This second property makes our measurements particularly amenable to perturbative models, which use the separation of spatial scales to construct effective theories of structure formation that are non-local in time: the fixed line-of-sight distance translates angular scales to fixed physical ones, enabling us to robustly distinguish between large and small scales, and the unknown time evolution of the effective-theory parameters is restricted by fitting in well-localized redshift bins.

We present the ``$2\times2$-point'' cosmological analysis of these data using perturbation theory techniques, yielding a constraint on the matter clustering amplitude $S_8 = 0.850^{+0.040}_{-0.051}$. For the first time, we include a full accounting of nonlinear contributions to the intrinsic alignments of galaxies to one-loop order in the effective theory using the Lagrangian formalism recently developed in ref.~\cite{Chen24}. This enables us to consistently model the matter density, galaxy densities, and galaxy shapes at each lens redshift on equal footing from first principles within the formalism of Lagrangian perturbation theory. In addition, we augment the LPT predictions for matter and galaxy clustering using hybrid effective field theory (HEFT) by combining the predictions of Lagrangian bias with non-linear dark matter displacements from N-body simulations using the \texttt{Aemulus} $\nu$ emulators \cite{DeRose2023} for the first time, along with \cite{Sailer24}. We test our modeling pipeline on mock data constructed from the \texttt{Buzzard} simulations \cite{DeRose2019,DeRose2022}, finding that both allow us to place unbiased constraints on cosmological parameters including $S_8$ out to angular scales corresponding to $k_{\rm max} = 0.4\ h \Mpc^{-1}$, substantially beyond the range of validity for linear theory, and improving upon the $S_8$ constraint of an analysis using $k_{\rm max} = 0.2\ h \Mpc^{-1}$ by $20\%$. This is comparable to the improvement found in the first application of a linear Taylor series HEFT model to data in \cite{hadzhiyska2021hefty}, although a thorough comparison with that analysis is beyond the scope of this work. Since the two-dimensional data mostly constrain the lensing amplitude, we also obtain constraints combining our measurements with baryon-acoustic oscillation data, finding a small improvement in $S_8 = 0.838^{+0.038}_{-0.045}$. 

Our $S_8$ constraints in both cases are in excellent agreement with those inferred from the primary cosmic microwave background (\cite{Aghanim:2018eyx,Tristam2023,Aiola2020} and auto power spectrum of CMB lensing (\cite{Madhavacheril2023}), within the $\Lambda$CDM model. This is in contrast to many works in the literature which have found lower lensing amplitudes than predicted by the CMB at $1-2\sigma$ level, as shown in Figure~\ref{fig:s8_summary} \cite{Abbott2023,Li2023,Abbott2022,kids1000,Sailer24,Kim24}, though our error bars are somewhat wider, and assumptions somewhat more conservative, than many of these constraints. 

One particularly relevant exception is the $2\times2$-point analysis of ref.~\cite{Porredon2021}, who conduct a similar analysis as we do but using the DES \texttt{Maglim} sample for their lenses. Their fiducial analysis, using only linear bias and modeling to a minimum scale of 8 and 6 $h^{-1}\, \rm Mpc$ for clustering and galaxy--galaxy lensing respectively, infers a lower value of $S_8$ than ours by approximately $1.5\sigma$, but an analysis of the same data using a nonlinear bias model \cite{Pandey2020,Pandey2021}, infers a value of $S_8$ that is $\sim 0.5\sigma$ higher than the DES linear bias analysis. Furthermore, an analysis using the DES nonlinear bias model and fitting to 4 $h^{-1}\, \rm Mpc$ infers a very similar value of $S_8$ to this work, with differences in constraining power due to our use of the more conservative intrinsic-alignments (IA) scheme (Fig.~\ref{fig:des_2x2}). 

The large shift observed between the DES linear and nonlinear bias results underscores the importance of a proper treatment of nonlinear galaxy bias. The source of the additional shift to higher $S_8$ when analyzing smaller scales in the DES analysis is not readily apparent, and a full investigation of this is beyond the scope of this work. We note that there is also an observed $\sim 0.5\sigma$ shift to lower $S_8$ when analyzing even larger scales than the fiducial linear bias analysis (see Figure 12 of \cite{Porredon2021}), lending some evidence to an explanation involving residual angular systematics on large scales in the \texttt{Maglim} sample, however we have not investigated the statistical significance of this shift. In this scenario, analyzing smaller scales with the nonlinear bias model may downweight these residual systematics, leading to the observed shift. It is also possible that the nonlinear bias model used in the DES analysis and that used in our analysis, which were validated on the same set of simulations as used here \cite{DeRose2022}, are insufficiently flexible to fit the data leading to shifts to higher $S_8$ when analyzing smaller scales, although this possibility is less likely given the stability of our constraints when varying our scale cuts as shown in Figure~\ref{fig:scalecut_sens}.

We also note that our $S_8$ constraints derive fundamentally from large scales, as marginalizing over nonlinear bias and counterterms effectively dilutes the clustering and lensing information at high $k$. Many references have suggested that the low $S_8$ measured by weak lensing surveys can be alleviated by invoking baryonic or beyond-$\Lambda$CDM physics at small scales  \cite{Schneider2021,Amon2022,He2023,Preston2023,Garcia-Garcia2024}---these scenarios cannot be easily ruled out by perturbative analyses such as ours, where these effects are absorbed into the counterterms ($b_{\nabla^2 x}, b_{\nabla^2 m}$), although the fact that our weak constraints on these effective-theory parameters, particularly in HEFT, are not in any tension with the input priors suggests that any small scale suppression of power in the data should be ``small'' at $k < 0.4 h \Mpc^{-1}$. On the other hand, our $S_8$ constraints should be quite degenerate with large-scale suppressions of power, for example due to neutrinos or ultralight axions \cite{Rogers2022}, and a bound on these effects can be inferred from our consistency with Planck $\Lambda$CDM. While our constraints are slightly weaker than many other weak-lensing results reported in the literature, their reliance on first-principles calculations and fundamental symmetries makes possible inferences about exotic physical scenarios rather robust, making it a priority to understand and hopefully reduce the sources of theoretical error in a rigorous way.

Indeed, the leading source of error, both statistical and theoretical, in our analysis setup is the uncertain contribution of galaxy intrinsic alignments to the GGL signal. At leading order, a change in the linear shape bias, $c_s$, and a change in the matter clustering amplitude, $S_8$, contribute equally to any given density-shape cross correlation, with the only difference being that $S_8$ changes cross-spectra coherently across redshifts, while the IA amplitude of each source sample at each lens redshift is an independent degree of freedom (Eqn~\ref{eq:ia_cont_ratio}). In our fiducial analysis we attempt to mitigate this degeneracy with a combined strategy of (a) putting priors on the size of the IA amplitude based conservatively on measurements of halos in simulations, which are known to be enhanced relative to galaxies and (b) deriving our constraints from pairs of lens and source samples with maximal redshift separation---specifically cross correlating the three lowest redshift lens samples with the two highest redshift source samples---such that the IA contamination is further suppressed by the overlap integral between their redshift distributions. Simultaneously, we put the shapes of galaxies on the same, consistent footing with their densities by allowing the equivalent degrees of freedom for each independent sample and redshift required of a---in this case spin 2---biased tracer at one-loop in perturbation theory. We show that, when all these degrees of freedom are properly accounted for in our \textit{maximally agnostic IA redshift dependence} (\texttt{MAIAR}) prescription, the subset of data used in our fiducial setup returns essentially identical constraints to the full set, because IA contaminated measurements now have their proper theory error accounted for (\S\ref{sec:projection_effects}).

Our modeling, as described above, is robust against the redshift evolution of intrinsic alignments, as well as variations between source bins, and we demonstrate this by injecting a number-density dependent IA signal into mock data. On the other hand, conventional modeling choices which describe IAs through a source-sample independent two-parameter power law in redshift yield biased constraints with tightened error bars by deriving their IA constraints from the most IA-dominated data points. If we were to adopt these conventional choices in our analysis, our constraints would artificially improve to $S_8 = 0.848^{+0.028}_{-0.033}$, a $51\%$ improvement. Further restricting to the NLA model in addition to using the more conventional redshift evolution model results in $S_8 = 0.836^{+0.026}_{-0.031}$, a $63\%$ improvement over our fiducial analysis. Conversely, by fixing the cosmological parameters in our fits to their Planck best fits, we can place tomographic constraints on the IA parameters of the DES source galaxies, particularly in the IA dominated bins we drop in our fiducial analysis. Such an analysis shows that the source-independent two-parameter model is strongly disfavored by the data, which show variations in the cross-spectrum amplitudes across sources at the same lens redshift that cannot be due to deviations from the Planck predictions for growth alone.

The methods introduced in this paper have immediate implications for future surveys and GGL analyses. On the data side, the careful selection of galaxy lens and source samples that can be well-localized and separated in redshift, e.g. through spectroscopic calibration of redshift distributions as we have done here, will be critical. Alternatively, the impact of the IA signal could be mitigated by placing priors based on direct measurements. Importantly, these measurements will have to probe the same effective redshifts and samples from which the lensing constraints are derived, rather than the IA dominated cross correlations from which past lensing surveys have tended to constrain their more restrictive IA models. One promising avenue is to measure the three-dimensional clustering of galaxy shapes through spectroscopic surveys---these measurements have recently attracted renewed interested as probes of fundamental physics in their own right \cite{Mandelbaum06,Singh2015,Kurita23,Okumura23,Xu23}.Without the significant expenditure of resources, particularly in the form of wide field spectroscopy, IAs will continue to be a dominant source of uncertainty in galaxy lensing analyses.

On the modeling side, our work can be directly extended to include the clustering of the lens galaxies in redshift space. Including these three-dimensional data will allow us to break degeneracies in our constraints by accessing cosmological information in the shape of the linear power spectrum, and improve clustering constraints through redshift-space distortions. Indeed, the Lagrangian effective theory models used in this work have been extensively used to analyze spectroscopic clustering \cite{Chen22a,Maus24a,Maus24b} and can be straightforwardly and consistently combined with the analysis in this work with minimal new free parameters. In addition, comparing measurements of the matter clustering amplitude in lensing and peculiar velocities will allow us to test the predictions of general relativity and study the effect of anisotropic selection bias, which are a major ``known unknown'' in spectroscopic galaxy samples \cite{Zhang07,Hirata09,Chen22b,Wenzl24b}.

The modeling in this paper can also be extended to the auto-correlation of galaxy shapes, utilizing the effective theory of galaxy shapes in order to extract the matter clustering information in cosmic shear, although modeling connected trispectrum terms may be important for this extension if considering smaller surveys than DES Y3. Unlike the GGL signal modeled in this paper, the auto-correlation is not well-localized in redshift, such that the \texttt{MAIAR} prescription used in this work will not be sufficient. However, in this case the redshift dependence can be treated using the spline formalism described in Section~\ref{sec:bias_evolution}, where the number free parameters are determined by the correlation length of bias parameters across redshift (i.e., the separation scale of the spline basis functions), and the values of the spline coefficients can be tomographically constrained by the GGL signal as we have done here. This analysis can also be extended to include cross correlations of galaxy shapes with CMB lensing, whose kernels likewise span a broad range of redshift. Since the constraining power of these setups will likely be dependent on the redshift correlations assumed, it will be critical to study the robustness of the resulting constraints varying the spline spacing. Conversely, such an analysis will be an important stress test of current cosmic shear constraints. We intend to return to this topic in the near future.

\section*{Data Availability}

Data from the plots in this paper are available on at \url{https://zenodo.org/records/12642934} as part of DESI's Data Management Plan. We intend to make our likelihood code, measurement pipeline and fiducial data vectors publicly available at \url{https://github.com/j-dr/DESIxDES} upon publication.

\section*{Acknowledgements}

The authors thank Nickolas Kokron for discussions related to galaxy bias and intrinsic alignment modeling, Elisabeth Krause and Anna Porredon for useful discussions regarding comparisons with DES Y3 results and Alejandro Aviles and Cristhian Garcia-Quintero for their helpful comments as DESI internal reviewers of this work. SC acknowledges the support of the National Science Foundation at the Institute for Advanced Study. JD is supported by the Chamberlain Fellowship at Lawrence Berkeley National Lab.

This material is based upon work supported by the U.S. Department of Energy (DOE), Office of Science, Office of High-Energy Physics, under Contract No. DE–AC02–05CH11231, and by the National Energy Research Scientific Computing Center, a DOE Office of Science User Facility under the same contract. This work used \texttt{Stampede2} at the Texas Advanced Computing Center and \texttt{Bridges2} at the Pittsburgh Supercomputing Center through allocation PHY200083 from the Extreme Science and Engineering Discovery Environment (XSEDE) \cite{towns2014xsede}, which was supported by National Science Foundation grant number 1548562. Additional support for DESI was provided by the U.S. National Science Foundation (NSF), Division of Astronomical Sciences under Contract No. AST-0950945 to the NSF’s National Optical-Infrared Astronomy Research Laboratory; the Science and Technology Facilities Council of the United Kingdom; the Gordon and Betty Moore Foundation; the Heising-Simons Foundation; the French Alternative Energies and Atomic Energy Commission (CEA); the National Council of Humanities, Science and Technology of Mexico (CONAHCYT); the Ministry of Science and Innovation of Spain (MICINN), and by the DESI Member Institutions: \url{https://www.desi.lbl.gov/collaborating-institutions}. Any opinions, findings, and conclusions or recommendations expressed in this material are those of the author(s) and do not necessarily reflect the views of the U. S. National Science Foundation, the U. S. Department of Energy, or any of the listed funding agencies.

The DESI Legacy Imaging Surveys consist of three individual and complementary projects: the Dark Energy Camera Legacy Survey (DECaLS), the Beijing-Arizona Sky Survey (BASS), and the Mayall $z$-band Legacy Survey (MzLS). DECaLS, BASS and MzLS together include data obtained, respectively, at the Blanco telescope, Cerro Tololo Inter-American Observatory, NSF’s NOIRLab; the Bok telescope, Steward Observatory, University of Arizona; and the Mayall telescope, Kitt Peak National Observatory, NOIRLab. NOIRLab is operated by the Association of Universities for Research in Astronomy (AURA) under a cooperative agreement with the National Science Foundation. Pipeline processing and analyses of the data were supported by NOIRLab and the Lawrence Berkeley National Laboratory. Legacy Surveys also uses data products from the Near-Earth Object Wide-field Infrared Survey Explorer (NEOWISE), a project of the Jet Propulsion Laboratory/California Institute of Technology, funded by the National Aeronautics and Space Administration. Legacy Surveys was supported by: the Director, Office of Science, Office of High Energy Physics of the U.S. Department of Energy; the National Energy Research Scientific Computing Center, a DOE Office of Science User Facility; the U.S. National Science Foundation, Division of Astronomical Sciences; the National Astronomical Observatories of China, the Chinese Academy of Sciences and the Chinese National Natural Science Foundation. LBNL is managed by the Regents of the University of California under contract to the U.S. Department of Energy. The complete acknowledgments can be found at \url{https://www.legacysurvey.org/}.

The authors are honored to be permitted to conduct scientific research on Iolkam Du’ag (Kitt Peak), a mountain with particular significance to the Tohono O’odham Nation.

The author list in this work was compiled using \texttt{mkauthlist}\footnote{\url{https://github.com/DarkEnergySurvey/mkauthlist}}. We made use of \texttt{bibmanager}\footnote{\url{https://bibmanager.readthedocs.io/en/latest/}} to edit our bibliography. This research has made use of NASA's Astrophysics Data System and adstex (\url{https://github.com/yymao/adstex}).

\appendix 

\section{BGS and LRG Sample Calibration}
\label{app:lens_calibration}

\begin{table*}
    \centering
    \begin{tabular}{|c|c|c|c|c|c|c|}
         \hline
         \hline
         & $z_{\rm eff}$ (Full) & $z_{\rm eff}$ (DES) & $\sigma(z)$ (Full) & $\sigma(z)$ (DES) & $f_{\rm star}$ (Full) & $f_{\rm star}$ (DES)\\
         \hline
         BGS0 & 0.228 & 0.229 (0.33\%) & 0.0591 & 0.0597 (0.952\%) & 0.00495 & 0.00278 (-43.8\%) \\
         BGS1 & 0.362 & 0.363 (0.345\%) & 0.0603 & 0.0621 (2.88\%) & 0.0025 & 0.00216 (-13.8\%) \\
         LRG0 & 0.47 & 0.469 (-0.222\%) & 0.0632 & 0.0636 (0.677\%) & 0.00141 & 0.000634 (-55.1\%) \\
         LRG1 & 0.628 & 0.626 (-0.311\%) & 0.0734 & 0.0715 (-2.48\%) & 0.00156 & 0.000602 (-61.3\%) \\
         LRG2 & 0.791 & 0.794 (0.371\%) & 0.0781 & 0.0766 (-1.91\%) & 0.00194 & 0.00146 (-24.5\%) \\
         LRG3 & 0.924 & 0.932 (0.82\%) & 0.0956 & 0.0913 (-4.48\%) & 0.00376 & 0.00218 (-41.9\%) \\
         \hline
         
    \end{tabular}
    \caption{Redshift distribution summary statistics and stellar contamination fractions for our lens samples measured over the full DESI Y1 footprint, and over the overlap area between DESI Y1 and DES Y3 footprints.}
    \label{tab:des_v_full}
\end{table*}

\begin{figure}
    \centering
    \includegraphics[width=\columnwidth]{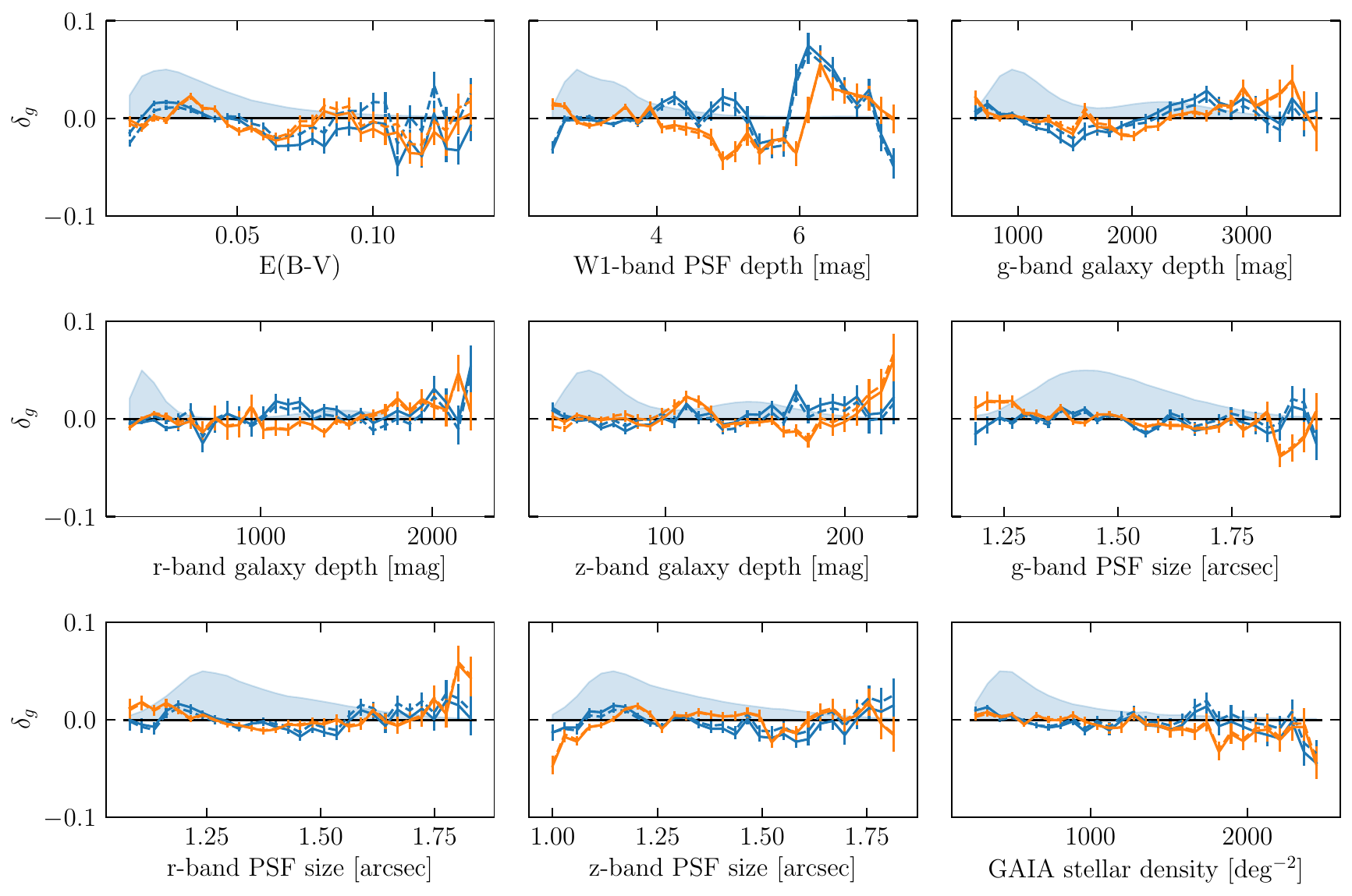}
    \caption{Overdensity trends with relevant potential systematics for the two BGS photometric bins used in this work. The solid lines use maps without weights, while the dashed lines include weights to correct for the observed trends, assuming a linear trend between the systematic and observed galaxy density.}
    \label{fig:bgs_sys_trends}
\end{figure}

\begin{figure}
    \centering
    \includegraphics[width=\columnwidth]{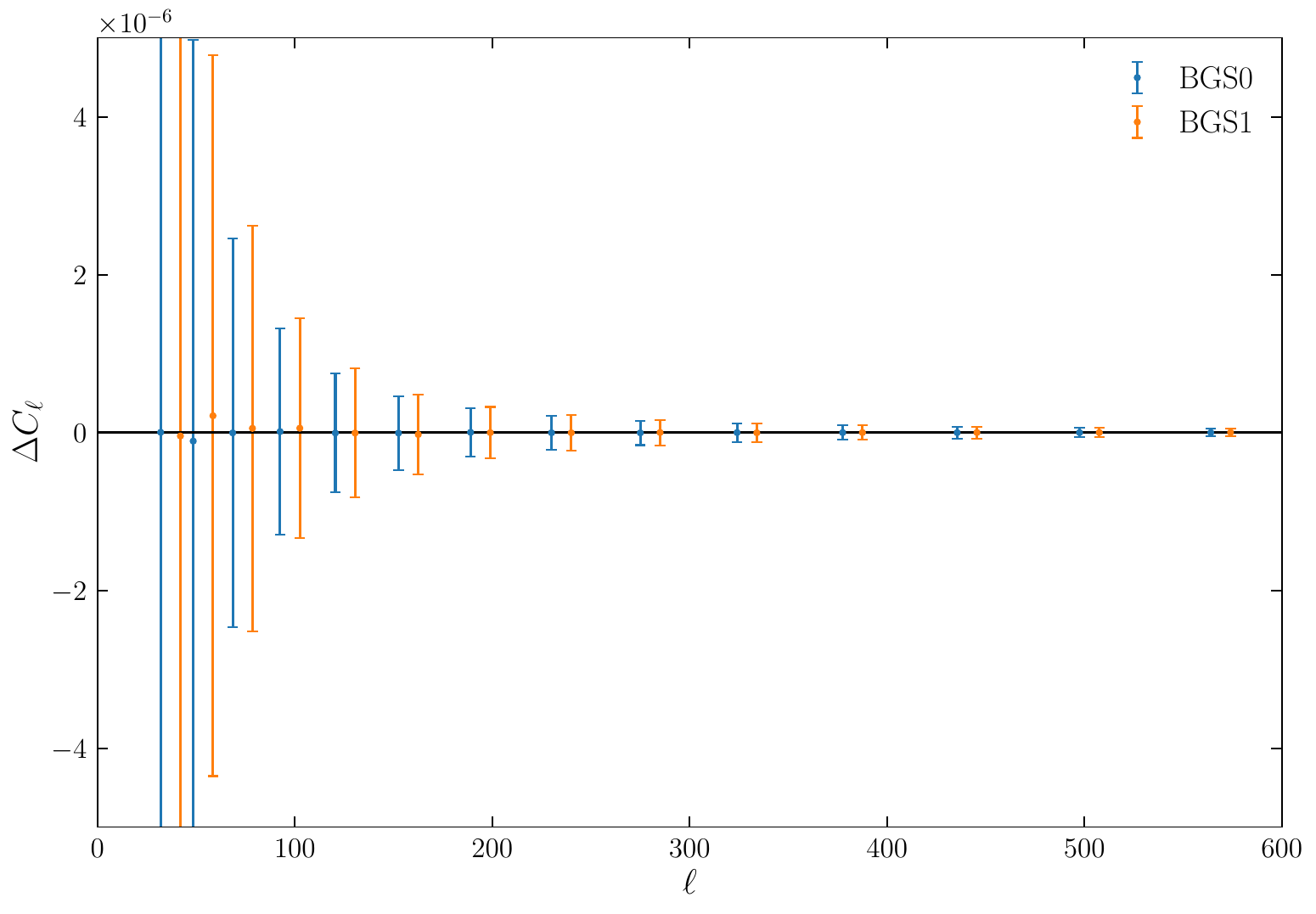}
    \caption{Difference between BGS auto-power spectra with and without applying linear systematic weights.}
    \label{fig:bgs_sys_cells}
\end{figure}

\begin{figure}
    \centering
    \includegraphics[width=\columnwidth]{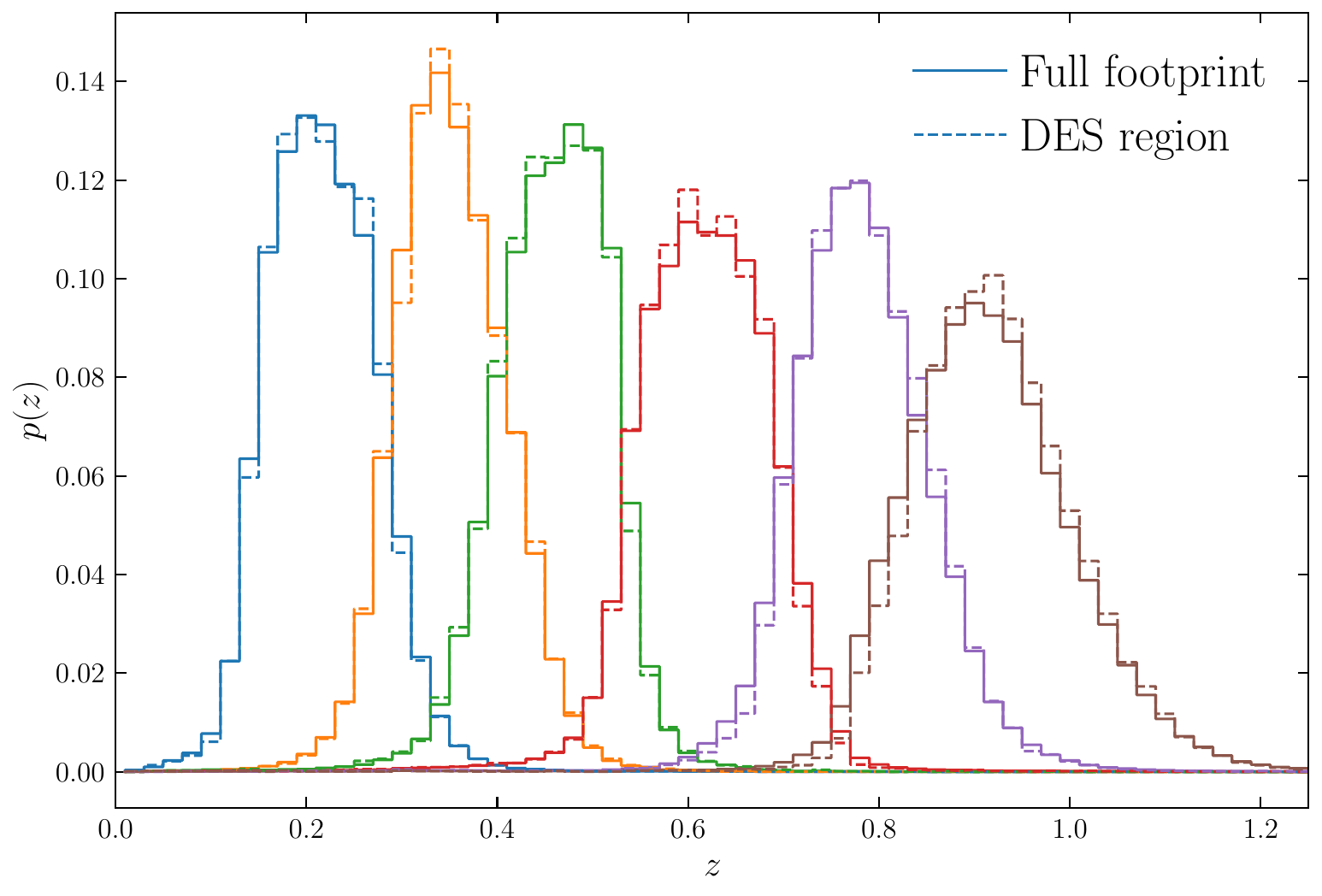}
    \caption{Comparison of the redshift distributions measured using the full Iron data set (solid) and only the overlap region with the DES Y3 footprint (dashed) for our all of our lens bins. The differences in the mean and standard deviation of the redshift distributions for these two footprints as well as differences in stellar contamination are listed in Table \ref{tab:des_v_full}.}
    \label{fig:nz_des_v_full}
\end{figure}

In this appendix we describe the systematics tests that we performed on the BGS and LRG redshift bins used in this work. Many of these systematics tests for the LRG samples are detailed in ref.~\cite{Zhou2023b}, and we repeat them for the BGS samples here. Firstly, we consider the trends in angular galaxy number density as a function of various survey properties, and potential contaminants. Figure \ref{fig:bgs_sys_trends} shows this for $E(B-V)$, the depth and psf size in each band used in the BGS sample selection, as well as stellar density as measured using GAIA. The blue line shows these trends for the first BGS bin, while the orange line shows the same for the second. In general, we see that these correlations are quite small in the regime where there are appreciable numbers of galaxies, as shown by the blue histograms. We fit a linear function to these trends, and use this to construct weights to remove them. The correlations between weighted galaxy number densities and these fields are shown by dashed lines. We note that there is very little change when applying these weights. 

In order to investigate the impact of the systematic weights that we have constructed, we examine the difference in $C^{\delta_g, \delta_g}_{\ell}$ measured with and without these systematic weights. Figure \ref{fig:bgs_sys_cells} shows this difference, compared to the statistical error on $C^{\delta_g, \delta_g}_{\ell}$ given by our fiducial covariance matrix. We see that any shift that is derived from applying weights for the BGS sample is well within these errors, and as such we have chosen to not apply systematics weights for this sample, since it is well known that such weights can potentially remove cosmological power, and there is no evidence that they are correcting for any detectable systematic trends in our data for this sample.

An additional concern that one might have about our analysis is that the redshift distributions that we use in our analysis are computed over the overlap region between DES Y3 and DESI Y1, which comprises just  of the 4143 sq. degrees used in our analysis. While the area used to measure our redshift distributions is large enough to have negligible statistical error, it is possible that the average $n(z)$ over the full footprint differs from the $n(z)$ measured over the overlap region due to varying survey conditions. In order to set a reasonable upper bound on this variation, we compare the $n(z)$ estimated for each of the samples used in this work measured over the overlap region between DESI Y1 and DES Y3 to that measured over the full DESI Y1 area. The BGS and LRG selections in the NGC of the DESI footprint use photometry from the Bok and Mayall telescopes, rather than DeCAM, and as such the $n(z)$'s could potentially differ significantly between the NGC and the overlap region if the selections were particularly sensitive to variations in photometry. 

These two sets of $n(z)$s are shown in Figure \ref{fig:nz_des_v_full}, and differences in these samples are further summarized in Table \ref{tab:des_v_full}. In general we see measure statistically significant, but small shifts in the mean and widths of the redshift distributions of our lenses. The largest shifts are observed in the last LRG bin, which is unsurprising given that it is comprised of the faintest galaxies, whose selection will be most influenced by changes in photometry. We propagate these differences to cosmology as shown in Figure \ref{fig:s8_consistency} for our fiducial lens bins, i.e., the BGS bins and first LRG bin and find these differences to be entirely negligible in terms of constraining power. As such, we conclude that this systematic is negligible at the constraining power of this analysis, although this and related concerns will need to be revisited should we combine these measurements with anisotropic power spectra of these samples measured with DESI spectroscopy.

\section{Fitting All Source and Lens Bin Combinations}
\begin{figure*}
    \centering
    \includegraphics[width=\textwidth]{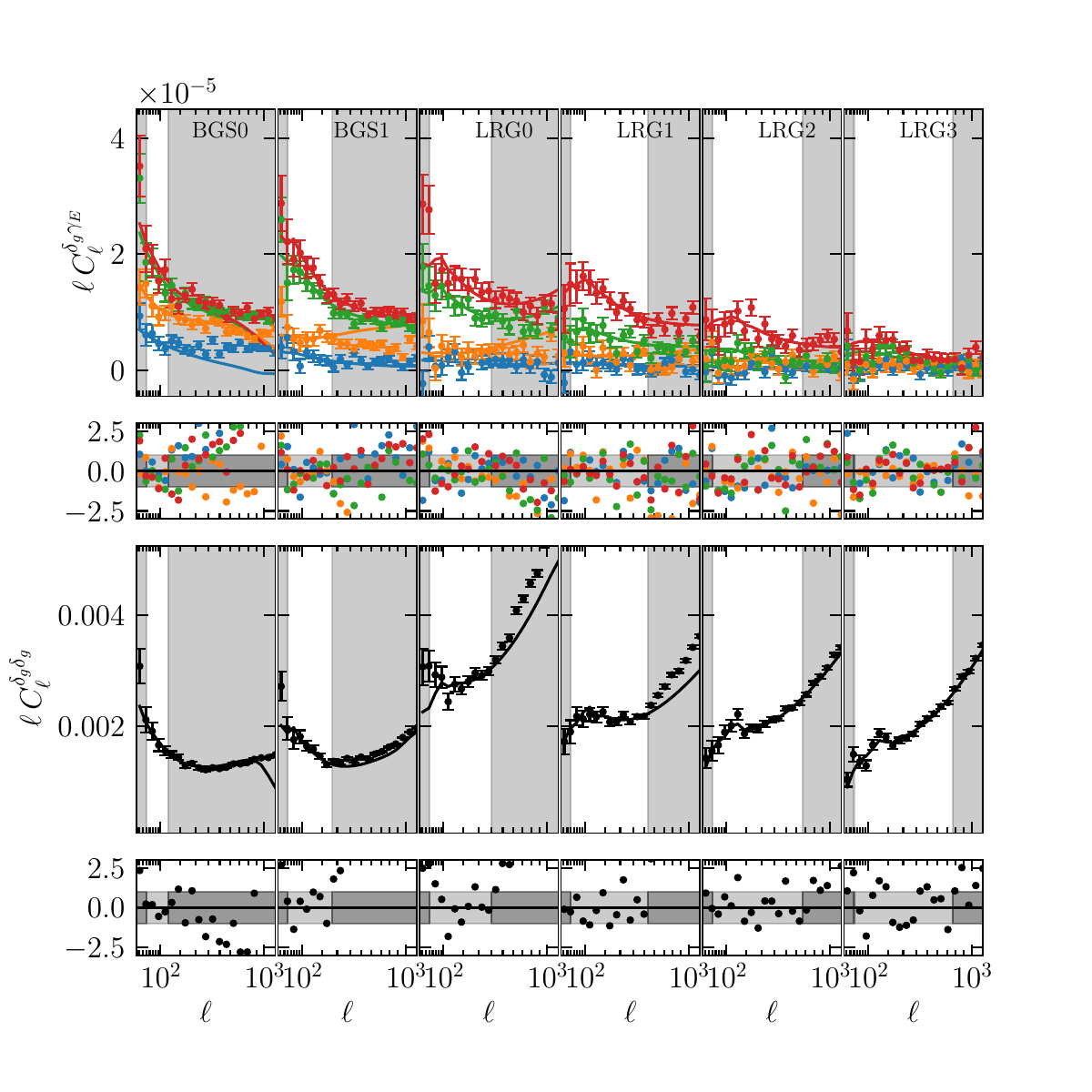} 
    \caption{Same as Figure~\ref{fig:bestfit_model}, but using our fiducial model and the full set of source and lens bins. We find $\chi^2=201.2$ for 280 data points using 178 free parameters, 132 of which are analytically marginalized over. Assuming again that IA, magnification and source uncertainities are prior-dominated and thus do not enter into the reduced $\chi^2$ calculation we obtain $\chi^2_{\rm red}=0.86$, though we note that this is likely an undercounting of the free parameters in the model.}
    \label{fig:bestfit_model_allbins}
\end{figure*}

For the results that fit all source and lens bin combinations, we use a covariance that takes as input the best fit of our fiducial model analyzing all of these bin combinations. This best-fit model is shown in Figure~\ref{fig:bestfit_model_allbins}.

\bibliography{main}
\end{document}